\documentclass[reprint,prd,nofootinbib,superscriptaddress]{revtex4}
\usepackage{amssymb,wrapfig,amsmath,hyperref,graphicx,epsfig,bm,color,verbatim,slashed,mathtools,xparse,adjustbox,dcolumn,float,relsize,tikz,cancel}
\usepackage{caption}
\usepackage{subcaption}
\usepackage{float}
\usepackage[compat=1.1.0]{tikz-feynman}
\usepackage{subcaption}
\usepackage[english]{babel}
\usepackage[utf8]{inputenc}
\usepackage[mathletters]{ucs}

\pdfoutput=1
\newcommand{\Steve}[1]{{#1}}

\newcommand{\func}{\operatorname}
\DeclareUnicodeCharacter{2212}{-}
\input{tcilatex}
\begin{document}

\title{Fermion mass hierarchies from vector-like families with an extended 2HDM and a possible explanation
for the electron and muon anomalous magnetic moments}
\author{A. E. C\'{a}rcamo Hern\'{a}ndez}
\email{antonio.carcamo@usm.cl}
\affiliation{Departamento de F\'{\i}sica, Universidad T\'{e}cnica Federico Santa Mar\'{\i}a,\\
 Casilla 110-V, Valpara\'{\i}so, Chile }

\author{S. F. King}
\email{king@soton.ac.uk}
\affiliation{School of Physics and
Astronomy, University of Southampton,\\
SO17 1BJ Southampton, United Kingdom }

\author{H. Lee}
\email{hl2n18@soton.ac.uk}
\affiliation{School of Physics and
Astronomy, University of Southampton,\\
SO17 1BJ Southampton, United Kingdom }

\date{\today }

\begin{abstract}
We study an extended 2 Higgs
doublet model (2HDM) in which the Standard Model (SM) Yukawa interactions are forbidden
due to a global $U(1)^\prime$ symmetry, but may arise via mixing with vector-like families. In this model, 
the hierarchical structure of Yukawa couplings of quarks and leptons in the SM
arises from the heavy masses of the fourth and fifth vector-like families. 
Within this model, we consider various non-standard contributions to the electron and muon anomalous magnetic moments.
We first consider the $W$ exchange at one-loop level, consistent
with the $\mu \rightarrow e \gamma$ constraint, and show that it yields a negligible contribution to both 
electron and muon anomalous magnetic moments. We then consider Higgs 
scalar exchange, together with vector-like leptons, at one-loop level and show that it is possible to
have non-standard contributions to the electron and muon anomalous magnetic moments within
the $1\sigma$ constraint of certain experiments.
We present some benchmark points for both the muon and the
electron anomalies, together with some numerical scans around these points,
which indicate the 
mass regions of the Higgs scalars of the 2HDM in this scenario.

\footnotesize
DOI:\href{https://doi.org/10.1103/PhysRevD.103.115024}{10.1103/PhysRevD.103.115024}
\normalsize
\end{abstract}

\maketitle


\section{INTRODUCTION}

\label{sec:Introduction}

The Standard Model (SM) has made many successful predictions for the phenomenology 
of both quark and lepton sectors with very high accuracy.
However there are long-established anomalies which are not addressed by 
the SM such as muon and electron anomalous magnetic moments
$a_{\mu }=\left( g-2 \right)_\mu/2, a_e=\left( g-2 \right)_e/2$. 
The muon
anomalous magnetic moment reported by the Brookhaven E821 experiment at BNL%
\cite{Bennett:2006fi} and the electron anomaly have confirmed $+3.5 \sigma$
and $-2.5 \sigma$ deviations from the SM, respectively. Detailed data analysis of the Standard Model predictions for the muon anomalous magnetic moment are provided in \cite{Hagiwara:2011af,Davier:2017zfy,Blum:2018mom,Keshavarzi:2018mgv,Aoyama:2020ynm}. The experimentally
observed values for the muon and electron anomalies at $1\sigma$ of
experimental error bars, respectively, read \footnote{It is worth mentioning that the experimental value of the anomalous magnetic moment of the electron is sensitive to the measurement of the fine-structure constant $\alpha$. The experimental value of $\Delta a_e=a_{e,exp}-a_e(\alpha_{\textit{Berkeley}})$ used in this work and given in Equation \ref{eqn:deltaamu_deltaae_at_1sigma} is obtained 
using $\alpha_{\textit{Berkeley}}$ from caesium recoil measurements by the Berkeley 2018 experiment \cite{Parker:2018vye}.
As this paper was being completed a different experiment \cite{Morel:2020dww} reported a result that implies 
$\Delta a_e = a_e^{\func{Exp}} - a_e^{\func{SM}} = \left( 0.48 \pm
0.30 \right) \times 10^{-12}$ which differs from the SM by $+1.6\sigma$.
The two experiments appear to be inconsistent with each other, and our results here are based on the earlier result in 
Equation \ref{eqn:deltaamu_deltaae_at_1sigma}.
}:

\begingroup
\begin{equation}
\begin{split}
\Delta a_\mu &= a_\mu^{\func{Exp}} - a_\mu^{\func{SM}} = \left( 26.1 \pm
8.0 \right) \times 10^{-10} \\
\Delta a_e &= a_e^{\func{Exp}} - a_e^{\func{SM}} = \left( -0.88 \pm
0.36 \right) \times 10^{-12}.  \label{eqn:deltaamu_deltaae_at_1sigma}
\end{split}%
\end{equation}
\endgroup

When trying to explain both anomalies to within $1\sigma $, a main
difficulty arises 
from the sign of each anomaly: the muon anomaly requires positive definite non-standard contributions, whereas the electron anomaly requires such contributions to contribute with a negative sign~ \cite{Morel:2020dww}.
Without loss of generality, the Feynman diagrams corresponding to the
contributions for the muon and electron anomalies take the same internal
structure at one-loop except from the fact that the external particles are
different. 
The similar structure 
of the one-loop level contributions to 
the muon and electron anomalous magnetic moments might be able to be
explained by the same new physics, but accounting for the relative negative sign is challenging. 
For example, considering the one-loop exchange of $W$ or $%
Z^{\prime }$ gauge bosons results 
in theoretical predictions for the muon and electron anomalies 
having the same sign.

In this paper we take the view that both anomalies should be
explained to $1\sigma $ using the same internal structure at the 
one-loop level by some new physics which is capable of accounting for the correct signs of the anomalies.
To explain the muon and electron anomalies, we focus on a well motivated model which is also capable 
of accounting for origin of Yukawa
couplings and hierarchies in the SM. 
The model we consider will account for the Yukawa coupling
constant for the top quark being nearly $1$ while that for the electron is around $10^{-6}$, as well as all the other fermion hierarchies
in between, as well as the neutrino masses and mixing.
In order to achieve this we shall 
introduce vector-like particles, which are charged under a 
global $U(1)^{\prime }$ symmetry. In a related previous work~\cite{King:2018fcg}, with a gauged $U(1)^{\prime }$ symmetry, the first
family of quarks and leptons remained massless when only one vector-like family is included.
Here we shall modify the model to include two vector-like families charged under a global $U(1)^{\prime }$ to allow 
also the first family to be
massive and avoid $Z'$ constraints. Then we shall apply the resulting model to the problem of muon and electron anomalous magnetic moments.
The considered model is based on 
a 2 Higgs doublet model (2HDM) extension of the SM, supplemented 
by a global $U(1)^\prime$ symmetry, where the particle spectrum is enlarged by
the inclusion of 
two 
vector-like fermion families, as well as one singlet Higgs to break the $U(1)^\prime$ symmetry \footnote{An example of a multiHiggs doublet model that uses a flavor dependent global $U(1)^{\prime }$ symmetry to explain the SM charged fermion mass hierarchy by hierarchies of the vacuum expectation values of the Higgs doublets is provided in \cite{CentellesChulia:2020bnf}}. The SM Yukawa interactions are forbidden, but the Yukawa interactions with vector-like families charged under the $U(1)^\prime$ symmetry are allowed. Once the flavon develops a vev and the heavy vector-like fermions are integrated out, 
the effective SM Yukawa interactions are generated, as indicated in 
Figure \ref%
{fig:mass_insertion_diagrams}. Furthermore, this model also highlights the
shape of the 2HDM model type II, since in our proposed model, one Higgs
doublet (which in the alignment limit corresponds to the SM Higgs doublet)
couples with the up type quarks whereas the other one features Yukawa
interactions with down type quarks and SM charged leptons. 
Regarding the neutrino sector, since we consider 
the SM neutrinos as Majorana particles, we have that this sector 
requires another approach relying on the inclusion of 
a new five dimensional Weinberg-like operator, which is allowed in this model
and which requires 
both SM Higgs doublets to be present, namely the so called Type Ib seesaw model~\cite{Hernandez-Garcia:2019uof}.
\\~\\
We shall show that the heavy vector-like leptons are useful and necessary to explain the
anomalous electron and muon magnetic moment deviations from the SM, of magnitude and opposite
signs given in Equation \ref{eqn:deltaamu_deltaae_at_1sigma}. A study of such $g-2$ anomalies in terms
of New Physics and a possible UV complete explanation via vector-like
leptons was performed in \cite{Crivellin:2018qmi}, although the model presented here is quite different, since our model is 
motivated by the requirement of accounting also for the fermion mass hierarchies. Other theories with extended
symmetries and particle spectrum have also been proposed to find an explanation
for the muon and electron anomalous magnetic moments \cite%
{Appelquist:2004mn,Giudice:2012ms,Freitas:2014pua,Falkowski:2018dsl,Crivellin:2018qmi,Allanach:2015gkd,Chen:2016dip,Raby:2017igl,Chiang:2017tai,Chen:2017hir,Davoudiasl:2018fbb,Liu:2018xkx,CarcamoHernandez:2019xkb,Nomura:2019btk,Kawamura:2019rth,Bauer:2019gfk,Han:2018znu,Dutta:2018fge,Badziak:2019gaf,Endo:2019bcj,Hiller:2019mou,CarcamoHernandez:2019ydc,CarcamoHernandez:2019lhv,Kawamura:2019hxp,Cornella:2019uxs,CarcamoHernandez:2020pxw,Arbelaez:2020rbq,Hiller:2020fbu,Jana:2020pxx,deJesus:2020ngn,deJesus:2020upp,Hati:2020fzp,Botella:2020xzf,Dorsner:2020aaz,Calibbi:2020emz,Dinh:2020pqn,Jana:2020joi,Chun:2020uzw,Chua:2020dya,Daikoku:2020nhr,Banerjee:2020zvi,Chen:2020jvl,Bigaran:2020jil,Kawamura:2020qxo,Endo:2020mev,Chakrabarty:2020jro,Li:2020dbg}%
. 
In the following we provide a brief comparison of our model to other works, starting
with the model proposed in \cite{Chun:2020uzw}
where vector-like leptons are also present. The model of \cite{Chun:2020uzw} corresponds to an extended type X lepton specific 2HDM model of \cite{Chun:2020uzw} having a $Z_2$ discrete symmetry under which one of the scalar doublets and the leptonic fields are charged. In such model the vector-like leptons induces a one-loop level contribution to the electron anomalous magnetic moment whereas the muon anomalous magnetic moment is generated at two-loop via the exchange of a light pseudoscalar. On the other hand, in our proposed model a spontaneously broken global $U(1)^\prime$ symmetry is considered instead of the $Z_2$ symmetry and the vector-like leptons generate one-loop level contributions to the muon and electron anomalous magnetic moments and at the same type produce the SM charged lepton masses, thus providing a connection of the charged lepton mass generation mechanism and the $g-2$ anomalies,
which is not given in the model of \cite{Chun:2020uzw}. It is also worth emphasising that our model is  
very different from other models proposed in the literature based on the Universal Seesaw mechanism
\cite{Davidson:1987mh,Davidson:1987mi,Berezhiani:1991ds,Sogami:1991yq,Gu:2010zv,Alvarado:2012xi,Hernandez:2013mcf,Kawasaki:2013apa,Mohapatra:2014qva,Dev:2015vjd,Borah:2017inr,Patra:2017gak,Babu:2018vrl,deMedeirosVarzielas:2018bcy,CarcamoHernandez:2018aon,CarcamoHernandez:2019vih,CarcamoHernandez:2019pmy,Hernandez:2021uxx}.
Universal Seesaw models are typically based on the left-right symmetric model with electroweak singlet fermions only,
while our vector-like fermions involves complete families, including electroweak doublets which are typically the lightest ones.
Some examples of theories relying on the Universal Seesaw mechanism to explain the SM charged fermion mass hierarchy are provided in 
~\cite{Davidson:1987mh,Davidson:1987mi,Berezhiani:1991ds,Sogami:1991yq,Gu:2010zv,Alvarado:2012xi,Hernandez:2013mcf,Kawasaki:2013apa,Mohapatra:2014qva,Dev:2015vjd,Borah:2017inr,Patra:2017gak,Babu:2018vrl,deMedeirosVarzielas:2018bcy,CarcamoHernandez:2018aon,CarcamoHernandez:2019vih,CarcamoHernandez:2019pmy,Hernandez:2021uxx}.\newline

In the approach followed in this paper the large third family quark and lepton Yukawa couplings are effectively generated via mixing with a vector-like fourth family of electroweak doublet fermions, which are assumed to be relatively light, with masses around the TeV scale. The smallness of the second family quark and lepton Yukawa couplings is due to their coupling to heavier vector-like fourth family electroweak singlet fermions. Similar considerations apply to the lightest first family quarks and leptons which couple to heavy fifth family vector-like fermions. 
It may seem that the problem of the hierarchies of SM fermions is not solved but simply reparameterised in terms of unknown vector-like fermion masses. However, there are four advantages to this approach. Firstly, the approach is dynamical, since the vector-like masses are new physical quantities which could in principle be determined by a future theory. Secondly, it has experimental consequences, since the new vector-like fermions can be discovered either directly, or (as in this paper) indirectly via their loop contributions. Thirdly, this approach can also account for small quark mixing angles~\cite{King:2018fcg}, as well as large lepton mixing angles via the type Ib seesaw mechanism~\cite{Hernandez-Garcia:2019uof}. Fourthly, the effective Yukawa couplings are proportional to a product of two other dimensionless couplings, so a small hierarchy in those couplings can give a quadratically larger hierarchy in the effective couplings. For all these reasons, the approach we follow in this paper is both well motivated and interesting.

Returning to 
our proposed model framework, we first consider the contribution of $W$ boson exchange with neutrinos to the
electron and muon anomalous magnetic moments at the one-loop level. Since this model involves the vector-like
neutrinos, the sensitivity of the branching ratio of $\mu \rightarrow e\gamma $
decay can be enhanced with respect to the observable level and the muon and
electron anomalous magnetic moments are studied while keeping the $\mu
\rightarrow e\gamma $ constraint. As a result, we find that the impact of
our predictions with $W$ exchange at one-loop level is 
negligible when compared to their experimental bound.
We then consider the contributions from the 2HDM scalar exchange.
To study the implications of the one-loop level 
scalar exchange in the muon and electron anomalous magnetic moments, we
first construct a scalar potential and derive the mass squared matrix for
CP-even, CP-odd and charged Higgses assuming there is no mixing between the
SM Higgs $h$ and two non-SM physical scalars $H_{1,2}$. A 
diagonal Yukawa matrix for charged leptons implies 
the absence of mixing between charged leptons, resulting in vanishing 
branching ratio for the 
$\mu \rightarrow e\gamma $ decay, which in turn leads to a fulfillment of
the charged lepton flavor violating constraints 
in this scenario. In such a framework we show that both anomalies can successfully explain both 
anomalies, including their opposite signs, 
at the $1\sigma $ level.
We present some benchmark points for both the muon and the
electron anomalies, together with some numerical scans around these points,
which indicate the 
mass regions of the Higgs scalars of the 2HDM in this scenario.
We also provide some analytic arguments to augment the numerical results.

The layout of the remainder of the paper is as follows.
In Section~\ref{II} we discuss the origin of Yukawa couplings from a fourth and fifth vector-like family, within a mass insertion formalism.
In Section~\ref{III} we construct the effective Yukawa matrices using a more detailed mixing formalism which goes beyond the 
mass insertion formalism.
In Section~\ref{IV} we consider $W$ exchange contributions to 
$\left( g-2 \right)_\protect\mu, \left( g-2 \right)_e$ and $\func{BR}\left(\protect\mu \rightarrow e \protect\gamma \right)$
based on the type Ib seesaw mechanism within our model and show that the contributions are too small.
In Section~\ref{sec:Analytic_arguments_muon_electron_g2_scalars} we turn to Higgs scalar exchange contributions to
$\left( g-2 \right)_\protect\mu, \left( g-2
\right)_e$ and $\func{BR}\left(\protect\mu \rightarrow e \protect\gamma \right)$, focussing on analytical formulae.
Then in Section~\ref{sec:Numerical_analysis_of_scalars} we give a full numerical analysis of such contributions, showing that they can 
successfully explain the anomalies, presenting some benchmark points for both the muon and the
electron anomalies, together with some numerical scans around these points,
which indicate the 
mass regions of the Higgs scalars of the 2HDM in this scenario.
Section~\ref{sec:Conclusion} concludes the main body of the paper.
Appendix~\ref{A} provides a discussion of the quark mass matrices in two bases.
Appendix~\ref{B} includes a brief discussion of heavy scalar production at a proton-proton collider.

\section{THE ORIGIN OF YUKAWA COUPLINGS FROM A FOURTH AND FIFTH VECTOR-LIKE
FAMILY}
\label{II}

\label{sec:The_origin_of_Yukawa_coupling}

\begin{figure}[tbp]
\begin{subfigure}{0.48\textwidth}
	\includegraphics[width=1.0\textwidth]{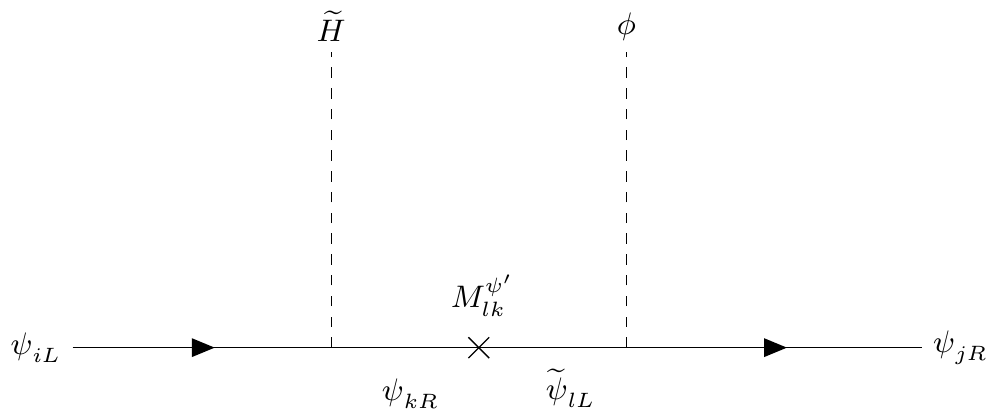}
\end{subfigure} \hspace{0.1cm} 
\begin{subfigure}{0.48\textwidth}
	\includegraphics[width=1.0\textwidth]{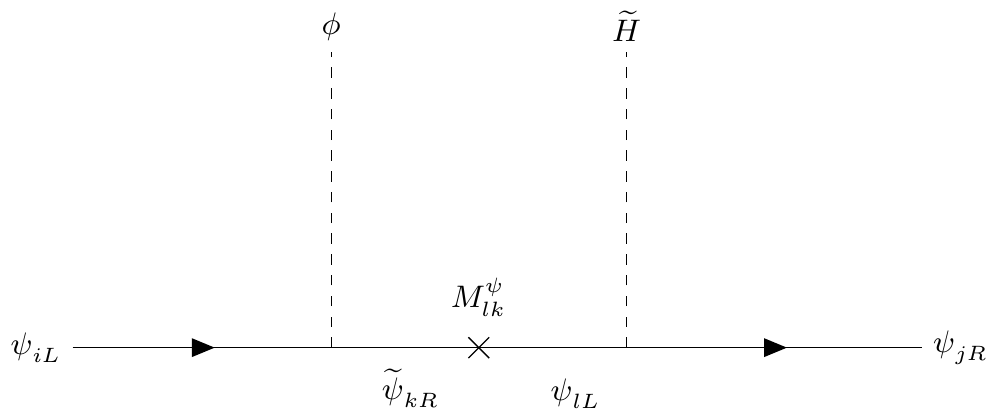}
\end{subfigure}
\caption{Diagrams in this model which lead to the effective Yukawa
interactions, where $\protect\psi,\protect\psi^\prime = Q,u,d,L,e$(neutrinos
will be treated separately) $i,j=1,2,3$, $k,l=4,5$, $M_{lk}$ is vector-like
mass and $\widetilde{H} = i\protect\sigma_2 H^*, H = H_{u,d}$}
\label{fig:mass_insertion_diagrams}
\end{figure}

We start by asking a question: what is the origin of the SM Yukawa
couplings? 
In addressing such question, we assume that 
the SM Yukawa Lagrangian is the low energy limit of an 
extended theory with enlarged symmetry and particle spectrum, and arises
after the spontanous breaking of an $U(1)^{\prime }$ global symmetry at an
energy scale 
as low as $\func{TeV}$. Therefore, understanding the origin of the Yukawa
interaction naturally leads to the presence of another Higgses whose masses
are higher than 
the mass of the SM Higgs. Furthermore, the SM Yukawa interactions are
forbidden by the global $U(1)^{\prime }$ symmetry, however the Yukawa
interaction with the vector-like particles are allowed. With these
considerations in place, the possible diagrams generating 
the Yukawa interactions can be drawn as indicated in 
Figure \ref{fig:mass_insertion_diagrams}.

There are two key features in Figure \ref{fig:mass_insertion_diagrams}, one
of which is the presence of the assumed flavon $\phi$ and the other one is
the vector-like mass $M$. Once the flavon $\phi$ develops its vev, the
effective Yukawa interactions $\overline{\psi}_{iL} \widetilde{H} \psi_{jR}$ are generated with a
coupling constant proportional to 
$\left\langle \phi \right\rangle/M$, \Steve{assumed to be less than unity}, which appears in front of the usual
Yukawa constant. The proportional factor $\left\langle \phi \right\rangle/M$
plays a crucial role 
in explaining why one Yukawa constant can be relatively smaller or bigger
than the other ones 
since the magnitude of each Yukawa constant is accompanied by the mass of
the vector-like particles. The effective Lagrangian in this diagram reads in the mass insertion formalism:

\begin{equation}
\mathcal{L}_{\func{eff}}^{\func{Yukawa}} =y_{ik}^\psi (M_{\psi^\prime}^{-1})_{kl} {x_{lj}^{\psi^\prime}
\left\langle \phi \right\rangle} \overline{%
\psi}_{iL} \widetilde{H} \psi_{jR} + {x_{ik}^{\psi} \left\langle \phi
\right\rangle}(M_{\psi}^{-1} )_{kl}  y_{lj}^\psi  \overline{\psi}_{iL} \widetilde{H}
\psi_{jR} + \func{h.c.}  \label{eqn:the_effective_Yukawa_Lagrangian}
\end{equation}

where $\psi,\psi^\prime = Q,u,d,L,e$ (neutrinos will be treated separately)
and $x$ is a Yukawa constant in the interaction with $\phi$ and $y$ is in
the interaction with $\widetilde{H}$ as per Figure \ref%
{fig:mass_insertion_diagrams}. Throughout this work, we take a view that the
Yukawa constant $y$ can be ideally of order unity while the $x$ is
small compared to the $y$. We shall also use a mixing formalism rather than the mass insertion formalism.

\subsection{The model with $U(1)^{\prime }$ global symmetry}

For an analysis of the phenomenology described above, we extend the SM fermion sector by adding two vector-like fermions, the SM gauge symmetry by including the global $U(1)^\prime$ symmetry and the scalar sector of the 2HDM model is enlarged by considering a gauge scalar singlet, whose VEV triggers the spontaneous breaking of the $U(1)^\prime$ symmetry. 
The scalar sector of the model is composed of 
by two $SU(2)$ doublet scalars 
$H_{u,d}$ and one flavon $\phi$. 
 Our extended 2HDM with enlarged particle spectrum and symmetries has the interesting feature that 
the SM Yukawa interactions are forbidden due to the global $U(1)^\prime$ symmetry
whereas the Yukawa interactions of SM fermions with vector-like families are allowed. Furthermore, such vector-like families have mass terms which are allowed by the symmetry. Thus, the SM charged fermions masses are generated from a Universal Seesaw mechanism mediated by heavy vector-like fermions.
 Unlike the $U(1)^\prime $ model proposed in \cite%
{King:2017anf}, we assume that the $U(1)^\prime $ symmetry is global instead
of local. This allows us more flexibility in the allowed range for the scale where the $U(1)^\prime $ symmetry is broken.
On top of that, the up-type
quarks feature Yukawa interaction with the up-type Higgs 
whereas the down-type ones interact with down-type Higgs. In this BSM
model, the SM particles are neutral under the $U(1)^\prime$ symmetry, while the vector-like particles and all other scalars are charged under the symmetry. The particle content and symmetries of the model are 
shown in Table \ref{tab:model_content}.

\begin{table}[]
\centering\renewcommand{\arraystretch}{1.3} 
\begin{tabular}{|c||c|c|c|c|c||c|c|c|c|c|c||c|c|c|c|c|c||c||c|c|}
\hline
Field & $Q_{iL}$ & $u_{iR}$ & $d_{iR}$ & $L_{iL}$ & $e_{iR}$ & $Q_{kL}$ & $%
u_{kR}$ & $d_{kR}$ & $L_{kL}$ & $e_{kR} $ & $\nu_{kR}$ & $\widetilde{Q}_{kR}$
& $\widetilde{u}_{kL}$ & $\widetilde{d}_{kL}$ & $\widetilde{L}_{kR}$ & $%
\widetilde{e}_{kL}$ & $\widetilde{\nu}_{kR}$ & $\phi$ & $H_u$ & $H_d$ \\ 
\hline\hline
$\func{SU(3)_C}$ & $\mathbf{3}$ & $\mathbf{3}$ & $\mathbf{3}$ & $\mathbf{1}$
& $\mathbf{1}$ & $\mathbf{3}$ & $\mathbf{3}$ & $\mathbf{3}$ & $\mathbf{1}$ & 
$\mathbf{1}$ & $\mathbf{1}$ & $\mathbf{3}$ & $\mathbf{3}$ & $\mathbf{3}$ & $%
\mathbf{1}$ & $\mathbf{1}$ & $\mathbf{1}$ & $\mathbf{1}$ & $\mathbf{1}$ & $%
\mathbf{1}$ \\ \hline
$\func{SU(2)_L}$ & $\mathbf{2}$ & $\mathbf{1}$ & $\mathbf{1}$ & $\mathbf{2}$
& $\mathbf{1}$ & $\mathbf{2}$ & $\mathbf{1}$ & $\mathbf{1}$ & $\mathbf{2}$ & 
$\mathbf{1}$ & $\mathbf{1}$ & $\mathbf{2}$ & $\mathbf{1}$ & $\mathbf{1}$ & $%
\mathbf{2}$ & $\mathbf{1}$ & $\mathbf{1}$ & $\mathbf{1}$ & $\mathbf{2}$ & $%
\mathbf{2}$ \\ \hline
$\func{U(1)_Y}$ & $\frac{1}{6}$ & $\frac{2}{3}$ & $-\frac{1}{3}$ & $-\frac{1%
}{2}$ & $1$ & $\frac{1}{6}$ & $\frac{2}{3}$ & $-\frac{1}{3}$ & $-\frac{1}{2}$
& $-1$ & $0$ & $\frac{1}{6}$ & $\frac{2}{3}$ & $-\frac{1}{3}$ & $-\frac{1}{2}
$ & $-1$ & $0$ & $0$ & $\frac{1}{2}$ & $-\frac{1}{2}$ \\ \hline
$\func{U(1)^\prime}$ & $0$ & $0$ & $0$ & $0$ & $0$ & $1$ & $-1$ & $-1$ & $1$
& $-1$ & $-1$ & $1$ & $-1$ & $-1$ & $1$ & $-1$ & $-1$ & $1$ & $-1$ & $-1$ \\ 
\hline
\end{tabular}%
\caption{This model is an extended 2HDM by the global $U(1)^\prime$ symmetry
with two vector-like families plus one flavon and reflects the property that the SM Yukawa interactions are forbidden. All SM particles $\protect\psi_i(i=1,2,3)$
are neutral under the $U(1)^\prime$ symmetry and the right neutrinos $%
\protect\nu_{iR}$ are not considered. Notice that this model involves two
right-handed vector-like neutrinos $\protect\nu_{kR}, \widetilde{\protect\nu}%
_{kR}$. The SM particles are extended by two vector-like families where $%
k=4,5$ and two SM Higgses $H_{u,d}$ are charged negatively under $%
U(1)^\prime $ to forbid the renormalizable SM Yukawa interactions. The flavon
field $\protect\phi$ plays a role of braking the $U(1)^\prime$ symmetry at $%
\func{TeV}$ scale.}
\label{tab:model_content}
\end{table}
The right-handed neutrinos $\nu_{iR}$ are absent in this model since we treat
the left-handed neutrinos in the lepton doublet as Majorana particles and
they are only extended by vector-like neutrinos. The vector-like particles
and their partners have exact opposite charge to each other under the
extended gauge symmetry to cancel out chiral anomaly. Lastly, the SM Higgses 
$H_{u,d}$ are negatively charged under the $U(1)^\prime$ symmetry to forbid
the renormalizable SM Yukawa interactions. 

\subsection{mass insertion formalism}

The renormalizable Yukawa interactions and mass terms 
for both up and down quark sectors read: 
\begin{equation}
\begin{split}
\mathcal{L}_{q}^{\func{Yukawa+Mass}} &= y_{ik}^{u} \overline{Q}_{iL} 
\widetilde{H}_u u_{kR} + x_{ki}^{u} \phi \overline{\widetilde{u}}_{kL}
u_{iR} + x_{ik}^Q \phi \overline{Q}_{iL} \widetilde{Q}_{kR} + y_{ki}^u 
\overline{Q}_{kL} \widetilde{H}_u u_{iR} \\
&+ y_{ik}^{d} \overline{Q}_{iL} \widetilde{H}_d d_{kR} + x_{ki}^{d} \phi 
\overline{\widetilde{d}}_{kL} d_{iR} + y_{ki}^d \overline{Q}_{kL} \widetilde{%
H}_d d_{iR} \\
&+ M_{kl}^{u} \overline{\widetilde{u}}_{lL} u_{kR} + M_{kl}^{d} \overline{%
\widetilde{d}}_{lL} d_{kR} + M_{kl}^Q \overline{Q}_{kL} \widetilde{Q}_{lR} + 
\func{h.c.}  \label{eqn:general_Quark_Yukawa__Mass_Lagrangian}
\end{split}%
\end{equation}

where $i,j=1,2,3$, $k,l=4,5$ and $\widetilde{H}=i\sigma _{2}H^{\ast }$.
The possible diagrams contributing to the low energy quark Yukawa interaction
are given in Figure \ref{fig:diagrams_quark_mass_insertion}:

\begin{figure}[]
\centering
\begin{subfigure}{0.48\textwidth}
	\includegraphics[width=1.0\textwidth]{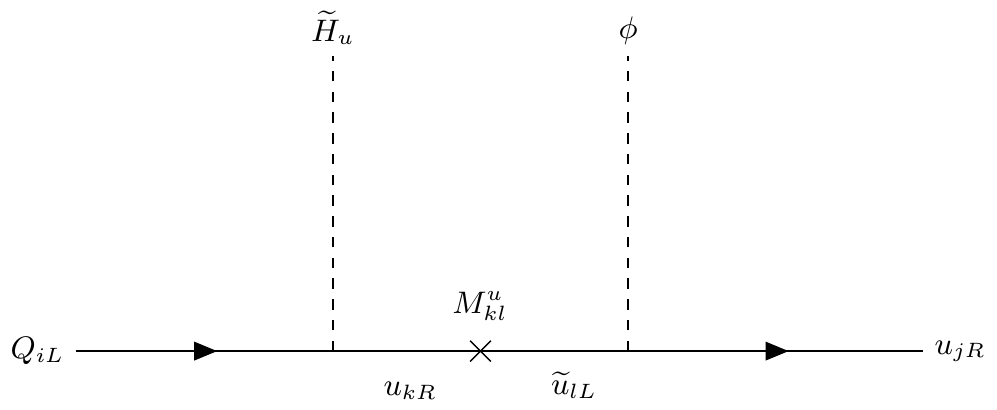}
\end{subfigure} \hspace{0.1cm} 
\begin{subfigure}{0.48\textwidth}
	\includegraphics[width=1.0\textwidth]{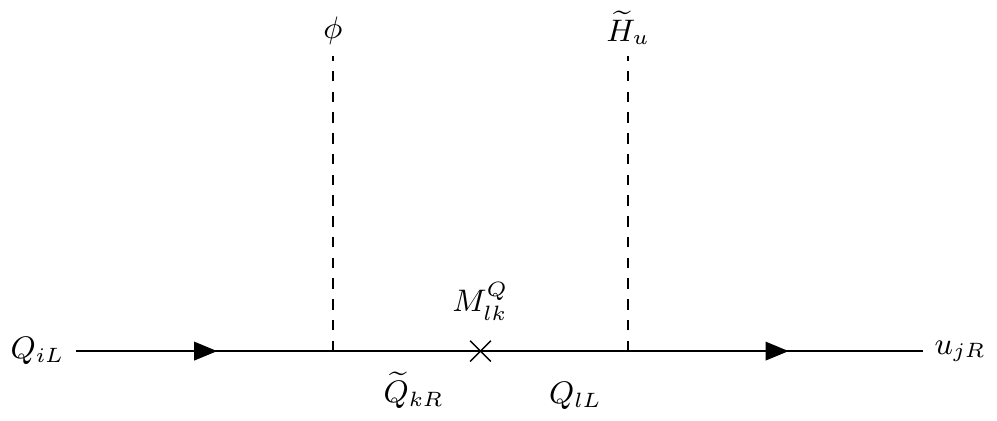}
\end{subfigure}
\begin{subfigure}{0.48\textwidth}
	\includegraphics[width=1.0\textwidth]{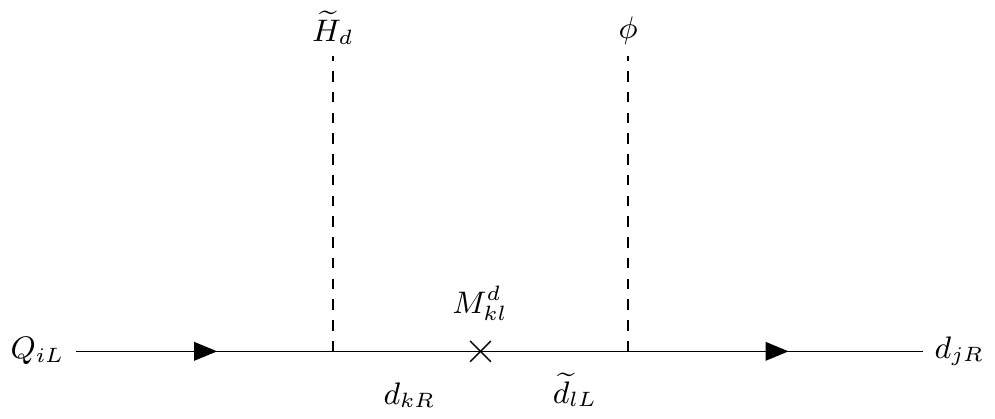}
\end{subfigure} \hspace{0.1cm} 
\begin{subfigure}{0.48\textwidth}
	\includegraphics[width=1.0\textwidth]{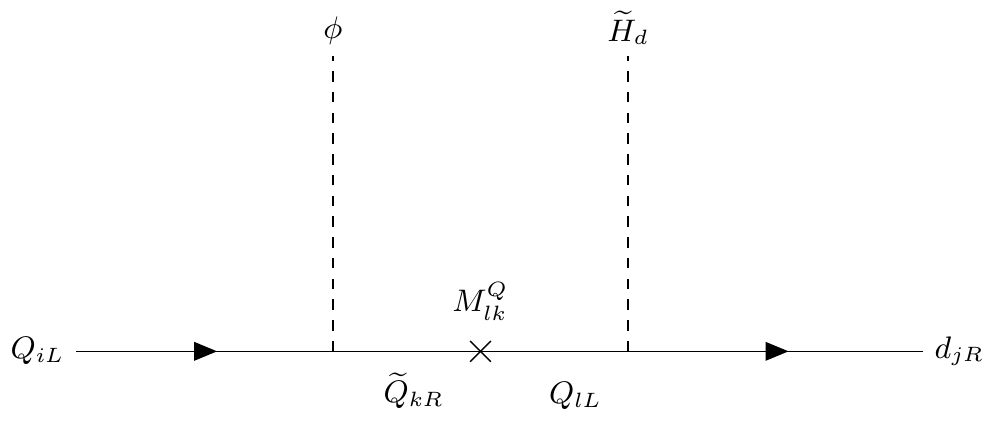}
\end{subfigure}
\caption{Diagrams in this model which lead to the effective Yukawa
interactions for the up quark sector(two above diagrams) and the down quark
sector(two below diagrams) in mass insertion formalism, where $i,j=1,2,3$
and $k,l=4,5$ and $M_{lk}$ is vector-like mass.}
\label{fig:diagrams_quark_mass_insertion}
\end{figure}

The above two diagrams correspond to the up-type quark sector 
whereas the below two diagrams correespond 
to the down-type quark sector. The model under consideration is an extended 2HDM where the 
up-type Higgs $H_u$ is relevant for the up-type quark sector whereas the%
down-type Higgs $H_d$ is suitable for the down-type quark and charged lepton sectors. Like in the quark sector, the Yukawa interactions
and mass terms 
for charged leptons can be written in a similar way:

\begin{equation}
\begin{split}
\mathcal{L}_{e}^{\func{Yukawa+Mass}} &= y_{ik}^{e} \overline{L}_{iL} 
\widetilde{H}_{d} e_{kR} + x_{ki}^{e} \phi \overline{\widetilde{e}}_{kL}
e_{iR} + x_{ik}^L \phi \overline{L}_{iL} \widetilde{L}_{kR} + y_{ki}^e 
\overline{L}_{kL} \widetilde{H}_{d} e_{iR} + M_{kl}^{e} \overline{\widetilde{%
e}}_{lL} e_{kR} + M_{kl}^L \overline{L}_{kL} \widetilde{L}_{lR} + \func{h.c.}
\end{split}
\label{eqn:general_charged_lepton_Yukawa_Mass_Lagrangian}
\end{equation}

Then, the possible diagrams giving rise to the 
charged lepton Yukawa interactions are shown in Figure \ref%
{fig:diagrams_charged_leptons_mass_insertion}: 
\begin{figure}[tbph]
\centering
\begin{subfigure}{0.48\textwidth}
	\includegraphics[width=1.0\textwidth]{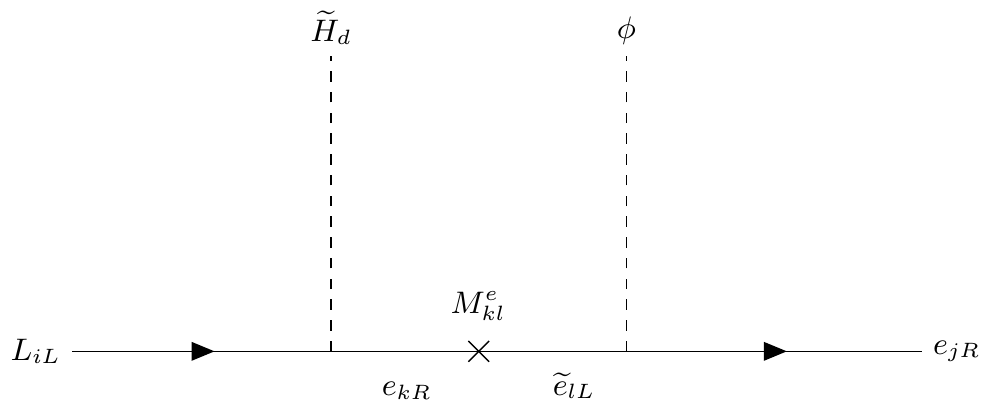}
\end{subfigure} \hspace{0.1cm} 
\begin{subfigure}{0.48\textwidth}
	\includegraphics[width=1.0\textwidth]{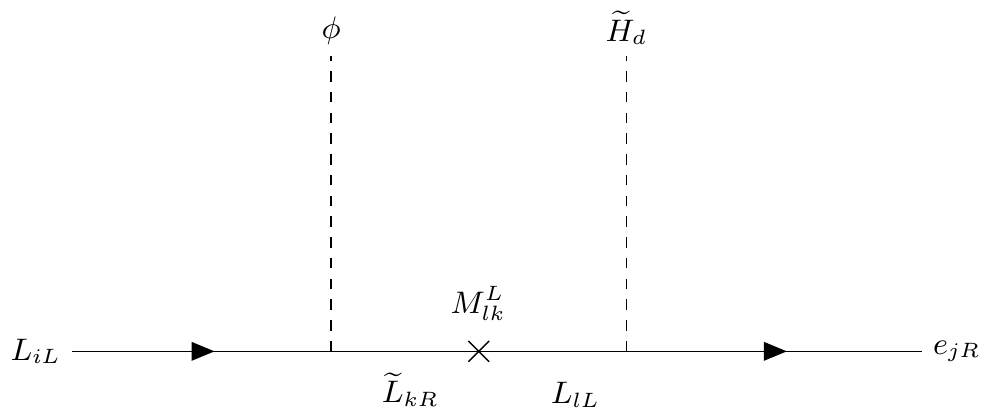}
\end{subfigure}
\caption{Diagrams in this model which lead to the effective Yukawa
interactions for the charged lepton sector in mass insertion formalism,
where $i,j=1,2,3$ and $k,l=4,5$ and $M_{lk}$ is vector-like mass.}
\label{fig:diagrams_charged_leptons_mass_insertion}
\end{figure}

As for the neutrinos, its behaviour is different as compared to the quarks or
charged leptons since there exists only Majorana neutrinos in this model so
initial and final neutrinos in mass insertion formalism diagrams must be same. The
Yukawa interactions and mass terms for the neutrino sector 
are given by:

\begin{equation}
\begin{split}
\mathcal{L}_{\nu}^{\func{Yukawa+Mass}} = y_{ik}^{\nu} \overline{L}_{iL} 
\widetilde{H}_u \nu_{kR} + x_{ik}^L \overline{L}_{iL} H_d \overline{%
\widetilde{\nu}}_{kR} + M_{kl}^{M} \overline{\widetilde{\nu}}_{lR} \nu_{kR}
+ \func{h.c.}
\end{split}
\label{eqn:general_neutrinos_Yukawa_Mass_Lagrangian}
\end{equation}

\begin{figure}[tbph]
\centering
\begin{subfigure}{0.48\textwidth}
	\includegraphics[width=1.0\textwidth]{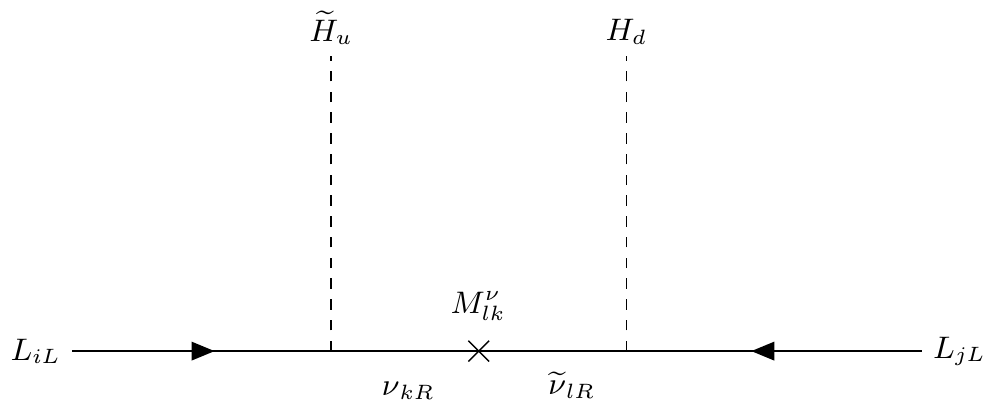}
\end{subfigure}
\caption{Type Ib seesaw diagram~\cite{Hernandez-Garcia:2019uof} which leads to the effective Yukawa
interactions for the Majorana neutrinos in mass insertion formalism,
where $i,j=1,2,3$ and $k,l=4,5$ and $M_{lk}$ is vector-like mass.}
\label{fig:diagrams_neutrinos_mass_insertion}
\end{figure}

Here, one important feature in Equation \ref%
{eqn:general_neutrinos_Yukawa_Mass_Lagrangian} is the presence of the 
vector-like mass $M$. From the two Yukawa interactions in Equation \ref%
{eqn:general_neutrinos_Yukawa_Mass_Lagrangian}, it follows that both $\nu_R$
and $\widetilde{\nu}_R $ have a lepton number $+1$ and they are different
particles. And then taking a look at the vector-like mass term in Equation %
\ref{eqn:general_neutrinos_Yukawa_Mass_Lagrangian}, it can be confirmed that
the vector-like mass is not a strict Majorana mass because $\nu_R$ and $%
\widetilde{\nu}_R$ are different particles but plays a role of Majorana mass
since the mass term violates the lepton number conservation. The
corresponding diagram for the 
neutrino sector in the mass insertion formalism is given in Figure \ref{fig:diagrams_neutrinos_mass_insertion}.
However for our calculations we use a mixing formalism (see next section).

The operator $ \overline{L}_i  \overline{L}_j \widetilde{H}_u H_d$ resulting from Figure \ref%
{fig:diagrams_neutrinos_mass_insertion} gives rise to the so 
called type Ib seesaw mechanism~\cite{Hernandez-Garcia:2019uof}
which differs from the usual 
type Ia seesaw mechanism corresponding to the Weinberg operator 
$ \overline{L}_i  \overline{L}_j \widetilde{H}_u\widetilde{H}_u$ and will be discussed later in detail.

\section{EFFECTIVE YUKAWA MATRICES USING A MIXING FORMALISM}
\label{III}

\label{sec:higgs_and_flavour_mixing}

As seen from Equation \ref{eqn:the_effective_Yukawa_Lagrangian}, we need to
mix Higgses with the 
flavon to generate the 
effective Yukawa Lagrangian required to produce the SM fermion mass
hierarchy. Since there is no an extra symmetry or constraint to keep the
mixing between Higgses and flavon from taking place, it is natural to assume
their mixing. 

\subsection{The $7\times7$ matrix}

\label{subsec:the_77_matrix}

Consider the $7\times 7$ mass matrix for Dirac fermions:

\begin{equation}
M^{\psi }=\left( 
\begin{array}{c|ccccccc}
& \psi _{1R} & \psi _{2R} & \psi _{3R} & \psi _{4R} & \psi _{5R} & 
\widetilde{\psi }_{4R} & \widetilde{\psi }_{5R} \\ \hline
\overline{\psi }_{1L} & 0 & 0 & 0 & y_{14}^{\psi }\left\langle \widetilde{H}%
^{0}\right\rangle & y_{15}^{\psi }\left\langle \widetilde{H}^{0}\right\rangle
& x_{14}^{\psi }\left\langle \phi \right\rangle & x_{15}^{\psi }\left\langle
\phi \right\rangle \\ 
\overline{\psi }_{2L} & 0 & 0 & 0 & y_{24}^{\psi }\left\langle \widetilde{H}%
^{0}\right\rangle & y_{25}^{\psi }\left\langle \widetilde{H}^{0}\right\rangle
& x_{24}^{\psi }\left\langle \phi \right\rangle & x_{25}^{\psi }\left\langle
\phi \right\rangle \\ 
\overline{\psi }_{3L} & 0 & 0 & 0 & y_{34}^{\psi }\left\langle \widetilde{H}%
^{0}\right\rangle & y_{35}^{\psi }\left\langle \widetilde{H}^{0}\right\rangle
& x_{34}^{\psi }\left\langle \phi \right\rangle & x_{35}^{\psi }\left\langle
\phi \right\rangle \\ 
\overline{\psi }_{4L} & y_{41}^{\psi }\left\langle \widetilde{H}%
^{0}\right\rangle & y_{42}^{\psi }\left\langle \widetilde{H}^{0}\right\rangle
& y_{43}^{\psi }\left\langle \widetilde{H}^{0}\right\rangle & 0 & 0 & 
M_{44}^{\psi } & M_{45}^{\psi } \\ 
\overline{\psi }_{5L} & y_{51}^{\psi }\left\langle \widetilde{H}%
^{0}\right\rangle & y_{52}^{\psi }\left\langle \widetilde{H}^{0}\right\rangle
& y_{53}^{\psi }\left\langle \widetilde{H}^{0}\right\rangle & 0 & 0 & 
M_{54}^{\psi } & M_{55}^{\psi } \\ 
\overline{\widetilde{\psi }}_{4L} & x_{41}^{\psi ^{\prime }}\left\langle
\phi \right\rangle & x_{42}^{\psi ^{\prime }}\left\langle \phi \right\rangle
& x_{43}^{\psi ^{\prime }}\left\langle \phi \right\rangle & M_{44}^{\psi
^{\prime }} & M_{45}^{\psi ^{\prime }} & 0 & 0 \\ 
\overline{\widetilde{\psi }}_{5L} & x_{51}^{\psi ^{\prime }}\left\langle
\phi \right\rangle & x_{52}^{\psi ^{\prime }}\left\langle \phi \right\rangle
& x_{53}^{\psi ^{\prime }}\left\langle \phi \right\rangle & M_{54}^{\psi
^{\prime }} & M_{55}^{\psi ^{\prime }} & 0 & 0
\end{array}%
\right) ,  \label{eqn:general_77_mass_matrix}
\end{equation}

with the coefficients $y$ and $x$ being 
Yukawa constants where the former is
expected to be of order unity whereas 
the latter is 
smaller than 
$y$. Furthermore, the $125$ GeV SM like Higgs boson $H$ will corresponds to the lightest of the CP even neutral scalar states arising from $H_u$, $H_d$ and $\phi$, whereas
$M$ is the
vector-like mass. The column vector located at the lower left block in
Equation \ref{eqn:general_77_mass_matrix} consists of left-handed particles
while the row vector at the upper right block are made up of right-handed
particles. The zeros in the $3\times3$ upper block in Equation \ref%
{eqn:general_77_mass_matrix} mean that no SM Yukawa interactions take place
due to 
 charge conservation as well as zeros in two $2\times2$ blocks. Since
we are interested in explaining the muon and electron anomalous magnetic
moments in this model, we first focus on the lepton sector 
in the next subsection and the method used for obtaining the low energy SM Yukawa matrices in the lepton sector can be
applied to the quark sector in the same way with a slight change so that the
quark sector will be discussed in Appendix~\ref{A}.

\subsection{A convenient basis for charged leptons}

\label{subsec:a_convenient_basis_for_charged_leptons}

From Equation \ref{eqn:general_77_mass_matrix}, we can take a specified
basis by rotating some fields as below:

\begin{equation}
M^{e}=\left( 
\begin{array}{c|ccccccc}
& e_{1R} & e_{2R} & e_{3R} & e_{4R} & e_{5R} & \widetilde{L}_{4R} & 
\widetilde{L}_{5R} \\ \hline
\overline{L}_{1L} & 0 & 0 & 0 & 0 & y_{15}^{e}v_{d} & 0 & x_{15}^{L}v_{\phi }
\\ 
\overline{L}_{2L} & 0 & 0 & 0 & y_{24}^{e}v_{d} & y_{25}^{e}v_{d} & 0 & 
x_{25}^{L}v_{\phi } \\ 
\overline{L}_{3L} & 0 & 0 & 0 & y_{34}^{e}v_{d} & y_{35}^{e}v_{d} & 
x_{34}^{L}v_{\phi } & x_{35}^{L}v_{\phi } \\ 
\overline{L}_{4L} & 0 & 0 & y_{43}^{e}v_{d} & 0 & 0 & M_{44}^{L} & M_{45}^{L}
\\ 
\overline{L}_{5L} & y_{51}^{e}v_{d} & y_{52}^{e}v_{d} & y_{53}^{e}v_{d} & 0
& 0 & 0 & M_{55}^{L} \\ 
\overline{\widetilde{e}}_{4L} & 0 & x_{42}^{e}v_{\phi } & x_{43}^{e}v_{\phi }
& M_{44}^{e} & 0 & 0 & 0 \\ 
\overline{\widetilde{e}}_{5L} & x_{51}^{e}v_{\phi } & x_{52}^{e}v_{\phi } & 
x_{53}^{e}v_{\phi } & M_{54}^{e} & M_{55}^{e} & 0 & 0
\end{array}%
\right) ,  \label{eqn:a_specified_basis_charged_leptons_first}
\end{equation}

where $v_{d}=\left\langle H_{d}^{0}\right\rangle $ and $\nu _{\phi }=$ $%
\left\langle \phi \right\rangle $. We start by pointing out the reason why
we take this specific basis for the charged leptons. The reason is that the
strong 
hierarchical structure of the SM fermion Yukawa couplings can be implemented
by the rotations with a simple assumption in this model to be specified
below. 
In order to arrive from Equation \ref{eqn:general_77_mass_matrix} to
Equation \ref{eqn:a_specified_basis_charged_leptons_first}, we rotate the leptonic fields $%
L_{4L} $ and $L_{5L}$ to turn off $M_{54}^{L}$ and rotate $e_{4R}$ and $%
e_{5R}$ to turn off $M_{45}^{e}$. Then, we can rotate $L_{1L}$ and $L_{3L}$
to set $x_{14}^{L}v_{\phi }$ to zero and then rotate $L_{2L}$ and $L_{3L}$
to set $x_{24}^{L}v_{\phi }$ to zero. The same rotation can be applied to $%
e_{1R,2R,3R}$ to set $y_{41,42}^{e}v_{d}$ to zero. Finally, we can further
rotate $L_{1L}$ and $L_{2L}$ to switch off $y_{14}^{e}v_{d}$ and this
rotation also goes for $e_{1R,2R}$ to switch off $x_{41}^{e}v_{\phi }$. 
The above given mass matrix includes three distinct mass scales which are the vev $v_{d}$
of the neutral component of the Higgs doublet $H_{d}$, the vev $v_{\phi }$
of the flavon $\phi $ and the vector-like masses $M$, whose orders of magnitude can be in principle be different.  
Therefore, the mass matrix will be diagonalized by the seesaw
mechanism step-by-step instead of diagonalising it at once. This mechanism
is also known as Universal Seesaw, and was proposed for the first time, in
the context of a left-right symmetric model in \cite{Davidson:1987mh}.

\subsection{A basis for decoupling heavy fourth and fifth vector-like family}

\label{subsec:a_decoupling_basis}

As mentioned in the previous section \ref%
{subsec:a_convenient_basis_for_charged_leptons}, the mass matrix in Equation \ref%
{eqn:a_specified_basis_charged_leptons_first}
involves
three distinct mass scales $v_{d}$, $v_{\phi }$ and $M$ so it is possible to
split 
this whole mass matrix by partial blocks to group mass terms with vev of $H_{d}$ as
in Equation \ref{eqn:a_specified_basis_charged_leptons_third}

\begin{equation}
M^{e}=\left( 
\begin{array}{c|ccccc|cc}
& e_{1R} & e_{2R} & e_{3R} & e_{4R} & e_{5R} & \widetilde{L}_{4R} & 
\widetilde{L}_{5R} \\ \hline
\overline{L}_{1L} & 0 & 0 & 0 & 0 & y_{15}^{e}v_{d} & 0 & x_{15}^{L}v_{\phi }
\\ 
\overline{L}_{2L} & 0 & 0 & 0 & y_{24}^{e}v_{d} & y_{25}^{e}v_{d} & 0 & 
x_{25}^{L}v_{\phi } \\ 
\overline{L}_{3L} & 0 & 0 & 0 & y_{34}^{e}v_{d} & y_{35}^{e}v_{d} & 
x_{34}^{L}v_{\phi } & x_{35}^{L}v_{\phi } \\ 
\overline{L}_{4L} & 0 & 0 & y_{43}^{e}v_{d} & 0 & 0 & M_{44}^{L} & M_{45}^{L}
\\ 
\overline{L}_{5L} & y_{51}^{e}v_{d} & y_{52}^{e}v_{d} & y_{53}^{e}v_{d} & 0
& 0 & 0 & M_{55}^{L} \\ \hline
\overline{\widetilde{e}}_{4L} & 0 & x_{42}^{e}v_{\phi } & x_{43}^{e}v_{\phi }
& M_{44}^{e} & 0 & 0 & 0 \\ 
\overline{\widetilde{e}}_{5L} & x_{51}^{e}v_{\phi } & x_{52}^{e}v_{\phi } & 
x_{53}^{e}v_{\phi } & M_{54}^{e} & M_{55}^{e} & 0 & 0
\end{array}%
\right) ,  \label{eqn:a_specified_basis_charged_leptons_third}
\end{equation}

and then elements of the blocks involving 
$\phi$ can be rotated away to make
those zeros by unitary mixing matrices of Equation \ref%
{eqn:unitary_mixing_matrix_charged_leptons} as per Equation \ref%
{eqn:basis_decoupling_heavy_VL_fermions}:

\begin{equation}
M^{e }=\left( 
\begin{array}{c|ccccccc}
& e _{1R} & e _{2R} & e _{3R} & e _{4R} & e _{5R} & 
\widetilde{L }_{4R} & \widetilde{L }_{5R} \\ \hline
\overline{L }_{1L} &  &  &  &  &  & 0 & 0 \\ 
\overline{L }_{2L} &  &  &  &  &  & 0 & 0 \\ 
\overline{L }_{3L} &  &  & \widetilde{y}_{\alpha \beta }^{\prime e}v_{d}
&  &  & 0 & 0 \\ 
\overline{L }_{4L} &  &  &  &  &  & \widetilde{M}_{44}^{L} & 
M_{45}^{\prime L} \\ 
\overline{L }_{5L} &  &  &  &  &  & 0 & \widetilde{M}_{55}^{L} \\ 
\overline{\widetilde{e }}_{4L} & 0 & 0 & 0 & \widetilde{M}_{44}^{e} & 0 & 
0 & 0 \\ 
\overline{\widetilde{e }}_{5L} & 0 & 0 & 0 & M_{54}^{\prime \prime e} & \widetilde{M}_{55}^{e} & 0 & 0%
\end{array}%
\right) ,  \label{eqn:basis_decoupling_heavy_VL_fermions}
\end{equation}

where the indices $\alpha,\beta$ run from 1 to 5, and tilde, primes repeated
in the mass matrix mean that the parameters are rotated. The unitary $%
5\times5$ matrices are defined to be

\begin{equation}
V_L = V_{45}^L V_{35}^L V_{25}^L V_{15}^L V_{34}^L V_{24}^L V_{14}^L, 
\hspace{0.5cm} V_{e} = V_{45}^{e} V_{35}^{e} V_{25}^{e} V_{15}^{e}
V_{34}^{e} V_{24}^{e} V_{14}^{e},
\label{eqn:unitary_mixing_matrix_charged_leptons}
\end{equation}

where each of the unitary matrices $V_{i4,5}$ are parameterized by a single
angle $\theta_{i4,5}$ describing the mixing between the $i$th chiral family
and the $4,5$th vector-like family. The $5 \times 5$ Yukawa constant matrix
in a mass basis (primed) can be diagonalized by the unitary rotation
matrices as below:

\begin{equation}
\widetilde{y}_{\alpha\beta}^{e} = V_L \widetilde{y}_{\alpha\beta}^{\prime e}
V_{e}^{\dagger}  \label{eqn:diagonalization_Yukawa_matrices}
\end{equation}

From Equation \ref{eqn:a_specified_basis_charged_leptons_first}, we can read
off the $5 \times 5$ upper block and confirm that the $(3,4), (1,5), (2,5),
(3,5) $ mixings in the $L$ sector and $(2,4), (3,4), (1,5), (2,5), (3,5)$
mixings in the $e$ sector are required to go to the decoupling basis. The
unitary matrices of Equation \ref{eqn:unitary_mixing_matrix_charged_leptons} and
mixing angles appearing in the unitary matrices are parameterized by

\begin{equation}
\begin{split}
V_L &= V_{35}^{L} V_{25}^{L} V_{15}^{L} V_{34}^{L} \\
&= 
\begin{pmatrix}
1 & 0 & 0 & 0 & 0 \\ 
0 & 1 & 0 & 0 & 0 \\ 
0 & 0 & c_{35}^L & 0 & s_{35}^L \\ 
0 & 0 & 0 & 1 & 0 \\ 
0 & 0 & -s_{35}^L & 0 & c_{35}^L%
\end{pmatrix}
\begin{pmatrix}
1 & 0 & 0 & 0 & 0 \\ 
0 & c_{25}^L & 0 & 0 & s_{25}^L \\ 
0 & 0 & 1 & 0 & 0 \\ 
0 & 0 & 0 & 1 & 0 \\ 
0 & -s_{25}^L & 0 & 0 & c_{25}^L%
\end{pmatrix}
\begin{pmatrix}
c_{15}^L & 0 & 0 & 0 & s_{15}^L \\ 
0 & 1 & 0 & 0 & 0 \\ 
0 & 0 & 1 & 0 & 0 \\ 
0 & 0 & 0 & 1 & 0 \\ 
-s_{15}^L & 0 & 0 & 0 & c_{15}^L%
\end{pmatrix}
\begin{pmatrix}
1 & 0 & 0 & 0 & 0 \\ 
0 & 1 & 0 & 0 & 0 \\ 
0 & 0 & c_{34}^L & s_{34}^L & 0 \\ 
0 & 0 & -s_{34}^L & c_{34}^L & 0 \\ 
0 & 0 & 0 & 0 & 1%
\end{pmatrix}
\\
&\approx 
\begin{pmatrix}
1 & 0 & 0 & 0 & s_{15}^L \\ 
0 & 1 & 0 & 0 & s_{25}^L \\ 
0 & 0 & 1 & s_{34}^L & s_{35}^L \\ 
0 & 0 & -s_{34}^L & 1 & 0 \\ 
-s_{15}^L & -s_{25}^L & -s_{35}^L & 0 & 1%
\end{pmatrix}%
, \\
&s_{34}^L = \frac{x_{34}^L \left\langle \phi \right\rangle}{\sqrt{\left(
x_{34}^L \left\langle \phi \right\rangle \right)^2 + \left( M_{44}^L
\right)^2}}, \hspace{0.5cm} s_{15}^L = \frac{x_{15}^L \left\langle \phi
\right\rangle}{\sqrt{\left( x_{15}^L \left\langle \phi \right\rangle
\right)^2 + \left( M_{55}^L \right)^2}}, \\
&s_{25}^L = \frac{x_{25}^L \left\langle \phi \right\rangle}{\sqrt{\left(
x_{25}^L \left\langle \phi \right\rangle \right)^2 + \left( M_{55}^{\prime
L} \right)^2}}, \hspace{0.5cm} s_{35}^L = \frac{x_{35}^{\prime L}
\left\langle \phi \right\rangle}{\sqrt{\left( x_{35}^{\prime L} \left\langle
\phi \right\rangle \right)^2 + \left( M_{55}^{\prime\prime L} \right)^2}}, \\
& x_{35}^{\prime L} \left\langle \phi \right\rangle = c_{34}^{L} x_{35}^{L}
\left\langle \phi \right\rangle + s_{34}^{L} M_{45}^{L}, \hspace{0.5cm}
M_{45}^{\prime L} = -s_{34}^{L} x_{35}^{L} \left\langle \phi \right\rangle +
c_{34}^{L} M_{45}^{L} \\
&\widetilde{M}_{44}^{L} = \sqrt{\left( x_{34}^L \left\langle \phi
\right\rangle \right)^2 + \left( M_{44}^L \right)^2}, \\
&M_{55}^{\prime L} = \sqrt{\left( x_{15}^L \left\langle \phi \right\rangle
\right)^2 + \left( M_{55}^L \right)^2}, \hspace{0.1cm} M_{55}^{\prime\prime
L} = \sqrt{\left( x_{25}^L \left\langle \phi \right\rangle \right)^2 +
\left( M_{55}^{\prime L} \right)^2}, \hspace{0.1cm} \widetilde{M}_{55}^{L} = 
\sqrt{\left( x_{35}^{\prime L} \left\langle \phi \right\rangle \right)^2 +
\left( M_{55}^{\prime\prime L} \right)^2}
\end{split}
\label{eqn:unitary_rotation_VQ}
\end{equation}

\begin{equation}
\begin{split}
V_{e} &= V_{35}^{e} V_{25}^{e} V_{15}^{e} V_{34}^{e} V_{24}^{e} \\
&= 
\begin{pmatrix}
1 & 0 & 0 & 0 & 0 \\ 
0 & 1 & 0 & 0 & 0 \\ 
0 & 0 & c_{35}^{e} & 0 & s_{35}^{e} \\ 
0 & 0 & 0 & 1 & 0 \\ 
0 & 0 & -s_{35}^{e} & 0 & c_{35}^{e}%
\end{pmatrix}
\begin{pmatrix}
1 & 0 & 0 & 0 & 0 \\ 
0 & c_{25}^{e} & 0 & 0 & s_{25}^{e} \\ 
0 & 0 & 1 & 0 & 0 \\ 
0 & 0 & 0 & 1 & 0 \\ 
0 & -s_{25}^{e} & 0 & 0 & c_{25}^{e}%
\end{pmatrix}
\begin{pmatrix}
c_{15}^{e} & 0 & 0 & 0 & s_{15}^{e} \\ 
0 & 1 & 0 & 0 & 0 \\ 
0 & 0 & 1 & 0 & 0 \\ 
0 & 0 & 0 & 1 & 0 \\ 
-s_{15}^{e} & 0 & 0 & 0 & c_{15}^{e}%
\end{pmatrix}
\\
& \times 
\begin{pmatrix}
1 & 0 & 0 & 0 & 0 \\ 
0 & 1 & 0 & 0 & 0 \\ 
0 & 0 & c_{34}^{e} & s_{34}^{e} & 0 \\ 
0 & 0 & -s_{34}^{e} & c_{34}^{e} & 0 \\ 
0 & 0 & 0 & 0 & 1%
\end{pmatrix}
\begin{pmatrix}
1 & 0 & 0 & 0 & 0 \\ 
0 & c_{24}^{e} & 0 & s_{24}^{e} & 0 \\ 
0 & 0 & 1 & 0 & 0 \\ 
0 & -s_{24}^{e} & 0 & c_{24}^{e} & 0 \\ 
0 & 0 & 0 & 0 & 1%
\end{pmatrix}
\approx 
\begin{pmatrix}
1 & 0 & 0 & 0 & \theta_{15}^{e} \\ 
0 & 1 & 0 & \theta_{24}^{e} & \theta_{25}^{e} \\ 
0 & 0 & 1 & \theta_{34}^{e} & \theta_{35}^{e} \\ 
0 & -\theta_{24}^{e} & -\theta_{34}^{e} & 1 & 0 \\ 
-\theta_{15}^{e} & -\theta_{25}^{e} & -\theta_{35}^{e} & 0 & 1%
\end{pmatrix}%
, \\
&s_{24}^{e} \approx \frac{x_{42}^{e} \left\langle \phi \right\rangle}{%
M_{44}^{e}}, \hspace{0.5cm} s_{34}^{e} \approx \frac{x_{43}^{e} \left\langle
\phi \right\rangle}{M_{44}^{\prime e}}, \hspace{0.5cm} s_{15}^{e} \approx \frac{x_{51}^{e}
\left\langle \phi \right\rangle}{M_{55}^{e}}, \hspace{0.5cm} s_{25}^{e}
\approx \frac{x_{52}^{\prime e} \left\langle \phi \right\rangle}{%
M_{55}^{\prime e}}, \hspace{0.5cm} s_{35}^{e} \approx \frac{x_{53}^{e}
\left\langle \phi \right\rangle}{M_{55}^{\prime\prime e}}, \\
& x_{52}^{\prime e} \left\langle \phi \right\rangle = c_{24}^{e} x_{52}^{e}
\left\langle \phi \right\rangle + s_{24}^{e} M_{54}^{e}, \hspace{0.5cm}
M_{54}^{\prime e} = -s_{24}^{e} x_{52}^{e} \left\langle \phi \right\rangle +
c_{24}^{e} M_{54}^{e}, \\
& x_{53}^{\prime e} \left\langle \phi \right\rangle = c_{34}^{e} x_{53}^{e}
\left\langle \phi \right\rangle + s_{34}^{e} M_{54}^{\prime e}, \hspace{0.5cm%
} M_{54}^{\prime\prime e} = -s_{34}^{e} x_{53}^{e} \left\langle \phi
\right\rangle + c_{34}^{e} M_{54}^{\prime e}, \\
&M_{44}^{\prime e} = \sqrt{\left( x_{42}^{e} \left\langle \phi \right\rangle
\right)^2 + \left( M_{44}^{e} \right)^2} \hspace{0.5cm}, \widetilde{M}%
_{44}^{e} = \sqrt{\left( x_{43}^{e} \left\langle \phi \right\rangle
\right)^2 + \left( M_{44}^{e} \right)^2}, \\
&M_{55}^{\prime e} = \sqrt{\left( x_{51}^{e} \left\langle \phi \right\rangle
\right)^2 + \left( M_{55}^{e} \right)^2}, \hspace{0.1cm} M_{55}^{\prime%
\prime e} = \sqrt{\left( x_{52}^{\prime e} \left\langle \phi \right\rangle
\right)^2 + \left( M_{55}^{\prime e} \right)^2}, \hspace{0.1cm} \widetilde{M}%
_{55}^{e} = \sqrt{\left( x_{53}^{\prime e} \left\langle \phi \right\rangle
\right)^2 + \left( M_{55}^{\prime\prime e} \right)^2}.
\end{split}
\label{eqn:unitary_rotation_Vec}
\end{equation}

Given the above unitary rotations, the $5\times5$ Yukawa matrices are
computed in terms of the mixing angles and the upper $3\times3$ block would
be the effective SM Yukawa matrix. Assuming all $\cos\theta$ to be 1 and
neglecting order of $\theta$ square or more than that, we have a simple $3
\times 3$ Yukawa matrix of Equation \ref%
{eqn:effective_Yukawa_constant_decoupling_basis_charged_leptons}.

\begin{equation}
\begin{split}
y_{ij}^e &= 
\begin{pmatrix}
s_{15}^L y_{51}^e + y_{15}^e \theta_{15}^{e} & s_{15}^L y_{52}^e + y_{15}^e
\theta_{25}^{e} & s_{15}^L y_{53}^e + y_{15}^e \theta_{35}^{e} \\ 
s_{25}^L y_{51}^e + y_{25}^e \theta_{15}^{e} & s_{25}^L y_{52}^e + y_{24}^e
\theta_{24}^{e} + y_{25}^e \theta_{25}^{e} & s_{25}^L y_{53}^e + y_{24}^e
\theta_{34}^{e} + y_{25}^e \theta_{35}^{e} \\ 
s_{35}^L y_{51}^e + y_{35}^e \theta_{15}^{e} & s_{35}^L y_{52}^e + y_{34}^e
\theta_{24}^{e} + y_{35}^e \theta_{25}^{e} & s_{34}^L y_{43}^e + s_{35}^L
y_{53}^e + y_{34}^e \theta_{34}^{e} + y_{35}^e \theta_{35}^{e}%
\end{pmatrix}%
\end{split}
\label{eqn:effective_Yukawa_constant_decoupling_basis_charged_leptons}
\end{equation}

\subsection{A convenient basis for neutrinos}

\label{subsec:a_convenient_basis_neutrinos}

The relevant Yukawa and mass terms of the neutrino sector give rise to the
following neutrino mass matrix: 
\begin{equation}
M^{\nu }=\left( 
\begin{array}{c|ccc|cccc}
& L_{1L} & L_{2L} & L_{3L} & \overline{\nu }_{4R} & \overline{\nu }_{5R} & 
\widetilde{\nu }_{4R} & \widetilde{\nu }_{5R} \\ \hline
L_{1L} & 0 & 0 & 0 & y_{14}^{\nu }v_{u} & y_{15}^{\nu }v_{u} & 
x_{14}^{L}v_{d} & x_{15}^{L}v_{d} \\ 
L_{2L} & 0 & 0 & 0 & y_{24}^{\nu }v_{u} & y_{25}^{\nu }v_{u} & 
x_{24}^{L}v_{d} & x_{25}^{L}v_{d} \\ 
L_{3L} & 0 & 0 & 0 & y_{34}^{\nu }v_{u} & y_{35}^{\nu }v_{u} & 
x_{34}^{L}v_{d} & x_{35}^{L}v_{d} \\ \hline
\overline{\nu }_{4R} & y_{14}^{\nu }v_{u} & y_{24}^{\nu }v_{u} & y_{34}^{\nu
}v_{u} & 0 & 0 & M_{44}^{\nu } & M_{54}^{\nu } \\ 
\overline{\nu }_{5R} & y_{15}^{\nu }v_{u} & y_{25}^{\nu }v_{u} & y_{35}^{\nu
}v_{u} & 0 & 0 & M_{45}^{\nu } & M_{55}^{\nu } \\ 
\widetilde{\nu }_{4R} & x_{14}^{L}v_{d} & x_{24}^{L}v_{d} & x_{34}^{L}v_{d}
& M_{44}^{\nu } & M_{45}^{\nu } & 0 & 0 \\ 
\widetilde{\nu }_{5R} & x_{15}^{L}v_{d} & x_{25}^{L}v_{d} & x_{35}^{L}v_{d}
& M_{54}^{\nu } & M_{55}^{\nu } & 0 & 0
\end{array}%
\right)  \label{eqn:mass_matrix_neutrinos}
\end{equation}

Here, the zeros in the upper $3 \times 3$ block of Equation \ref%
{eqn:mass_matrix_neutrinos} mean that neutrinos remain massless in the SM.
Therefore, the SM neutrinos can be massive via the inclusion 
of two
vector-like families. In order to make this mass matrix as simple as
possible, the only choice left is to rotate $\nu_{4R}$ and $\nu_{5R}$ to
turn off $M_{45}^{\nu}$ since rotations between $L_{1L,2L,3L}$ are already
used in the charged lepton sector. 
\section{$W$ BOSON EXCHANGE CONTRIBUTIONS TO  $\left( g-2 \right)_\protect\mu, \left( g-2
\right)_e$ AND $\func{BR}\left(\protect\mu \rightarrow e \protect\gamma %
\right)$ }
\label{IV}

\label{sec:Analytic_arguments_muon_electron_g2_W} 
Within the framework of our proposed model, we start by investigating the
muon and electron anomalous magnetic moments with $W$ boson exchange first. 
Given that such $W$ boson exchange contribution also involves virtual
neutrinos in the internal lines of the loop, we revisit the mass matrix for
neutrinos. In this mass matrix, we remove fifth vector-like neutrinos $%
\nu_{5R}$ and $\widetilde{\nu}_{5R}$ since they are too heavy to contribute
to the phenomenology under study. As mentioned in the previous section, we
stick to a condition where the coefficient $y$ is expected to be of order
unity, 
whereas the coupling $x$ is expected to be smaller than $y$. 
Such condition can be easily seen by substituting the coefficients $%
y_{i4}^\nu$ by $y_{i}^\nu$ and the coefficients $x_{i4}^L$ by $\epsilon
y_{i}^{\nu \prime}$ where $\epsilon$ is a suppression factor. Putting all
these considerations together, the mass matrix for neutrinos in Equation \ref%
{eqn:mass_matrix_neutrinos} after electroweak symmetry breaking takes the
form:

\begin{equation}
M^\nu \approx \left( 
\begin{array}{c|ccc|cc}
& \nu_{1L} & \nu_{2L} & \nu_{3L} & \overline{\nu}_{4R} & \widetilde{\nu}_{4R}
\\ \hline
\nu_{1L} & 0 & 0 & 0 & y_{1}^\nu v_u & \epsilon y_{1}^{\nu \prime} v_d \\ 
\nu_{2L} & 0 & 0 & 0 & y_{2}^\nu v_u & \epsilon y_{2}^{\nu \prime} v_d \\ 
\nu_{3L} & 0 & 0 & 0 & y_{3}^\nu v_u & \epsilon y_{3}^{\nu \prime} v_d \\ 
\hline
\overline{\nu}_{4R} & y_{1}^\nu v_u & y_{2}^\nu v_u & y_{3}^\nu v_u & 0 & 
M_{44}^{\nu} \\ 
\widetilde{\nu}_{4R} & \epsilon y_{1}^{\nu \prime} v_d & \epsilon y_{2}^{\nu
\prime} v_d & \epsilon y_{3}^{\nu \prime} v_d & M_{44}^{\nu} & 0%
\end{array}
\right) \equiv \left( 
\begin{array}{cc}
0 & m_D \\ 
m_D^T & M_N%
\end{array}
\right) , \label{eqn:mass_matrix_neutrinos_reduced}
\end{equation}

where $v_u(v_d)$ is the vev of $\widetilde{H}_u(H_d)$, $v_u$ runs from $246/%
\sqrt{2}\func{GeV} \simeq 174 \func{GeV}$ to $246 \func{GeV}$ and $%
v_u^2+v_d^2 = \left( 246 \func{GeV} \right)^2$.

\subsection{Type 1b seesaw mechanism}

\label{subsec:type_1b_seesaw_mechanism}

Now that we constructed the neutrino mass matrix for this task, the next
step is to read off the operator which gives rise to the neutrino mass from
the mass matrix. Generally, the well-known operator for neutrino mass is the
Weinberg operator(type 1a seesaw mechanism) $\frac{1}{\Lambda}L_i L_j H H$.
A main feature of the Weinberg operator is the same SM Higgs should be repeated
in the operator, however that 
property is not present in our model 
since the Higgs doublets $H_{u,d}$ are negatively charged under the $%
U(1)^{\prime}$ symmetry, which implies 
the corresponding Weinberg operator having such fields will not be invariant
under the $U(1)^{\prime}$ unless an insertion of a quadratic power of the
gauge singlet scalar $\phi$ is considered. However we do not consider the
operators $\frac{1}{\Lambda^3}(\bar{L}_i\tilde{H}_u)(\tilde{H}%
_uL^C_j)(\phi^*)^2$ and $\frac{1}{\Lambda^3}(\bar{L}_iH_d)(H_dL^C_j)\phi^2$
in the neutrino sector, since they are very subleading and thus will give a
tiny contribution to the light active neutrino masses. 
Instead of relying on a seven dimensional Weinberg to generate the tiny
masses for the light active neutrinos, we take another approach named type
1b seesaw mechanism (we call the Weinberg operator ``type 1a seesaw
mechanism" to differentiate with) 
where the mixing of different $SU(2)$ Higgs doublets can appear satisfying
charge conservation. Diagrams for the operators are given in Figure \ref%
{fig:diagrams_type1ab_operators} for comparison:

\begin{figure}[]
\centering
\begin{subfigure}{0.48\textwidth}
	\includegraphics[width=1.0\textwidth]{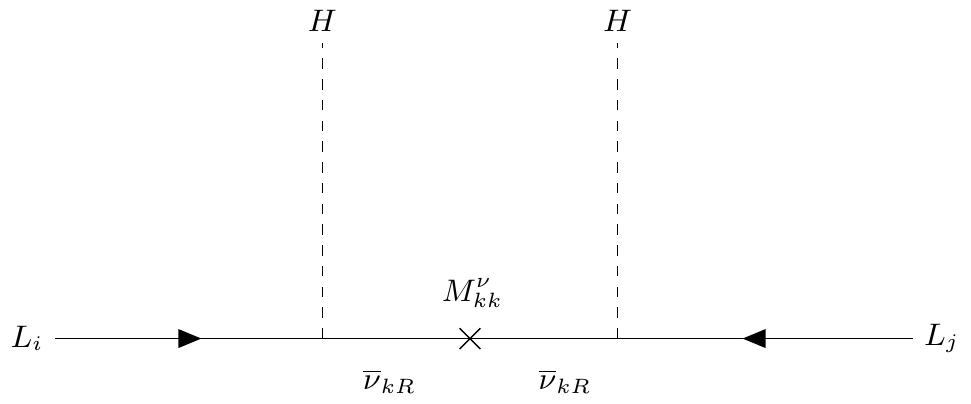}
\end{subfigure} \hspace{0.1cm} 
\begin{subfigure}{0.48\textwidth}
	\includegraphics[width=1.0\textwidth]{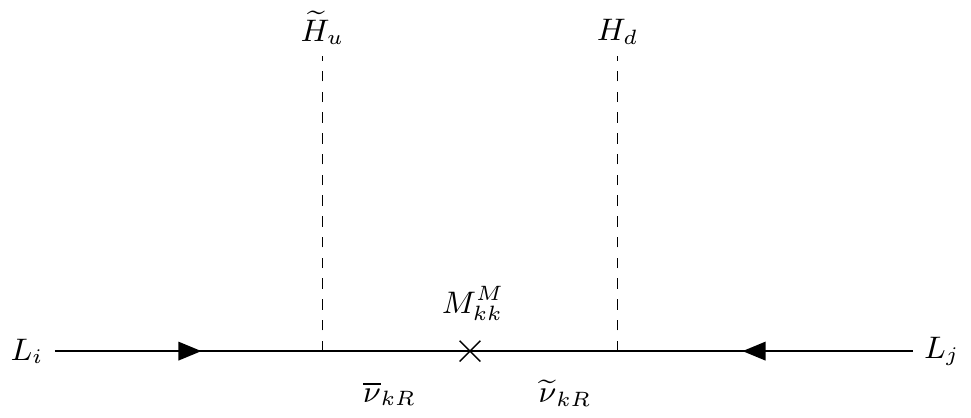}
\end{subfigure}
\caption{Diagrams which lead to effective Weinberg operators for the
Majorana and vector-like mass in the mass insertion formalism, where $%
i,j=1,2,3$ and $k=4$, respectively. The left is the Weinberg operator(or
type 1a seesaw mechanism) in which mass $M$ is Majorana mass and the right
is Weinberg-like operator(or type 1b seesaw mechanism) in which mass $M$ is
vector-like mass.}
\label{fig:diagrams_type1ab_operators}
\end{figure}

The diagrams in Figure \ref{fig:diagrams_type1ab_operators} clearly tell the
difference between Majorana mass and vector-like mass. They share a common
property that they violate the lepton number conservation, 
whereas the particles appearing in 
a Majorana mass term are same but those ones involved in vector-like mass
terms are different. As the type 1b seesaw mechanism only works in this
model, we make use of this seesaw mechanism for the analysis of neutrinos.
With the operator, the renormalizable Lagrangian for neutrinos can be
written as:

\begin{equation}
\begin{split}
\mathcal{L}_{\nu}^{\func{Yukawa+Mass}} = y_{i}^{\nu} \overline{L}_{iL} 
\widetilde{H}_u \nu_{kR} + \epsilon y_i^{\nu\prime} \overline{L}_{iL} H_d 
\overline{\widetilde{\nu}}_{kR} + M_{kk}^{M} \overline{\widetilde{\nu}}_{kR}
\nu_{kR} + \func{h.c.} , \label{eqn:Lagrangian_neutrinos}
\end{split}%
\end{equation}

where $i = 1,2,3$ and $k=4$. The renormalizable Lagrangian of Equation %
\ref{eqn:Lagrangian_neutrinos} above the electroweak scale 
generates an effective Lagrangian after decoupling the heavy vector-like
neutrinos, which is suitable for study of low energy neutrino phenomenology.
The effective Lagrangian for neutrino at electroweak scale is given by\cite%
{Hernandez-Garcia:2019uof}

\begin{equation}
\mathcal{L}^{d=5} = c_{ij}^{d=5} \left( \left( L_i^T \widetilde{H}_u \right)
\left( H_d^T L_j \right) + \left( L_i^T H_d \right) \left( \widetilde{H}_u^T
L_j \right) \right),  \label{eqn:effective_operators_neutrinos}
\end{equation}

where the coefficient $c_{ij}^{d=5}$ is suppressed by a factor of the
vector-like mass $M$. The neutrino mass matrix of Equation \ref%
{eqn:mass_matrix_neutrinos_reduced} can be diagonalized by the unitary
matrix $U$ as below:

\begin{equation}
U^T 
\begin{pmatrix}
0 & m_D^T \\ 
m_D & M_N%
\end{pmatrix}
U = 
\begin{pmatrix}
m_\nu^{\func{diag}} & 0 \\ 
0 & M_N^{\func{diag}}%
\end{pmatrix},
\label{eqn:diagonalization_neutrinos}
\end{equation}

where $m_\nu^{\func{diag}}$ is a diagonal matrix for the light left-handed
neutrinos $\nu_{iL}$ and $M_N^{\func{diag}}$ is that for the heavy
vector-like neutrinos $\nu_{4R}, \widetilde{\nu}_{4R}$. Here, the unitary
mixing matrix $U$ is defined by multiplication of two unitary matrices which
we call $U_A$ and $U_B$, respectively\cite{Blennow:2011vn}:

\begin{equation}
\begin{split}
U &= U_A \cdot U_B \\
U_A &= \func{exp}%
\begin{pmatrix}
0 & \Theta \\ 
-\Theta^\dagger & 0%
\end{pmatrix}
\simeq 
\begin{pmatrix}
I-\frac{\Theta \Theta^\dagger}{2} & \Theta \\ 
-\Theta^\dagger & I-\frac{\Theta \Theta^\dagger}{2}%
\end{pmatrix}
\quad \text{at leading order in $\Theta$} \\
U_B &= 
\begin{pmatrix}
U_{\func{PMNS}} & 0 \\ 
0 & I%
\end{pmatrix}%
\end{split}
\label{eqn:unitary_matrix_neutrinos}
\end{equation}

The unitary matrix $U_{\func{PMNS}}$ in $U_B$ is the well-known
Pontecorvo-Maki-Nakagawa-Sakata matrix and is parameterized by \cite%
{Hernandez-Garcia:2019uof,Chau:1984fp}

\begin{equation}
U_{\func{PMNS}} = 
\begin{pmatrix}
1 & 0 & 0 \\ 
0 & \cos\theta_{23} & \sin\theta_{23} \\ 
0 & -\sin\theta_{23} & \cos\theta_{23}%
\end{pmatrix}
\begin{pmatrix}
\cos\theta_{13} & 0 & \sin\theta_{13} e^{-i\delta_{\func{CP}}} \\ 
0 & 1 & 0 \\ 
-\sin\theta_{13} e^{i\delta_{\func{CP}}} & 0 & \cos\theta_{13}%
\end{pmatrix}
\begin{pmatrix}
\cos\theta_{12} & \sin\theta_{12} & 0 \\ 
-\sin\theta_{12} & \cos\theta_{12} & 0 \\ 
0 & 0 & 1%
\end{pmatrix}
\begin{pmatrix}
e^{-i\alpha^\prime/2} & 0 & 0 \\ 
0 & e^{-i\alpha/2} & 0 \\ 
0 & 0 & 1%
\end{pmatrix}%
,  \label{eqn:PMNS_mixing_matrix}
\end{equation}

where the Majorana phase $\alpha^\prime$ is set to zero in this model. The mixing matrices $%
U_{A,B}$ are unitary, however the $3\times 3$ upper block of the unitary
matrix $U$ is not unitary due to the 
factor $\left( I - \Theta \Theta^\dagger/2 \right)$ for the light neutrinos. An
interesting feature of the unitary matrix $U$ is it is unitary globally, but
non-unitary locally and this non-unitarity contributes to explain muon and
electron anomalous magnetic moments. Replacing the unitary matrices in
Equation \ref{eqn:unitary_matrix_neutrinos} back to Equation \ref%
{eqn:diagonalization_neutrinos}, the result is simplified with the
assumption $M_N \gg m_D$ to the conventional seesaw mechanism:

\begin{equation}
\begin{split}
\Theta &\simeq m_D^\dagger M_N^{-1} \\
U_{\func{PMNS}}^* m_{\nu}^{\func{diag}} U_{\func{PMNS}}^\dagger &\simeq
-m_D^T M_N^{-1} m_D \equiv -m \\
M_N^{\func{diag}} &\simeq M_N,  \label{eqn:parameterisation_neutrinos}
\end{split}%
\end{equation}

where $m$ is the effective mass matrix resulted from Equation \ref%
{eqn:mass_matrix_neutrinos_reduced}.

\begin{equation}
m_{ij} = \frac{\epsilon v_u v_d}{M_{44}^{\nu}} \left( y_i^\nu
y_j^{\nu\prime} + y_i^{\nu\prime} y_j^\nu \right)
\label{eqn:effective_neutrino_mass_matrix}
\end{equation}

Therefore, smallness of the light neutrino masses can be understood not only
from mass of vector-like mass $M_{44}^{\nu}$ but also from the suppression
factor $\epsilon$ and the presence of $\epsilon$ allows more flexibility in the allowed mass values of the vector-like neutrinos. Revisiting non-unitarity part for the light neutrinos from
the unitary matrix $U$\cite{Blennow:2011vn,FernandezMartinez:2007ms}, it
reads:

\begin{equation}
\begin{split}
&\left( I - \frac{\Theta_i \Theta_j^\dagger}{2} \right) U_{\func{PMNS}} =
\left( I - \eta_{ij} \right) U_{\func{PMNS}}
\label{eqn:non_unitarity_light_neutrinos}
\end{split}%
\end{equation}

The non-unitarity $\eta$ is associated with the presence of the heavy vector-like
neutrinos and can be derived from a coefficient of the effective Lagrangian
at dimension 6\cite{Broncano:2002rw}:

\begin{equation}
\mathcal{L}^{d=6} = c_{ij}^{d=6} \left( \left( L_i^\dagger \widetilde{H}_u
\right) i\slashed{\partial} \left( \widetilde{H}_u^\dagger L_j \right) +
\left( L_i^\dagger H_d \right) i\slashed{\partial} \left( H_d^\dagger L_j
\right) \right)  \label{eqn:effective_operators_dimension6}
\end{equation}

Once the SM Higgs doublets in Equation \ref%
{eqn:effective_operators_dimension6} develop its vev, the Lagrangian at
dimension 6 causes non-diagonal kinetic terms for the light neutrinos and it
gives rise to deviations of unitarity when it is diagonalized. The
deviations of unitarity can be expressed in terms of the coefficient at
dimension 6 $\eta_{ij} \equiv v^2 c_{ij}^{d=6}/2$.

\begin{equation}
\eta_{ij} = \frac{\Theta_i \Theta_j^\dagger}{2} = \frac{1}{2} \frac{%
m_D^\dagger m_D}{M_N^2} = \frac{1}{2M_{44}^{\nu 2}} \left( v_u^2 y_i^{\nu*}
y_j^\nu + \epsilon^2 v_d^2 y_i^{\nu\prime*} y_j^{\nu\prime} \right) \simeq 
\frac{v_u^2}{2M_{44}^{\nu 2}} y_i^{\nu *} y_j^{\nu}
\label{eqn:deviation_of_unitarity}
\end{equation}

From the fourth term in Equation \ref{eqn:deviation_of_unitarity}, the term
with $\epsilon^2$ can be safely ignored due to both relative
smallness of $v_d$ and the suppression factor $\epsilon$. Thus, the
deviation of unitarity $\eta$ consists of the vector-like mass $M_{44}^{\nu}$
and the Yukawa couplings $y_{i,j}^\nu$. As an interesting example, it is
possible that the Yukawa couplings $y_{i,j}^\nu$ can be obtained from the
observables such as the PMNS mixing matrix and two mass squared splitting, $%
\Delta m_{sol}^2$ and $\Delta m_{atm}^2$, in the neutrino oscillation
experiments. Since the hierarchy between the light neutrinos is not yet
determined, there are two possible scenarios, normal hierarchy(NH) and
inverted hierarchy(IH), and the lightest neutrino remains massless, whereas two
other neutrinos get massive. The Yukawa couplings $y_{i}^{\nu,\nu\prime}$
for the NH($m_1 = 0$) are determined by

\begin{equation}
\begin{split}
y_i^\nu &= \frac{y}{\sqrt{2}} \left( \sqrt{1+\rho} \left( U_{\func{PMNS}}^*
\right)_{i3} + \sqrt{1-\rho} \left( U_{\func{PMNS}}^* \right)_{i2} \right) \\
y_i^{\nu\prime} &= \frac{y^\prime}{\sqrt{2}} \left( \sqrt{1+\rho} \left( U_{%
\func{PMNS}}^* \right)_{i3} - \sqrt{1-\rho} \left( U_{\func{PMNS}}^*
\right)_{i2} \right),
\end{split}%
\end{equation}

where $y$ and $y^\prime$ are real numbers and $\rho=(1-\sqrt{r})/(1+\sqrt{r}%
) $ with $r \equiv \lvert \Delta m_{sol}^2 \rvert/\lvert \Delta m_{atm}^2
\rvert = \Delta m_{21}^2/\Delta m_{31}^2$, whereas the Yukawa couplings $y_i^{\nu,\nu\prime}$ for the IN($m_3=0$) are

\begin{equation}
\begin{split}
y_i^\nu &= \frac{y}{\sqrt{2}} \left( \sqrt{1+\rho} \left( U_{\func{PMNS}}^*
\right)_{i2} + \sqrt{1-\rho} \left( U_{\func{PMNS}}^* \right)_{i1} \right) \\
y_i^{\nu\prime} &= \frac{y^\prime}{\sqrt{2}} \left( \sqrt{1+\rho} \left( U_{%
\func{PMNS}}^* \right)_{i2} - \sqrt{1-\rho} \left( U_{\func{PMNS}}^*
\right)_{i1} \right),
\end{split}%
\end{equation}

where $\rho=(1-\sqrt{1+r})/(1+\sqrt{1+r})$ with $r \equiv \lvert \Delta
m_{sol}^2 \rvert/\lvert \Delta m_{atm}^2 \rvert = \Delta m_{21}^2/\Delta
m_{32}^2$.

\subsection{The charged lepton flavour violation(CLFV) $\protect\mu %
\rightarrow e \protect\gamma$ decay}

\label{subsec:CLFV_muegamma_decay}

\begin{figure}[]
\centering
\begin{subfigure}{0.48\textwidth}
	\includegraphics[width=1.0\textwidth]{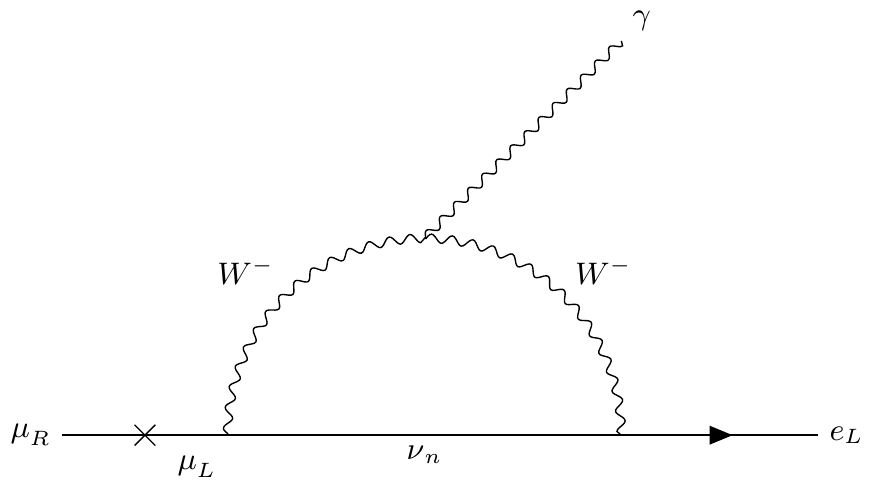}
\end{subfigure} \hspace{0.1cm}
\begin{subfigure}{0.48\textwidth}
	\includegraphics[width=1.0\textwidth]{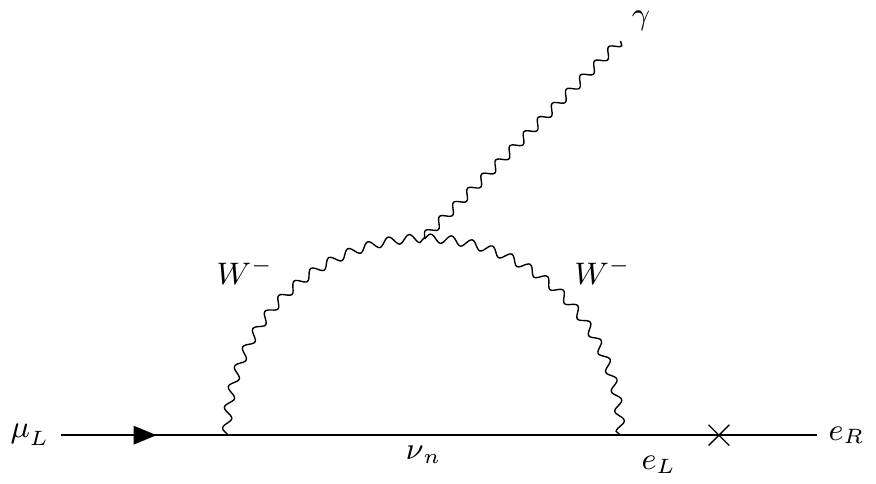}
\end{subfigure}
\caption{Diagrams for CLFV $\protect\mu \rightarrow e \protect\gamma$ decay
with all neutrinos. Here $n=1,2,3,4,5$.}
\label{fig:diagrams_muegamma_with_allneutrinos}
\end{figure}

Consider the three light neutrinos in the SM for the CLFV $\mu \rightarrow e \gamma$
decay first. In this case, the unitary mixing matrix becomes just the PMNS
mixing matrix and the GIM mechanism which suppresses flavour-changing
process works, therefore it leads quite suppressed sensitivity for $\func{BR}%
\left(\mu \rightarrow e \gamma\right)$ about $10^{-55}$\cite{Calibbi:2017uvl}%
, which is impossible to observe with the current sensitivity of $\mu
\rightarrow e \gamma$ decay. This impractical sensitivity can be enhanced to
the observable level by introducing the heavy vector-like neutrinos which give
rise to deviation of unitarity. With the presence of heavy vector-like
neutrinos, the GIM mechanism is gone and the factor suppressed by GIM
mechanism can survive with a factor of deviation of unitarity, which plays a
crucial role to increase significantly order of theoretical prediction for $%
\mu \rightarrow e \gamma$ decay\cite{Glashow:1970gm}. Therefore, the
strongest constraint for deviation of unitarity in the modified PMNS mixing
matrix comes from CLFV $\mu \rightarrow e \gamma$ decay. The possible one-loop
diagrams for the CLFV $\mu \rightarrow e \gamma$ with all neutrinos in this
model are given in Figure \ref{fig:diagrams_muegamma_with_allneutrinos}.

The amplitude from above diagrams in Figure \ref%
{fig:diagrams_muegamma_with_allneutrinos} reads\cite{Calibbi:2017uvl}:

\begin{equation}
\begin{split}
\mathcal{M}\left( \mu \rightarrow e \gamma \right) &= \overline{u}_e i
\sigma_{\mu\nu} q^{\nu} \left( F_1 + F_2 \gamma^5 \right) u_\mu
\epsilon^{*\mu} \\
&= \overline{u}_e i \sigma_{\mu\nu} q^{\nu} \left( A_R P_R + A_L P_L \right)
u_\mu \epsilon^{* \mu},  \label{eqn:amplitude_muegamma_W}
\end{split}%
\end{equation}

where $u$ is Dirac spinor for the muon and electron, $q$ is
four momentum of an outgoing photon, $F_{1,2}$ are form factors, $%
A_{L,R}$ are left- and right-handed amplitude defined to be $A_{L,R} = F_1
\pm F_2$ and lastly $P_{L,R}$ are projection operators. From the amplitude,
the helicity flip between initial particle and final particle should arise
and this makes the helicity flip process takes place on one of external legs
since the $W$ gauge boson couples only to left-handed fields. Comparing the
left diagram with the right, the left is proportional to the muon mass,
while the right is proportional to the electron mass, which means that
impact of the right is ignorable. The unpolarized squared amplitude $\lvert 
\mathcal{M} \rvert^2$ takes the form:

\begin{equation}
\lvert \mathcal{M} \rvert^2 = m_\mu^4 \left( A_R + A_L \right)^2 \simeq
m_\mu^4 \left( A_R \right)^2  \label{eqn:unpolarised_squared_amplitude}
\end{equation}

Then, the decay rate is given by

\begin{equation}
\Gamma\left( \mu \rightarrow e \gamma \right) = \frac{\lvert \mathcal{M}
\rvert^2}{16\pi m_\mu} = \frac{m_\mu^3}{16\pi} \lvert A_R \rvert^2
\label{eqn:decay_rate_muegamma}
\end{equation}

where $A_R$ is expressed by\cite{Cheng:1985bj,Calibbi:2017uvl}\footnote{%
Since the PMNS mixing matrix is multiplied by a factor of deviation of
unitarity, it is not unitary any more. Therefore, the first term of sum over
neutrino eigenstates in Equation (28) of \cite{Calibbi:2017uvl} does not
vanish and come in our prediction with a loop function $F(x_n)$.}

\begin{equation}
A_R = \frac{g^2 e}{128\pi^2} \frac{m_\mu}{M_W^2} \sum_{n=1,2,3,4,5} U_{2 n}
U_{1 n}^* F(x_n) \left[ 1 - \frac{1}{3} \frac{\ln \xi}{\xi-1} + \frac{1}{%
\xi-1} \left( \frac{\xi \ln \xi}{\xi-1} - 1 \right) \right]
\label{eqn:left_handed_amplitdue_AR}
\end{equation}

Taking the unitary gauge into account, $\xi \rightarrow \infty$, the
additional $\xi$-dependent terms in $A_R$ all are cancelled by contribution
of Goldstone bosons so $A_R$ is gauge invariant. Substituting the gauge
invariant $A_R$ back into the decay rate of Equation \ref%
{eqn:decay_rate_muegamma} and dividing the expanded decay rate by the total
muon decay rate $\Gamma\left(\mu \rightarrow e \nu \overline{\nu} \right) =
G_F^2 m_\mu^5/192\pi^3$, we have the prediction for $\mu \rightarrow e
\gamma $ decay\cite{Hernandez-Garcia:2019uof,Calibbi:2017uvl}:

\begin{equation}
\func{BR}\left(\mu \rightarrow e \gamma \right) = \frac{\Gamma \left( \mu
\rightarrow e \gamma \right)}{\Gamma \left( \mu \rightarrow e \nu_\mu 
\overline{\nu}_e \right)} = \frac{3\alpha}{32\pi} \frac{\lvert \sum_{n=1}^5
U_{2n} U_{n1}^{\dagger} F(x_n) \rvert^2}{\left( U U^{\dagger} \right)_{11}
\left( U U^{\dagger} \right)_{22}} , \label{eqn:prediction_muegamma_neutrinos}
\end{equation}

where $x_n = M_n^2/M_W^2$ and the loop function $F(x_n)$ is

\begin{equation}
F(x_n) = \frac{10 - 43 x_n + 78 x_n^2 - (49 - 18 \log x_n) x_n^3 + 4 x_n^4}{%
3(x_n-1)^4}.  \label{eqn:loop_function_F}
\end{equation}

Numerator in Equation \ref{eqn:prediction_muegamma_neutrinos} can be
simplified by separating the light neutrinos and heavy vector-like neutrinos as
below(Contribution of the fifth neutrino $\widetilde{\nu}_{4R}$ is safely ignored both by the
suppression factor $\epsilon$ and by relative smallness of $v_d$ compared to 
$v_u$):

\begin{equation}
\begin{split}
\lvert \sum_{n=1}^5 U_{2n} U_{n1}^{\dagger} F(x_n) \rvert^2 &\simeq \lvert
U_{2i} U_{i1}^\dagger F(0) + U_{24} U_{41}^\dagger F(x_4) \rvert^2 \\
U_{2i} U_{i1}^{\dagger} &= -\eta_{12}^* - \eta_{21} = -2\eta_{21} \\
U_{24} U_{41}^{\dagger} &= \Theta_{24} \Theta_{14}^* = 2\eta_{21} \\
\lvert \sum_{n=1}^5 U_{2n} U_{n1}^{\dagger} F(x_n) \rvert^2 &\simeq \lvert
4\eta_{21} \rvert^2 \left( F(x_4) - F(x_0) \right)^2
\label{eqn:numerator_simplified}
\end{split}%
\end{equation}

The final form for the CLFV $\mu \rightarrow e \gamma$ decay in this model reads:

\begin{equation}
\func{BR}\left(\mu \rightarrow e \gamma \right) = \frac{3\alpha_{\func{em}}}{%
8\pi} \lvert \eta_{21} \rvert^2 \left( F(x_4) - F(0) \right)^2,
\label{eqn:prediction_muegamma_final}
\end{equation}

where $\alpha_{\func{em}}$ is the fine structure constant. We find that our
theoretical prediction for the $\mu \rightarrow e \gamma$ decay can be expressed
in terms of the deviation of unitarity $\eta_{21}$.

\subsection{The anomalous muon magnetic moment $g-2$}

\label{subsec:muon_g2_Wboson}

We derive our prediction for the muon anomalous magnetic moment in this
section and confirm the derived expression can be consistent with an
expression of the theoretical prediction for $\mu \rightarrow e \gamma$ in
references\cite{Calibbi:2017uvl,Lindner:2016bgg}. Consider two possible
diagrams for muon anomalous magnetic moment at one-loop level in Figure \ref%
{fig:diagrams_muong2_with_allneutrinos}.

\begin{figure}[]
\centering
\begin{subfigure}{0.48\textwidth}
	\includegraphics[width=1.0\textwidth]{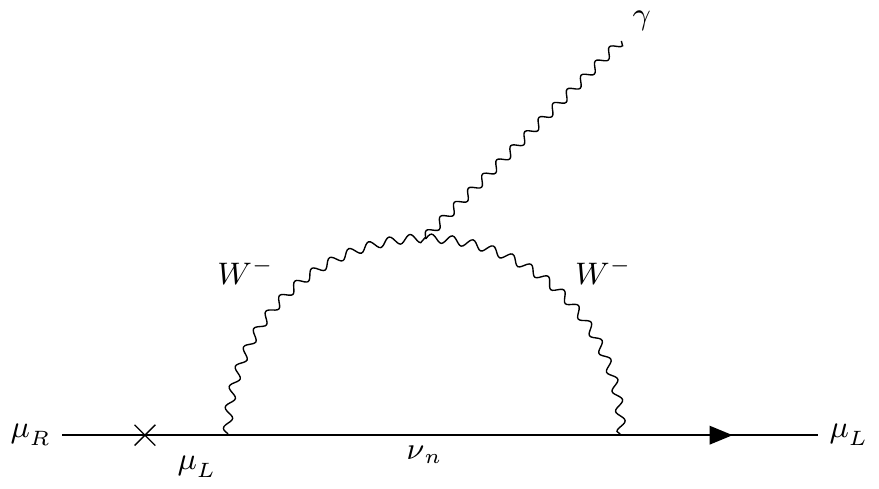}
\end{subfigure} \hspace{0.1cm}
\begin{subfigure}{0.48\textwidth}
	\includegraphics[width=1.0\textwidth]{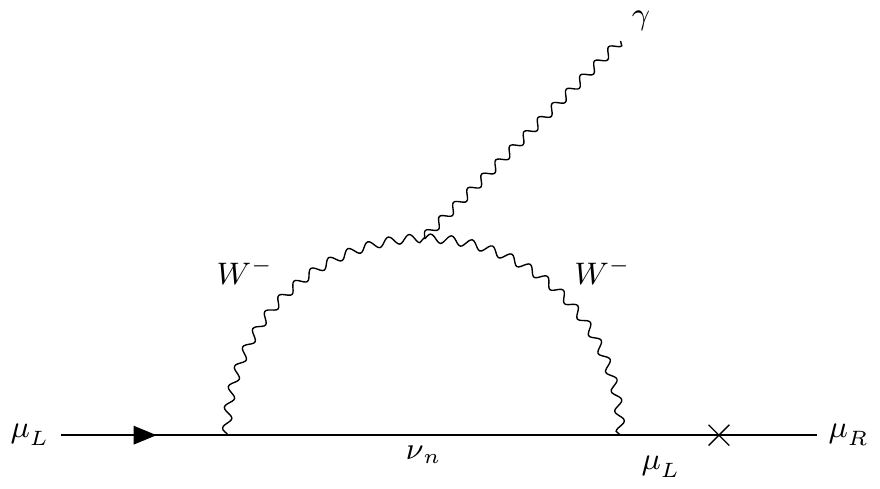}
\end{subfigure}
\caption{Diagrams for muon anomalous magnetic moment with all neutrinos. Here $n=1,2,3,4,5$.}
\label{fig:diagrams_muong2_with_allneutrinos}
\end{figure}

The amplitude for the muon anomalous magnetic moment at one-loop level is:

\begin{equation}
\begin{split}
\mathcal{M}\left( \Delta a_\mu \right) &= \overline{u}_\mu i \sigma_{\mu\nu}
q^{\nu} \left( F_1 + F_2 \gamma^5 \right) u_\mu \epsilon^{*\mu} \\
&= \overline{u}_\mu i \sigma_{\mu\nu} q^{\nu} \left( A_R P_R + A_L P_L
\right) u_\mu \epsilon^{* \mu}  \label{eqn:amplitude_muong2_W}
\end{split}%
\end{equation}

Unlike the CLFV $\mu \rightarrow e \gamma$ decay, muon anomaly diagrams have
the same structure for helicity flip process. So we conclude $A_R$ is equal
to $A_L$ and can make use of other expression of this amplitude to derive
our own expression for $\Delta a_\mu$\cite{Lindner:2016bgg}.

\begin{equation}
\begin{split}
V &= \overline{u}_\mu i \sigma_{\alpha\beta} q^\beta e m_\mu \left(
A_{\mu\mu}^M + \gamma_5 A_{\mu\mu}^E \right) u_\mu \epsilon^{*\alpha} \\
&= \overline{u}_\mu i \sigma_{\alpha \beta} q^{\beta} e m_\mu \left( \left(
A_{\mu\mu}^M + A_{\mu\mu}^E \right) P_R + \left( A_{\mu\mu}^M - A_{\mu\mu}^E
\right) P_L \right) u_\mu \epsilon^{*\alpha}
\end{split}
\label{eqn:amplitude_muong2_W_LiviewonLFV}
\end{equation}

Comparing Equation \ref{eqn:amplitude_muong2_W} with Equation \ref%
{eqn:amplitude_muong2_W_LiviewonLFV}, we confirm that

\begin{equation}
\begin{split}
A_R &= e m_\mu \left( A_{\mu\mu}^M + A_{\mu\mu}^E \right) \\
A_L &= e m_\mu \left( A_{\mu\mu}^M - A_{\mu\mu}^E \right)
\label{eqn:equality_ALR_AME}
\end{split}%
\end{equation}

Here, we can use the condition that $A_R = A_L$ identified in Figure \ref%
{fig:diagrams_muong2_with_allneutrinos} and can rearrange $A_{L,R}$ in terms
of $A_{\mu\mu}^{M,E}$, which are essential to derive our theoretical muon
anomaly prediction. Then, we find our desirable form $A_{\mu\mu}^{M,E}$ for
the muon anomalous magnetic moment.

\begin{equation}
\begin{split}
A_{\mu\mu}^M &= \frac{1}{e m_\mu} A_R = \frac{g^2}{128\pi^2} \frac{1}{M_W^2}
\sum_{n=1,2,3,4,5} U_{2n} U_{2n}^* F(x_n) \\
A_{\mu\mu}^E &= 0  \label{eqn:redefinition_AME}
\end{split}%
\end{equation}

Using the definition for both the muon anomalous magnetic moment and branching
ratio of $\mu \rightarrow e \gamma$ decay in \cite{Lindner:2016bgg}, we can
check our analytic argument for the observable and constraint are correct.

\begin{equation}
\begin{split}
\Delta a_\mu &= A_{\mu\mu}^M m_\mu^2 \\
\func{BR}\left( \mu \rightarrow e \gamma \right) &= \frac{3(4\pi)^3 \alpha_{%
\func{em}}}{4G_F^2} \left( \lvert A_{\mu e}^M \rvert^2 + \lvert A_{\mu e}^E
\rvert^2 \right)  \label{eqn:definition_muong2_muegamma_reivewonLFV}
\end{split}%
\end{equation}

One difference between $A_{\mu\mu}^{M,E}$ and $A_{L,R}$ is $%
that A_{\mu\mu}^{M,E} $ is only determined by the internal structure of the loop in
Figure \ref{fig:diagrams_muong2_with_allneutrinos}, whereas $A_{L,R}$ is the
extended factor by multiplying $A_{\mu\mu}^{M,E}$ by the helicity flip mass
in one of the external legs. Therefore, it is natural to think $A_{\mu\mu}^{M,E}$
is the same as $A_{\mu e}^{M,E}$ since their internal structure of loop are
exactly same\footnote{%
One can concern the coefficient at the vertex with electron. However, this
change is already reflected on the loop integration $A_R$ of Equation \ref%
{eqn:left_handed_amplitdue_AR} by $U_{1n}$. For the muon anomaly, the
coefficient is simply replaced by $U_{2n}$, therefore, modification of the
coefficient at the vertex does not harm our argument.}. The muon anomalous
magnetic moment and the branching ratio of $\mu \rightarrow e \gamma$ take
the form:

\begin{equation}
\begin{split}
\Delta a_\mu &= \frac{\alpha_W}{32\pi} \frac{m_\mu^2}{M_W^2}
\sum_{n=1,2,3,4,5} U_{2n} U_{2n}^* F(x_n) \\
\func{BR}\left(\mu \rightarrow e \gamma \right) &= \frac{3\alpha_{\func{em}}%
}{32\pi} \lvert \sum_{n=1,2,3,4,5} U_{2n} U_{1n}^* F(x_n) \rvert^2
\label{eqn:muong2_muegamma_W_ourprediction}
\end{split}%
\end{equation}

where the $\alpha_W$ is the weak coupling constant. As for the branching
ratio of $\mu \rightarrow e \gamma$ in Equation \ref%
{eqn:muong2_muegamma_W_ourprediction}, we showed that substituting $A_{\mu
e} $ back into the branching ratio in Equation \ref%
{eqn:definition_muong2_muegamma_reivewonLFV} is exactly consistent with the
one in Equation \ref{eqn:prediction_muegamma_neutrinos}. Expanding the unitary
mixing matrices in the muon anomaly prediction in Equation \ref%
{eqn:muong2_muegamma_W_ourprediction}, yields the following relation:

\begin{equation}
\Delta a_\mu = \frac{\alpha_W}{32\pi} \frac{m_\mu^2}{M_W^2} \left( (1 -
2\eta_{22})F(0) + 2\eta_{22}F(x_4) \right).
\label{eqn:expansion_muong2_ourprediction}
\end{equation}

Looking at Equation \ref%
{eqn:expansion_muong2_ourprediction}, it is clear that the SM part which is without $\eta$
and the BSM having $\eta$ are entangled together. We arrive at the right
prediction for the muon anomaly at one-loop by removing the SM part from
Equation \ref{eqn:expansion_muong2_ourprediction}

\begin{equation}
\Delta a_\mu = \frac{\alpha_W}{16\pi} \frac{m_\mu^2}{M_W^2} \eta_{22} \left(
F(x_4) - F(0) \right).  \label{eqn:expansion_muong2_correct_prediction}
\end{equation}

Similarly to the branching ratio of $\mu \rightarrow e \gamma$ decay, it can
be confirmed that the prediction for the muon anomaly also consists of the
factor of deviation of unitarity $\eta$.

\subsection{The anomalous electron magnetic moment $g-2$}

\label{subsec:electron_g2_Wboson}

\begin{figure}[]
\centering
\begin{subfigure}{0.48\textwidth}
	\includegraphics[width=1.0\textwidth]{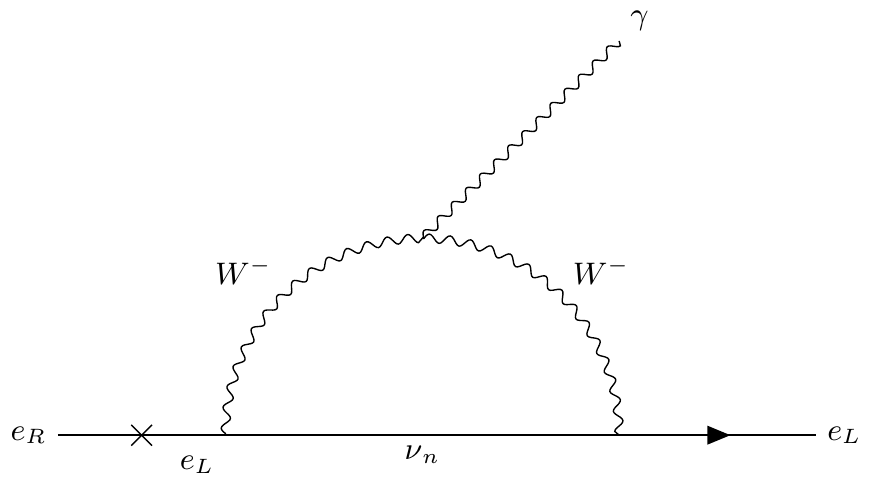}
\end{subfigure} \hspace{0.1cm}
\begin{subfigure}{0.48\textwidth}
	\includegraphics[width=1.0\textwidth]{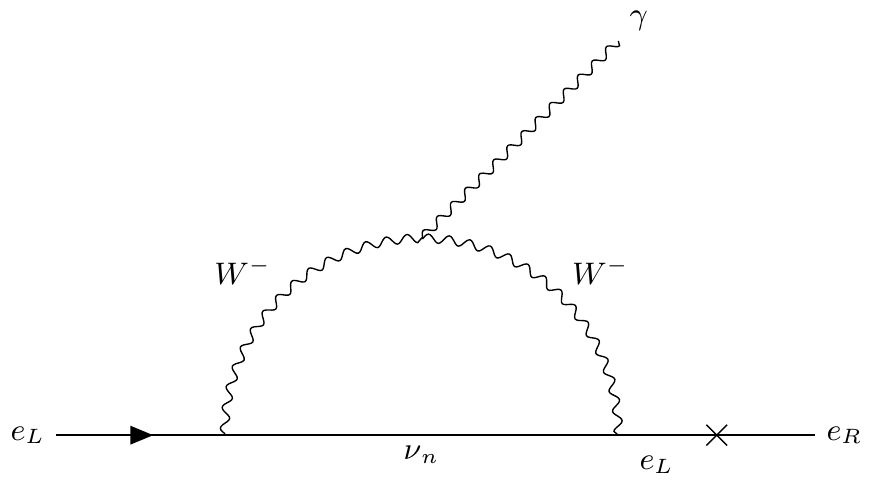}
\end{subfigure}
\caption{Diagrams for electron anomalous magnetic moment with all neutrinos. Here $n=1,2,3,4,5$.}
\label{fig:diagrams_electrong2_with_allneutrinos}
\end{figure}

As in the muon anomalous magnetic moment, the same diagrams with external
particles replaced by electrons can be generated in Figure \ref{fig:diagrams_electrong2_with_allneutrinos}.

Using the complete form of the muon anomaly prediction in Equation \ref%
{eqn:expansion_muong2_correct_prediction}, we can derive the right
prediction for the electron anomalous magnetic moment with slight modifications $%
m_\mu \rightarrow m_e, \eta_{22} \rightarrow \eta_{11}$.

\begin{equation}
\Delta a_e = \frac{\alpha_W}{16\pi} \frac{m_e^2}{M_W^2} \eta_{11} \left(
F(x_4) - F(0) \right).  \label{eqn:expansion_electrong2_correct_prediction}
\end{equation}

\subsection{Numerical analysis of $W$ exchange contributions}


The presence of heavy vector-like neutrinos leads to the deviation of
unitarity and the observables $\Delta a_{\mu,e}$ and constraint $\func{BR}%
\left(\mu \rightarrow e \gamma\right)$ can be written in terms of the factor
of non-unitarity $\eta$.

\begin{equation}
\begin{split}
\func{BR}\left(\mu \rightarrow e \gamma \right) &= \frac{3\alpha_{\func{em}}%
}{8\pi} \lvert \eta_{21} \rvert^2 \left( F(x_4) - F(0) \right)^2 \\
\Delta a_\mu &= \frac{\alpha_W}{16\pi} \frac{m_\mu^2}{M_W^2} \eta_{22}
\left( F(x_4) - F(0) \right) \\
\Delta a_e &= \frac{\alpha_W}{16\pi} \frac{m_e^2}{M_W^2} \eta_{11} \left(
F(x_4) - F(0) \right).  \label{eqn:prediction_muegamma_muong2_electrong2}
\end{split}%
\end{equation}

\subsubsection{The branching ratio of $\protect\mu \rightarrow e \protect\gamma$
decay}

We consider the branching ratio of $\mu \rightarrow e \gamma$ decay first.
Since we assume that mass of heavy vector-like neutrinos are heavier than $1%
\func{TeV}$, the value of $F(0)$ for the light neutrinos converges to approximately $%
3.3$, while that of $F(x_4)$ for the heavy vector-like neutrino converges to $%
1.3 $. Therefore, the branching ratio of $\mu \rightarrow e \gamma$ decay
can be reduced to\cite{Hernandez-Garcia:2019uof}

\begin{equation}
\func{BR}\left(\mu \rightarrow e \gamma \right) = \frac{3\alpha_{\func{em}}}{%
8\pi} \lvert \eta_{21} \rvert^2 \left( F(x_4) - F(0) \right)^2 \leq \frac{%
3\alpha_{\func{em}}}{2\pi} \lvert \eta_{21} \rvert^2.
\label{eqn:muegamma_simplified}
\end{equation}

\begin{figure}[]
\centering
\begin{subfigure}{0.48\textwidth}
	\includegraphics[keepaspectratio,width=\textwidth]{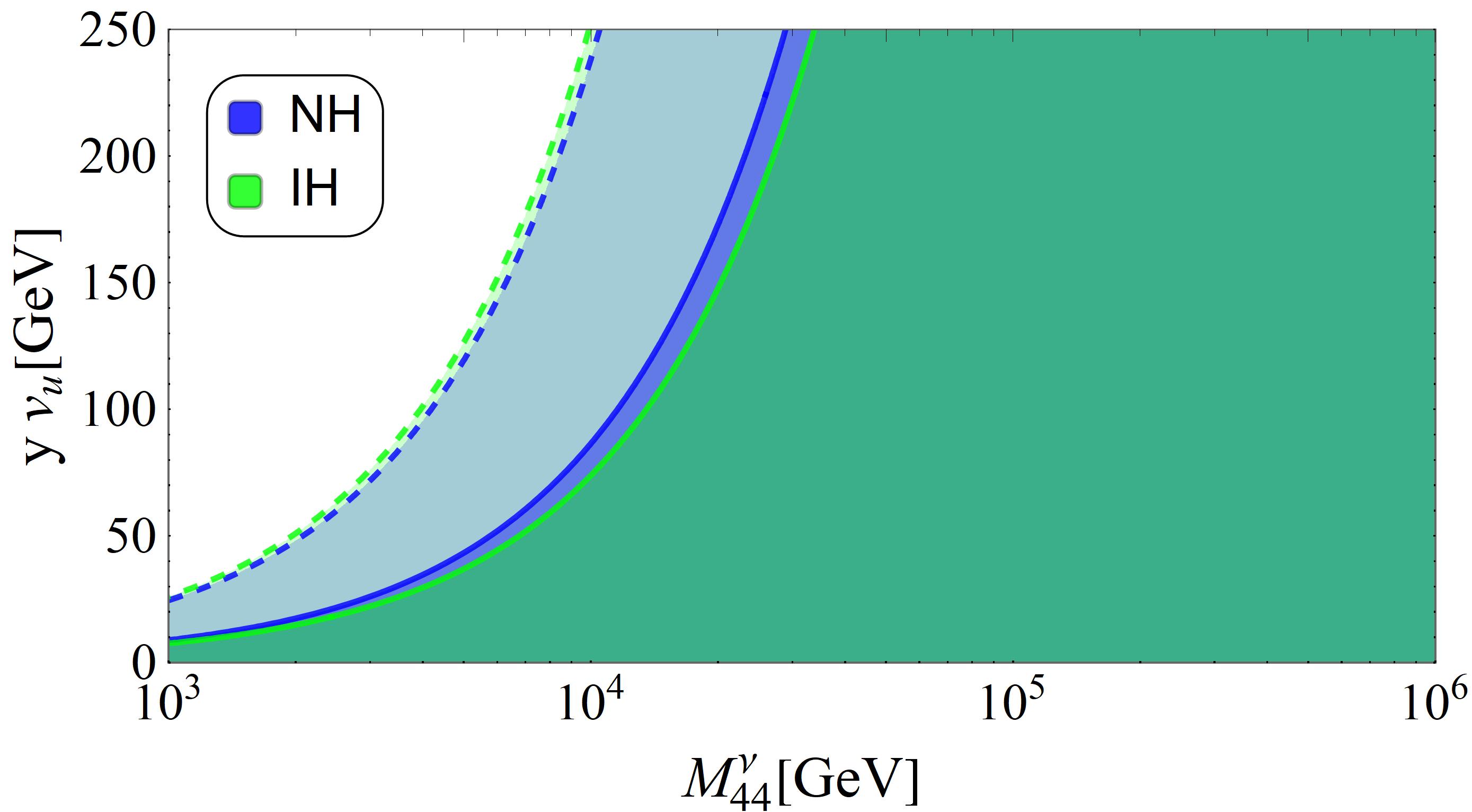}
\end{subfigure} \hspace{0.1cm}
\begin{subfigure}{0.48\textwidth}
	\includegraphics[keepaspectratio,width=\textwidth]{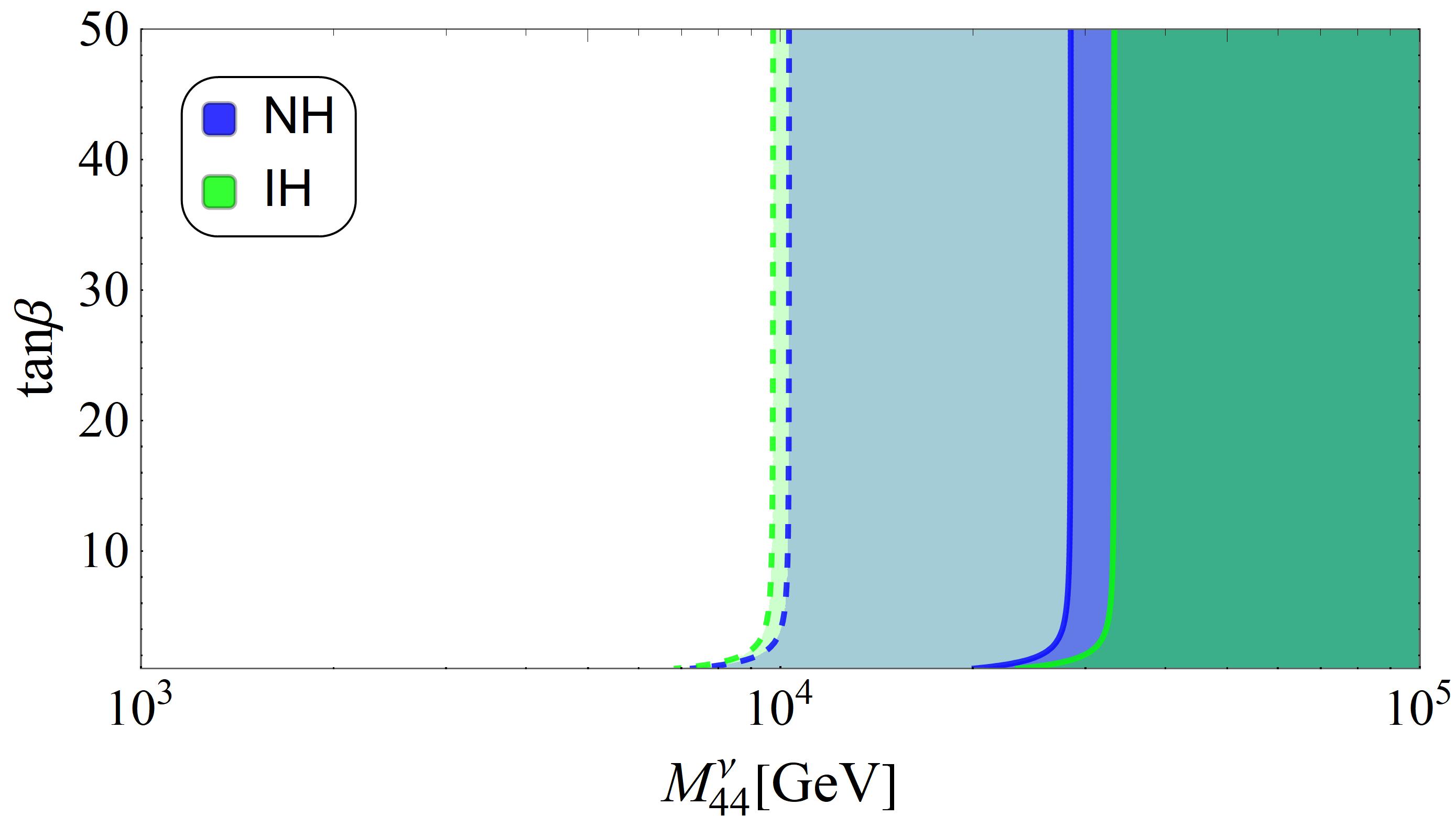}
\end{subfigure}
\caption{The left plot is an available parameter space for two free
parameters: mass of vector-like neutrino $M_{44}^{\protect\nu}$ and SM
up-type Higgs vev $v_u$. Here, the free parameter $y$ is set to $1$. The
right plot is the case where vev of the up-type Higgs is constrained from $%
246/\protect\sqrt{2} \simeq 174$ to $246\func{GeV}$ or from $\tan\protect%
\beta = 1$ to $50$ in a same way}
\label{fig:parameter_space_muegamma_M44_vu_tanbeta}
\end{figure}

The non-unitarity $\eta$ of Equation \ref{eqn:deviation_of_unitarity}
consists of four free parameters: mass of heavy vector-like neutrinos $%
M_{44}^{\nu}$, a real number $y$, a CP violation phase $\delta$, and a
Majorana phase $\alpha$. The experimental branching ratio of $\mu
\rightarrow e \gamma$ decay constrains the minimal parameter space in terms
of $M_{44}^{\nu}$ and $y$, while setting up two phases $\delta,\alpha$ which
maximize or minimize the branching ratio of $\mu \rightarrow e \gamma$\cite%
{Hernandez-Garcia:2019uof}, and the minimal parameter space is shown in Figure \ref{fig:parameter_space_muegamma_M44_vu_tanbeta}.

The left plot in Figure \ref{fig:parameter_space_muegamma_M44_vu_tanbeta} is
an available parameter space for mass of the vector-like neutrino versus the free
parameter $y$ times SM up-type Higgs vev $v_u$. The blue bold line
corresponds to bound of the branching ratio of $\mu \rightarrow e \gamma$
decay at the normal hierarchy with CP violation phase $\delta = 0$ and
Majorana phase $\alpha = 0$ and this line can be relaxed up to the blue
dotted line where $\delta = 0$, $\alpha = 2\pi$. The green bold(dotted) line
corresponds to the inverted hierarchy with $\delta = \pi/2(0)$ and $\alpha = 
\frac{9\pi}{10}(0)$. Since we are especially interested in the range of SM
up-type Higgs vev $v_u$ from $174$ to $246\func{GeV}$, the right plot
consistent with the interested range is extracted from the left after
replacing $v_u$ by $\tan\beta=v_u/v_d$ using the relation $v_u^2 + v_d^2 =
(246 \func{GeV})^2$. \newline
~\newline
As for the constraint of deviation of unitarity $\eta$ with the CLFV $\mu
\rightarrow e \gamma$ decay at $1 \sigma$, it is given by\cite%
{Fernandez-Martinez:2016lgt,Tanabashi:2018oca}

\begin{equation}
\lvert \eta_{21} \rvert \leq 8.4 \times 10^{-6}.
\label{eqn:constraint_nonunitarity_eta21}
\end{equation}

\subsubsection{The muon and electron anomalous magnetic moments $\Delta a_{%
\protect\mu,e}$}

\label{subsec:muon_and_electrong2_W}

As in the constraint for $\eta_{21}$ in Equation \ref%
{eqn:constraint_nonunitarity_eta21}, the other non-unitarities $\eta_{11,22}$
for the electron and muon anomalous mangetic moment are given by\cite%
{Hernandez-Garcia:2019uof,Fernandez-Martinez:2016lgt}

\begin{equation}
\begin{split}
\eta_{11} &< 4.2 \times 10^{-4} \text{ (for NH) }, \quad < 4.8 \times
10^{-4} \text{ (for IH) } \\
\eta_{22} &< 2.9 \times 10^{-7} \text{ (for NH) }, \quad < 2.4 \times
10^{-7} \text{ (for IH) }  \label{eqn:non_unitarity_eta11_eta_22}
\end{split}%
\end{equation}

With the constraints $\eta_{11,22}$ in Equation \ref%
{eqn:non_unitarity_eta11_eta_22}, we can calculate impact of the muon and
electron anomalous magnetic moments at NH(IH) using Equation \ref%
{eqn:prediction_muegamma_muong2_electrong2}.

\begin{equation}
\begin{split}
\Delta a_\mu &= \frac{\alpha_W}{16\pi} \frac{m_\mu^2}{M_W^2} \eta_{22}
\left( F(x_4) - F(0) \right) \simeq -6.6(-5.5) \times 10^{-16} \\
\Delta a_e &= \frac{\alpha_W}{16\pi} \frac{m_e^2}{M_W^2} \eta_{11} \left(
F(x_4) - F(0) \right) \simeq -2.2(-2.6) \times 10^{-17}
\label{eqn:magnitude_muon_electrong2_with_constraints}
\end{split}%
\end{equation}

There are two interesting features in the above prediction for the muon and
electron anomalous magnetic moments. One feature is sign of each prediction.
As mentioned in the introduction, this prediction with the $W$ exchange can not flip the sign of
each anomaly. In order to explain both anomalies at $1 \sigma$, the
prediction for both anomalies with $W$ exchange requires additional
contributions such as $Z^\prime$ or scalar exchange. Another feature is
magnitude of each prediction. For the muon anomaly, the experimental order
of magnitude at $1 \sigma$ is about $10^{-9}$, however our prediction is
much smaller than that of the experimental bound as well as the electron
anomaly, which means the non-unitarity derived from the presence of heavy
vector-like neutrino can not bring the anomalies to the observable level.
This inadequate prediction with $W$ exchange has been a good motivation to
search for another possibility such as scalar exchange.

\section{HIGGS EXCHANGE CONTRIBUTIONS TO $\left( g-2 \right)_\protect\mu, \left( g-2
\right)_e$ AND $\func{BR}\left(\protect\mu \rightarrow e \protect\gamma %
\right)$}

\label{sec:Analytic_arguments_muon_electron_g2_scalars}

The relevant sector for the muon and electron anomalous magnetic moments with
scalar exchange is the charged lepton Yukawa matrix which can be expressed in the mass insertion formalism as,
\begin{equation}
\begin{split}
y_{ij}^e &= 
\begin{pmatrix}
0 & 0 & 0 \\ 
0 & y_{24}^e x_{42}^{e} & y_{24}^e x_{43}^{e} \\ 
0 & y_{34}^e x_{42}^{e} & y_{34}^e x_{43}^{e}
\end{pmatrix}
\frac{\left\langle \phi \right\rangle}{M_{44}^{e}} + 
\begin{pmatrix}
y_{15}^e x_{51}^{e} & y_{15}^e x_{52}^{e} & y_{15}^e x_{53}^{e} \\ 
y_{25}^e x_{51}^{e} & y_{25}^e x_{52}^{e} & y_{25}^e x_{53}^{e} \\ 
y_{35}^e x_{51}^{e} & y_{35}^e x_{52}^{e} & y_{35}^e x_{53}^{e}
\end{pmatrix}
\frac{\left\langle \phi \right\rangle}{M_{55}^{e}} \\
&+ 
\begin{pmatrix}
y_{51}^e x_{15}^{L} & y_{52}^e x_{15}^{L} & y_{53}^e x_{15}^{L} \\ 
y_{51}^e x_{25}^{L} & y_{52}^e x_{25}^{L} & y_{53}^e x_{25}^{L} \\ 
y_{51}^e x_{35}^{L} & y_{52}^e x_{35}^{L} & y_{53}^e x_{35}^{L}
\end{pmatrix}
\frac{\left\langle \phi \right\rangle}{M_{55}^{L}} + 
\begin{pmatrix}
0 & 0 & 0 \\ 
0 & 0 & 0 \\ 
0 & 0 & x_{34}^L y_{43}^e
\end{pmatrix}
\frac{\left\langle \phi \right\rangle}{M_{44}^{L}}
\end{split}
\label{eqn:effective_Yukawa_matrix_charged_leptons_assumption_second}
\end{equation}

The effective Yukawa matrix of Equation \ref%
{eqn:effective_Yukawa_matrix_charged_leptons_assumption_second} in the
mass basis is diagonalized by the universal seesaw mechanism due to involving a
few of different mass scales. Therefore, the only diagonal
components should alive in the mass matrix. In order to make the mass matrix
diagonal, we assume that $y_{34}^e = x_{43}^{e} = y_{15,25,35}^e =
x_{51,52,53}^{e} = x_{25,35}^L = y_{52,53}^e = 0$. Then, the mass matrix is
reduced to

\begin{equation}
\begin{split}
y_{ij}^e &= 
\begin{pmatrix}
0 & 0 & 0 \\ 
0 & y_{24}^e x_{42}^{e} & 0 \\ 
0 & 0 & 0
\end{pmatrix}
\frac{\left\langle \phi \right\rangle}{M_{44}^{e}} + 
\begin{pmatrix}
0 & 0 & 0 \\ 
0 & 0 & 0 \\ 
0 & 0 & 0
\end{pmatrix}
\frac{\left\langle \phi \right\rangle}{M_{55}^{e}} + 
\begin{pmatrix}
y_{51}^e x_{15}^{L} & 0 & 0 \\ 
0 & 0 & 0 \\ 
0 & 0 & 0
\end{pmatrix}
\frac{\left\langle \phi \right\rangle}{M_{55}^{L}} + 
\begin{pmatrix}
0 & 0 & 0 \\ 
0 & 0 & 0 \\ 
0 & 0 & x_{34}^L y_{43}^e
\end{pmatrix}
\frac{\left\langle \phi \right\rangle}{M_{44}^{L}} \\
y_{ij}^e &= 
\begin{pmatrix}
y_{51}^e s_{15}^L & 0 & 0 \\ 
0 & y_{24}^e s_{24}^{e} & 0 \\ 
0 & 0 & y_{43}^e s_{34}^L%
\end{pmatrix},
\label{eqn:diagonalized_Yukawa_matrix_charged_leptons}
\end{split}%
\end{equation}

where $s_{15}^L \simeq x_{15}^L \left\langle \phi \right\rangle/M_{55}^{L}$, 
$s_{24}^{e} \simeq x_{42}^{e} \left\langle \phi \right\rangle/M_{44}^{e}$, $%
s_{34}^L \simeq x_{34}^L \left\langle \phi \right\rangle/M_{44}^L$ and the
diagonal elements from top-left to bottom-right should be responsible for
electron, muon and tau Yukawa constants, respectively. After removing all
irrelevant terms to both anomalies and applying the assumption, the $7
\times 7$ mass matrix in the interaction basis is also reduced to as below:

\begin{equation}
M^{e}=\left( 
\begin{array}{c|cccc}
& e_{1R} & e_{2R} & e_{4R} & \widetilde{L}_{5R} \\ \hline
\overline{L}_{1L} & 0 & 0 & 0 & x_{15}^{L}v_{\phi } \\ 
\overline{L}_{2L} & 0 & 0 & y_{24}^{e}v_{d} & 0 \\ 
\overline{L}_{5L} & y_{51}^{e}v_{d} & 0 & 0 & M_{55}^{L} \\ 
\overline{\widetilde{e}}_{4L} & 0 & x_{42}^{e}v_{\phi } & M_{44}^{e} & 0
\end{array}%
\right) \label{eqn:reduced_charged_lepton_mass_matrix}
\end{equation}

The reduced charged lepton mass matrix of Equation \ref%
{eqn:reduced_charged_lepton_mass_matrix} clearly tells
that no mixing between charged leptons arise so the branching ratio of $\mu
\rightarrow e \gamma$ is naturally satisfied under this scenario. The scalar
exchange for both anomalies can be realized by closing the Higgs sectors in
Figure \ref{fig:diagrams_charged_leptons_mass_insertion} as per Figure \ref%
{fig:muon_electrong2_scalar_exchange}.

\begin{figure}[]
\centering
\begin{subfigure}{0.48\textwidth}
	\includegraphics[width=1.0\textwidth]{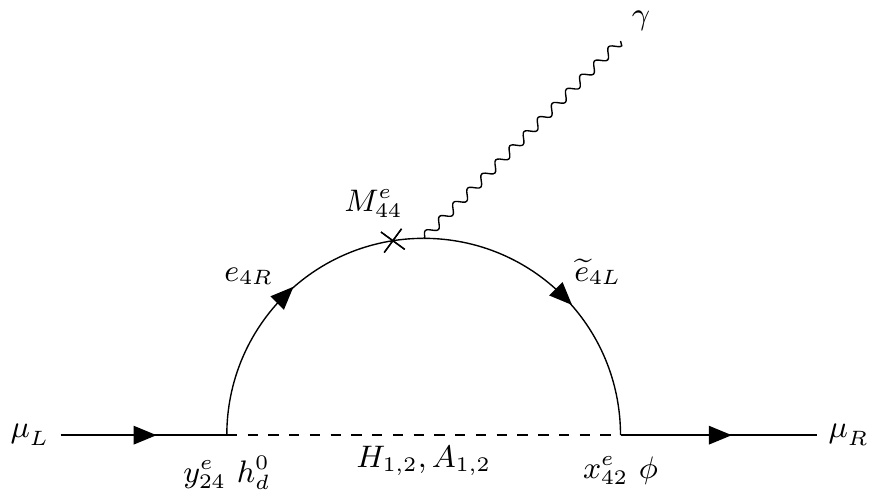}
\end{subfigure} \hspace{0.1cm}
\begin{subfigure}{0.48\textwidth}
	\includegraphics[width=1.0\textwidth]{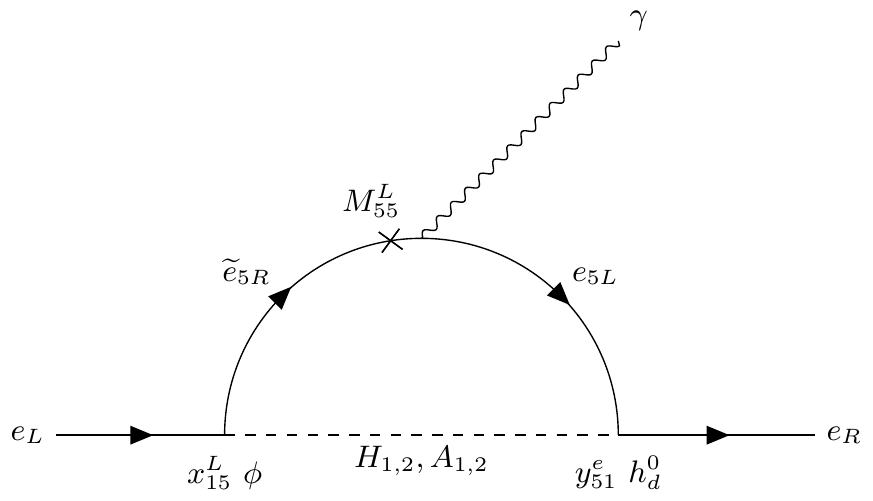}
\end{subfigure}
\caption{Diagrams contributing to the muon anomaly (left) and
the electron anomaly (right) where $H_{1,2}$ are CP-even non-SM
scalars and $A_{1,2}$ are CP-odd scalars in the physical basis}
\label{fig:muon_electrong2_scalar_exchange}
\end{figure}

In Figure \ref{fig:muon_electrong2_scalar_exchange}, the CP-even non-SM scalars $%
H_{1,2}$ and CP-odd scalars $A_{1,2}$ appear as a result of mixing between
Higgses $H_u,H_d$ and $\phi$ in the interaction basis. The Higgs sector in the interaction
basis is defined by

\begin{equation}
\begin{split}
H_u = 
\begin{pmatrix}
H_u^+ \\ 
v_u + \frac{1}{\sqrt{2}}\left( \func{Re}H_{u}^0 + i\func{Im}H_{u}^0 \right)%
\end{pmatrix}%
, \quad H_d = 
\begin{pmatrix}
v_d + \frac{1}{\sqrt{2}}\left( \func{Re}H_{d}^0 + i\func{Im}H_{d}^0 \right)
\\ 
H_d^-%
\end{pmatrix}%
, \quad \phi = \frac{1}{\sqrt{2}} \left( v_\phi + \func{Re}\phi + i\func{Im}%
\phi \right).  \label{eqn:Higgs_fields_mass_basis}
\end{split}%
\end{equation}

For consistency, we equate $v_u, v_d$ and $v_\phi$ to $v_1, v_2$ and $v_3$,
respectively.

\subsection{The 2HDM scalar potential}

\label{subsec:the_scalar_potential} The scalar potential of the model under
consideation takes the form:%
\begin{equation}
\begin{split}
V =&\mu _{1}^{2}\left( H_{u}H_{u}^{\dagger }\right) +\mu _{2}^{2}\left(
H_{d}H_{d}^{\dagger }\right) +\mu _{3}^{2}\left( \phi \phi ^{\ast }\right)
+\mu _{\func{sb}}^{2}\left[ \phi ^{2}+\left( \phi ^{\ast }\right) ^{2}\right]
+\lambda _{1}\left( H_{u}H_{u}^{\dagger }\right) ^{2}+\lambda _{2}\left(
H_{d}H_{d}^{\dagger }\right) ^{2} \\
&+\lambda _{3}\left( H_{u}H_{u}^{\dagger }\right) \left(
H_{d}H_{d}^{\dagger }\right) +\lambda _{4}\left( H_{u}H_{d}^{\dagger
}\right) \left( H_{d}H_{u}^{\dagger }\right) +\lambda _{5}\left( \varepsilon
_{ij}H_{u}^{i}H_{d}^{j}\phi ^{2}+\func{h.c}\right)  \\
&+\lambda _{6}\left( \phi \phi ^{\ast }\right) ^{2}+\lambda _{7}\left( \phi
\phi ^{\ast }\right) \left( H_{u}H_{u}^{\dagger }\right) +\lambda _{8}\left(
\phi \phi ^{\ast }\right) \left( H_{d}H_{d}^{\dagger }\right) ,
\end{split}
\end{equation}%
where the $\lambda _{i}$ ($i=1,2,\cdots ,8$) are dimensionless parameters
whereas the $\mu _{j}$ ($j=1,2,3$) are dimensionful parameters and $\mu _{%
\func{sb}}$ is a dimensionfull soft-breaking parameter. We consider the $%
U(1)^{\prime }$ symmetry as global in this model so our model does not
feature $Z^{\prime }$ boson and the scalar potential requires the inclusion
of the soft-breaking mass term $-\mu _{\func{sb}}^{2}\left[ \phi ^{2}+\left( \phi
^{\ast }\right) ^{2}\right] $ in order to prevent the appearance of a
massless scalar state arising from the imaginary part of $\phi $.

The minimization conditions of the scalar potential yield the following
relations: 
\begin{equation}
\begin{split}
\mu _{1}^{2} &=-2\lambda _{1}v_{1}^{2}-\lambda _{3}v_{2}^{2}-\frac{1}{2}%
\lambda _{7}v_{3}^{2}+\frac{\lambda _{5}v_{2}v_{3}^{2}}{2v_{1}}, \\
\mu _{2}^{2} &=-2\lambda _{2}v_{2}^{2}-\lambda _{3}v_{1}^{2}-\frac{1}{2}%
\lambda _{8}v_{3}^{2}+\frac{\lambda _{5}v_{3}^{2}v_{1}}{
2v_{2}}, \\
\mu _{3}^{2} &=-\lambda _{8}v_{2}^{2}-\lambda _{6}v_{3}^{2}+v_{1}\left(
2\lambda _{5}v_{2}-\lambda _{7}v_{1}\right) -2\mu _{\func{sb}}^{2}.
\end{split}
\end{equation}

\subsection{Mass matrix for CP-even, CP-odd neutral and charged scalars}

\label{subsec:mass_mamtrix_for_CP_even_odd_charged} The squared mass matrix
for the CP-even scalars in the basis $\left( \func{Re}H_{u}^{0},\func{Re}%
H_{d}^{0},\func{Re}\phi \right) $ takes the form:%
\begin{equation}
\mathbf{M}_{\func{CP-even}}^{2}=\left( 
\begin{array}{ccc}
4\lambda _{1}v_{1}^{2}+\frac{\lambda _{5}v_{2}v_{3}^{2}}{2v_{1}} & -\frac{1}{2%
}\lambda _{5}v_{3}^{2}+2\lambda _{3}v_{1}v_{2} & \sqrt{2}v_{3}\left( -\lambda
_{5}v_{2}+\lambda _{7}v_{1}\right) \\ 
-\frac{1}{2}\lambda _{5}v_{3}^{2}+2\lambda _{3}v_{1}v_{2} & 4\lambda
_{2}v_{2}^{2}+\frac{\lambda _{5}v_{1}v_{3}^{2}}{2v_{2}} & \sqrt{2}%
v_{3}\left( -\lambda _{5}v_{1}+\lambda _{8}v_{2}\right) \\ 
\sqrt{2}v_{3}\left( -\lambda _{5}v_{2}+\lambda _{7}v_{1}\right) & \sqrt{2}%
v_{3}\left( -\lambda _{5}v_{1}+\lambda _{8}v_{2}\right) & 2\lambda
_{6}v_{3}^{2}
\end{array}%
\right).  \label{eqn:mass_matrix_CP_even}
\end{equation}
From the mass matrix given above, we find that the CP-even scalar spectrum
is composed of the $125$ GeV\ SM-like Higgs $h$ and two non-SM CP-even
Higgses $H_{1,2}$. Furthermore, we assume that no mixing between the SM
physical Higgs $h$ and the two non-SM CP-even Higgses $H_{1,2}$ arise and
this assumption constrains the $(1,2)$, $(1,3)$, $(2,1)$ and $(3,1)$ elements of
CP-even mass matrix of Equation \ref{eqn:mass_matrix_CP_even}. The
constraints are given by the following decoupling limit scenario

\begin{equation}
\begin{split}
\lambda_5 &= \frac{4 v_1 v_2}{v_3^2} \lambda_3 \\
\lambda_7 &= \frac{v_2}{v_1} \lambda_5 = \frac{4 v_2^2}{v_3^2} \lambda_3,
\label{eqn:constraints_in_mass_matrix_CP_even}
\end{split}%
\end{equation}

and then the CP-even mass matrix of Equation \ref{eqn:mass_matrix_CP_even}
with the constraints is simplified to

\begin{equation}
\mathbf{M}_{\func{CP-even}}^{2}=\left( 
\begin{array}{ccc}
4\lambda _{1}v_{1}^{2} + 2v_{2}^{2} \lambda_{3} & 0 & 0 \\ 
0 & 4\lambda _{2}v_{2}^{2} + 2v_{1}^{2} \lambda_{3} & \sqrt{2}v_{3}\left( -%
\frac{4 v_{1}^{2} v_{2}}{v_{3}^{2}} \lambda_{3} +\lambda _{8}v_{2}\right) \\ 
0 & \sqrt{2}v_{3}\left( -\frac{4 v_{1}^{2} v_{2}}{v_{3}^{2}} \lambda_{3}
+\lambda _{8}v_{2}\right) & 2\lambda _{6}v_{3}^{2}%
\end{array}%
\right).  \label{eqn:mass_matrix_CP_even_simplified}
\end{equation}

In the above given decoupling limit scenario, chosen in order to simplify our analysis, 
the CP-even neutral scalar states contained in the $SU(2)$
doublet $H_{u}$ will not mix with the CP-even neutral ones contained in $%
H_{d}$. In such limit, the neutral CP-even states of $H_{u}$ will not
feature mixing with the gauge singlet scalar $\phi $. Thus, the lightest $125%
\func{GeV}$ CP-even scalar of our model will have couplings to the SM
particles close to the SM expectation, which is consistent with the current
experimental data.

Diagonalizing the simplified CP-even mass matrix, it reveals masses of the
physical SM Higgs $h$ and non-SM CP-even scalars $H_{1,2}$ in the physical
basis $\left( h, H_1, H_2 \right)$

\begin{equation}
R_{\func{CP-even}}^\dagger \mathbf{M}_{\func{CP-even}}^2 R_{\func{CP-even}}
= \func{diag} \left( m_h^2, m_{H_1}^2, m_{H_2}^2 \right).
\label{eqn:diagonalising_CP_even_mass_matrix}
\end{equation}

The SM Higgs $h$ is appeared as $\func{Re}H_{u}^{0}$ itself and the non-SM
CP-even scalars $H_{1,2}$ are the states which $\func{Re}H_{d}^{0}$ is mixed
with $\func{Re}\phi $. Regarding the CP-odd scalar sector, we find that the
squared mass matrix for the CP-odd scalars in the basis $\left( \func{Im}%
H_{u}^{0},\func{Im}H_{d}^{0},\func{Im}\phi \right) $ is given by: 
\begin{equation}
\mathbf{M}_{\func{CP-odd}}^{2}=\left( 
\begin{array}{ccc}
\frac{\lambda _{5}v_{2}v_{3}^{2}}{2v_{1}} & \frac{1}{2}\lambda
_{5}v_{3}^{2} & \sqrt{2}\lambda _{5}v_{2}v_{3} \\ 
\frac{1}{2}\lambda _{5}v_{3}^{2} & \frac{\lambda _{5}v_{1}v_{3}^{2}}{2v_{2}%
} & \sqrt{2}\lambda _{5}v_{1}v_{3} \\ 
\sqrt{2}\lambda _{5}v_{2}v_{3} & \sqrt{2}\lambda _{5}v_{1}v_{3} & 
4\lambda _{5}v_{1}v_{2}-4\mu _{\func{sb}}^{2}%
\end{array}%
\right).
\end{equation}

The squared CP-odd mass matrix is diagonalized in the same way as in the
CP-even mass matrix and the CP-odd physical basis is given by $\left(
G_{Z},A_{1},A_{2}\right) $ where $G_{Z}$ is the massless Goldstone bosons
associated with the longitudinal components of the $Z$ gauge boson, whereas $%
A_{1}$ and $A_{2}$ are massive non-SM CP-odd scalars

\begin{equation}
R_{\func{CP-odd}}^{\dagger }\mathbf{M}_{\func{CP-odd}}^{2}R_{\func{CP-odd}}=%
\func{diag}\left( 0,m_{A_{1}}^{2},m_{A_{2}}^{2}\right) .
\label{eqn:diagonalizing_CP_odd_mass_matrix}
\end{equation}

Furthermore, the squared mass matrix for the electrically charged scalars is
given by: 
\begin{equation}
\mathbf{M}_{\func{charged}}^{2}=\left( 
\begin{array}{cc}
\lambda _4 v_2^2+\frac{\lambda _5 v_2 v_3^2}{2 v_1} & \lambda _4 v_1 v_2+%
\frac{1}{2} \lambda _5 v_3^2 \\ 
\lambda _4 v_1 v_2+\frac{1}{2} \lambda _5 v_3^2 & \lambda _4 v_1^2+\frac{%
\lambda _5 v_1 v_3^2}{2 v_2}%
\end{array}
\right).
\end{equation}

The charged scalar mass matrix can be diagonalized in the basis $\left(
H_{1}^{\pm}, H_{2}^{\pm} \right)$ as in CP-even or -odd mass matrix:

\begin{equation}
R_{\func{charged}}^\dagger \mathbf{M}_{\func{charged}}^{2} R_{\func{charged}%
} = \func{diag} \left( 0, m_{H^\pm}^2 \right).
\label{eqn:diagonalizing_charged_mass_matrix}
\end{equation}

Then, the electrically charged scalar sector contains the massive scalars $%
H^{\pm }$ and the massless electrically charged scalars $G_{W}^{\pm }$ which
correspond to the Goldstone bosons associated with the longitudinal
components of the $W^{\pm }$ gauge bosons. In the following sections we will
analyze the phenomenological implications of our model in the Higgs diphoton
decay as well as in the muon and electron anomalous magnetic moments. 

\subsection{The Higgs diphoton signal strength}

\label{Diphpoton} The rate for the $h\rightarrow \gamma \gamma $ decay is
given by: 
\begin{equation}
\Gamma (h\rightarrow \gamma \gamma )=\dfrac{\alpha _{\func{em}}^{2}m_{h}^{3}%
}{256\pi ^{3}v^{2}}\left\vert \sum_{f}a_{hff}N_{C}Q_{f}^{2}F_{1/2}(\rho
_{f})+a_{hWW}F_{1}(\rho _{W})+\frac{C_{hH^{\pm }H^{\mp }}v}{2m_{H^{\pm }}^{2}%
}F_{0}(\rho _{H_{k}^{\pm }})\right\vert ^{2},
\end{equation}%
where $\rho _{i}$ are the mass ratios $\rho _{i}=\frac{m_{h}^{2}}{4M_{i}^{2}}
$ with $M_{i}=m_{f},M_{W}$; $\alpha _{\func{em}}$ is the fine structure
constant; $N_{C}$ is the color factor ($N_{C}=1$ for leptons and $N_{C}=3$
for quarks) and $Q_{f}$ is the electric charge of the fermion in the loop.
From the fermion-loop contributions we only consider the dominant top quark
term. Furthermore, $C_{hH^{\pm }H^{\mp }}$ is the trilinear coupling between
the SM-like Higgs and a pair of charged Higges, whereas $a_{htt}$ and $%
a_{hWW}$ are the deviation factors from the SM Higgs-top quark coupling and
the SM Higgs-W gauge boson coupling, respectively (in the SM these factors
are unity). Such deviation factors are close to unity in our model and they
are defined as below:

\begin{equation}
a_{htt} \simeq 1, \quad a_{hWW} = \frac{1}{\sqrt{v_1^2 + v_2^2}} \frac{%
\partial}{\partial h} \left( \sum_{i,j=1,2,3} v_i \left( R_{\func{CP-even}%
}^T \right)_{ij} \left( h, H_1, H_2 \right)_{j} \right) = \frac{v_1}{\sqrt{%
v_1^2 + v_2^2}}
\end{equation}

Furthermore, $F_{1/2}(z)$ and $F_{1}(z)$ are the dimensionless loop factors
for spin-$1/2$ and spin-$1$ particles running in the internal lines of the
loops. These loop factors take the form:

\begin{equation}
\begin{split}
F_{1/2}(z)& =2(z+(z-1)f(z))z^{-2}, \\
F_{1}(z)& =-2(2z^{2}+3z+3(2z-1)f(z))z^{-2}, \\
F_{0}(z)& =-(z-f(z))z^{-2},
\end{split}%
\end{equation}%
with 
\begin{equation}
f(z)=\left\{ 
\begin{array}{lcc}
\arcsin ^{2}\sqrt{2} & \text{for} & z\leq 1 \\ 
&  &  \\ 
-\frac{1}{4}\left( \ln \left( \frac{1+\sqrt{1-z^{-1}}}{1-\sqrt{1-z^{-1}}%
-i\pi }\right) ^{2}\right) & \text{for} & z>1
\end{array}%
\right.
\end{equation}
In order to study the implications of our model in the decay of the $125$
GeV Higgs into a photon pair, one introduces the Higgs diphoton signal
strength $R_{\gamma \gamma }$, which is defined as: 
\begin{equation}
R_{\gamma \gamma }=\frac{\sigma (pp\rightarrow h)\Gamma (h\rightarrow \gamma
\gamma )}{\sigma (pp\rightarrow h)_{\func{SM}}\Gamma (h\rightarrow \gamma
\gamma )_{\func{SM}}}\simeq a_{htt}^{2}\frac{\Gamma (h\rightarrow \gamma
\gamma )}{\Gamma (h\rightarrow \gamma \gamma )_{\func{SM}}}.  \label{eqn:hgg}
\end{equation}%
That Higgs diphoton signal strength, normalizes the $\gamma \gamma $ signal
predicted by our model in relation to the one given by the SM. Here we have
used the fact that in our model, single Higgs production is also dominated
by gluon fusion as in the Standard Model.

The ratio $R_{\gamma \gamma }$ has been measured by CMS and ATLAS
collaborations with the best fit signals \cite{Sirunyan:2018ouh,Aad:2019mbh}%
: 
\begin{equation}
R_{\gamma \gamma }^{\func{CMS}}=1.18_{-0.14}^{+0.17}\quad \text{and}\quad R_{\gamma
\gamma }^{\func{ATLAS}}=0.96\pm 0.14.  \label{eqn:rgg}
\end{equation}

As it will be shown in the next subsection, the constraints arising from the Higgs diphoton decay rate will be considered in our numerical analysis. 

\subsection{The muon and electron anomalous magnetic moments}

\label{gminus2} 

The Yukawa interactions relevant for the computation of the muon anomalous
magnetic moment are:

%
%
%

\begin{equation}
\mathcal{L}_{\Delta a_\mu} = y_{24}^e \mu \left( \func{Re} H_{d}^0 - i
\gamma^5 \func{Im} H_{d}^0 \right) \overline{e}_4 + x_{42}^e \widetilde{e}_4
\left( \func{Re} \phi - i \gamma^5 \func{Im} \phi \right) \overline{e}_2 +
M_{44}^e \widetilde{e}_4 \overline{e}_4 + \func{h.c.}
\label{eqn:Lagrangian_muon_anomaly_expansion}
\end{equation}
where the Yukawa coupling constants $y_{24}^e,x_{42}^e$ are assumed to be real, the scalar fieds have been expanded by their real and imaginary parts and the properties of the projection operators $P_{L,R}$ acting on the charged leptonic fields have been used.

By expressing the scalar fields in the interaction basis in terms of the scalar fields in the physical basis, the charged lepton Yukawa interactions relevant for the computation of the $g-2$ anomalies take the form:

\begin{equation}
\begin{split}
\mathcal{L}_{\Delta a_{\mu }}& =y_{24}^{e}\mu \left(
(R_{e}^{T})_{22}H_{1}+(R_{e}^{T})_{23}H_{2}-i\gamma
^{5}(R_{o}^{T})_{22}A_{1}-i\gamma ^{5}(R_{o}^{T})_{23}A_{2}\right) \overline{%
e}_{4} \\
& +x_{42}^{e}\widetilde{e}_{4}\left(
(R_{e}^{T})_{32}H_{1}+(R_{e}^{T})_{33}H_{2}-i\gamma ^{5}( R_{o}^{T})
_{32}A_{1}-i\gamma ^{5}(R_{o}^{T})_{33}A_{2}\right) \overline{e}%
_{2}+M_{44}^{e}\widetilde{e}_{4}\overline{e}_{4}+\func{h.c.}
\end{split}%
\end{equation}

where we are using the unitary gauge where the contributions arising from
unphysical Goldstone bosons to the muon anomaly are excluded and we shorten
the notations $R_{\func{CP}}$ by $R_{e(o)}$. Here $R_{e}$ and $R_{o}$ are the rotation matrices that diagonalize the squared mass matrices for the CP even and CP odd scalars, respectively. Then, it follows that the muon and electron anomalous magnetic moments in the scenario of diagonal SM charged lepton mass matrix take the form:

\begin{equation}
\begin{split}
\Delta a_{\mu }=y_{24}^{e}x_{42}^{e}\frac{m_{\mu }^{2}}{8\pi ^{2}}\Big[&
\left( R_{e}^{T}\right) _{22}\left( R_{e}^{T}\right) _{32}I_{S}^{(\mu
)}\left( m_{e_{4}},m_{H_{1}}\right) +\left( R_{e}^{T}\right) _{23}\left(
R_{e}^{T}\right) _{33}I_{S}^{(\mu )}\left( m_{e_{4}},m_{H_{2}}\right) \\
& -\left( R_{o}^{T}\right) _{22}\left( R_{o}^{T}\right) _{32}I_{P}^{(\mu
)}\left( m_{e_{4}},m_{A_{1}}\right) -\left( R_{o}^{T}\right) _{23}\left(
R_{o}^{T}\right) _{33}I_{P}^{\left( \mu \right) }\left(
m_{e_{4},}m_{A_{2}}\right) \Big] \\
\Delta a_{e}=y_{51}^{e}x_{15}^{L}\frac{m_{e}^{2}}{8\pi ^{2}}\Big[& \left(
R_{e}^{T}\right) _{22}\left( R_{e}^{T}\right) _{32}I_{S}^{(e)}\left(
m_{e_{5}},m_{H_{1}}\right) +\left( R_{e}^{T}\right) _{23}\left(
R_{e}^{T}\right) _{33}I_{S}^{(e)}\left( m_{e_{5}},m_{H_{2}}\right) \\
& -\left( R_{o}^{T}\right) _{22}\left( R_{o}^{T}\right)
_{32}I_{P}^{(e)}\left( m_{e_{5}},m_{A_{1}}\right) -\left( R_{o}^{T}\right)
_{23}\left( R_{e}^{T}\right) _{33}I_{P}^{\left( E\right) }\left(
m_{e_{5}},m_{A_{2}}\right) \Big],
\end{split}
\label{eqn:muon_electron_anomalous_magnetic_moments}
\end{equation}

where the loop integrals are given by \cite{Diaz:2002uk,Jegerlehner:2009ry,Kelso:2014qka,Lindner:2016bgg,Kowalska:2017iqv}: 
\begin{equation}
I_{S\left( P\right) }^{\left( e,\mu \right) }\left( m_{E_{4,5}},m_{S}\right)
=\int_{0}^{1}\frac{x^{2}\left( 1-x\pm \frac{m_{E_{4,5}}}{m_{e,\mu }}\right) }{%
m_{e,\mu }^{2}x^{2}+\left( m_{E_{4,5}}^{2}-m_{e,\mu }^{2}\right) x+m_{S,P}^{2}\left(
1-x\right) }dx  \label{eqn:loop_integrals}
\end{equation}

and $S(P)$ means scalar (pseudoscalar) and $E_{4,5}$ stands for the vector-like family. It is worth mentioning that $E_{4}$ and $E_{5}$ only contribute to the muon and electron anomalous magnetic moments, respectively.

%

\section{NUMERICAL ANALYSIS OF THE HIGGS EXCHANGE CONTRIBUTIONS}

\label{sec:Numerical_analysis_of_scalars}

For the sake of simplicity, we consider the scenario of absence of mixing between SM charged leptons, which automatically prevents charged lepton flavour violating decays.   
In our numerical analysis we have
found that the non-SM CP-even scalar mass can reach values around $200 \func{%
GeV}$. Despite the fact that the non SM CP-even scalar is quite light and can have
a sizeable decay mode into a bottom-anti bottom quark pair, its single LHC
production via gluon fusion mechanism is strongly suppressed since it is
dominated by the triangular bottom quark loop. Such non SM CP-even scalar $H$
can also be produced by vector boson fusion but such production is expected
to have a low total cross section due to small $HWW$ and $HZZ$ couplings,
which are proportional to $v_d$. In this section we will discuss the
implications of our model in the muon and electron anomalous magnetic
moments.

\subsection{The fitting function $\protect\chi^2$ and free parameter setup}

\label{subsec:fitting_function_chi_and_parameters}

For the first approach to both anomalies, we construct the fitting function $%
\chi^2$

\begin{equation}
\begin{split}
\chi^2 &= \frac{\left( m_h^{\func{Thy}}-m_h^{\func{Cen}} \right)^2}{\left(
\delta m_h^{\func{Dev}} \right)^2} + \frac{\left( a_{hWW}^{\func{Thy}%
}-a_{hWW}^{\func{Cen}} \right)^2}{\left( \delta a_{hWW}^{\func{Dev}}
\right)^2} + \frac{\left( R_{\gamma\gamma}^{\func{Thy}}-R_{\gamma\gamma}^{%
\func{Cen}} \right)^2}{\left( \delta R_{\gamma\gamma}^{\func{Dev}} \right)^2}
\\
&+ \frac{\left( \Delta a_{\mu}^{\func{Thy}}-\Delta a_{\mu}^{\func{Cen}}
\right)^2}{\left( \delta \Delta a_{\mu}^{\func{Dev}} \right)^2} + \frac{%
\left( \Delta a_{e}^{\func{Thy}}-\Delta a_{e}^{\func{Cen}} \right)^2}{\left(
\delta \Delta a_{e}^{\func{Dev}} \right)^2},
\end{split}%
\end{equation}

where the superscripts $\func{Thy}$, $\func{Cen}$ and $\func{Dev}$ mean
theoretical prediction, central value of experimental bound and deviation
from the central value at one of $1,2,3\sigma$, respectively. The parameters
used in this fitting function are defined as below(the integer number
multiplied in delta terms means $\sigma$):

\begin{equation}
\begin{split}
m_h^{\func{Cen}} = 125.38 \func{GeV}, &\quad \delta m_h^{\func{Dev}} = 3
\times 0.14 \func{GeV}, \\
a_{hWW}^{\func{Cen}} = 0.59, &\quad \delta a_{hWW}^{\func{Dev}} = 1 \times
0.35, \\
R_{\gamma \gamma}^{\func{Cen}} = \frac{1}{2} \left( R_{\gamma \gamma}^{\func{%
CMS}} + R_{\gamma \gamma}^{\func{ATLAS}} \right) = 1.07, &\quad \delta
R_{\gamma \gamma}^{\func{Dev}} = 1 \times 0.14, \\
\Delta a_\mu^{\func{Cen}} = 26.1 \times 10^{-10}, &\quad \delta\Delta a_\mu^{%
\func{Dev}} = 1 \times \left( 8.0 \times 10^{-10} \right) \\
\Delta a_e^{\func{Cen}} = -0.88 \times 10^{-12}, &\quad \delta\Delta a_e^{%
\func{Dev}} = 2 \times \left( 0.36 \times 10^{-12} \right)
\end{split}%
\end{equation}

For an initial scan, we set up the starting parameter region as below:

\begin{center}
{\renewcommand{\arraystretch}{1.5} 
\begin{tabular}{|c|c|}
\hline
\textbf{Parameter} & \textbf{Value/Scanned Region($\func{GeV}$)} \\ \hline
$v_u = v_1$ & $\frac{\tan\beta}{\sqrt{1+\tan\beta^2}} \times 246$ \\ 
$v_d = v_2$ & $\frac{1}{\sqrt{1+\tan\beta^2}} \times 246$ \\ 
$v_\phi = v_3$ & $\pm [0.01,1.00] \times 1000$ \\ \hline
$\tan\beta = v_u/v_d$ & $\left[ 5, 50 \right]$ \\ \hline
$\lambda_1$ & $\left( m_h^2 - \frac{v_2 v_3^2 \lambda_5}{2 v_1}
\right)/\left( 4 v_1^2 \right)$ \\ 
$\lambda_2$ & $\pm \left[ 0.50, 12.00 \right]$ \\ 
$\lambda_3$ & $\pm \left[ 0.50, 12.00 \right]$ \\ 
$\lambda_4$ & $\pm \left[ 0.50, 12.00 \right]$ \\ 
$\lambda_5$ & $4 v_1 v_2 \lambda_3/(v_3)^2$ \\ 
$\lambda_6$ & $\pm \left[ 0.50, 12.00 \right]$ \\ 
$\lambda_7$ & $v_2 \lambda_5/v_1$ \\ 
$\lambda_8$ & $\pm \left[ 0.50, 12.00 \right]$ \\ \hline
$M_{44}^{e}$ & $\left[ 2 \times 10^2, 2 \times 10^3 \right]$ \\ 
$M_{55}^{L}$ & $\left[ 2 \times 10^2, 2 \times 10^3 \right]$ \\ 
$\mu_{\func{sb}}$ & $i^{[0,1]} \times \left[300, 500 \right]$ \\ \hline
$y_{e}$ & $\sqrt{2} m_{e}/v_2$ \\ 
$y_{\mu}$ & $\sqrt{2} m_{\mu}/v_2$ \\ 
$y_{24}^{e} = y_2$ & $\pm \left[ 1.0, 3.5 \right]$ \\ 
$y_{51}^{e} = y_1$ & $\pm \left[ 1.0, 3.5 \right]$ \\ 
$x_{42}^{e} = x_2$ & $\lvert y_{\mu} M_{44}^{e} / \left( y_{24}^{e} v_3 \right)
\rvert$ \\ 
$x_{15}^{L} = x_1$ & $\lvert y_{e} M_{55}^{L} / \left( y_{51}^{e} v_3 \right)
\rvert$ \\ \hline
\end{tabular}
\captionof{table}{Initial parameter setup} \label%
{tab:parameter_region_initial_scan}}
\end{center}

\begin{enumerate}
\item For the Higgs vevs, we are interested in the range of $\tan\beta$ from 
$5$ to $50$ as in the $W$ boson exchange in Figure \ref%
{fig:parameter_space_muegamma_M44_vu_tanbeta}

\item For $\lambda_1$, we fixed mass of the SM physical Higgs $h$ to be $125%
\func{GeV}$ to save time and to make the calculation faster. For $%
\lambda_{5,7} $, the assumption that no mixing between the SM Higgs $h$ and
non-SM Higgses $H_{1,2}$ arise is reflected on these parameters. All quartic coupling constants $\lambda_{1,\cdots,8}$ are set up not to go over $%
4\pi$ for perturbativity.

\item For the vector-like masses $M_{44}^{e}$ and $M_{55}^{L}$, there is a
constraint that the lightest should be greater than $200\func{GeV}$ \cite%
{Xu:2018pnq}.

\item In our numerical analysis we consider solutions where the non SM
scalar masses are larger than about $200\func{GeV}$ as done in \cite%
{Hernandez-Sanchez:2020vax}.

\item The soft-breaking mass term $\mu_{\func{sb}}$ is a free parameter,
which does not generate any problem and appropiate values of this parameters
yields masses for scalars and vector-like fermions consistent with the
experimental constraints. 

\item The diagonal Yukawa constants appearing in Equation \ref%
{eqn:diagonalized_Yukawa_matrix_charged_leptons} should be the Yukawa
constant for electron, muon and tau, respectively. The Yukawa constants $%
y_{24,51}$ and $x_{42,15}$ interacting with vector-like families are defined
under this consideration. For perturbativity, the Yukawa constants $%
y_{24,51} $ are considered not to go over $\sqrt{4\pi}$.
\end{enumerate}
After saturating value of the $\chi^2$ function less than or nearly $2$
which we believe it is converged enough, we find a best peaked value for
each free parameter. For the given parameters, we rename them by adding an
index ``p" to the end of subscript of each parameter like $\func{\tan\beta_p}
$ and then the expansion factor $\kappa$ is multiplied to find a correlation
between the observables and the mass parameters. Then, the parameter region
is refreshed by both the specific value of each parameter and the expansion
factor $\kappa$ as per Table \ref{tab:parameter_region_subsequent_scan}.

\begin{center}
{\renewcommand{\arraystretch}{1.5} 
\begin{tabular}{|c|c|}
\hline
\textbf{Parameter} & \textbf{Value/Scanned Region($\func{GeV}$)} \\ \hline
$v_u = v_1$ & $\frac{\tan\beta_p}{\sqrt{1+\tan\beta_p^2}} \times 246$ \\ 
$v_d = v_2$ & $\frac{1}{\sqrt{1+\tan\beta_p^2}} \times 246$ \\ 
$v_\phi = v_3$ & $\left[ (1 - \kappa), (1 + \kappa) \right] \times v_{3p}$
\\ \hline
$\tan\beta = v_u/v_d$ & $\left[ (1 - \kappa), (1 + \kappa) \right] \times 
\func{\tan\beta}_p$ \\ \hline
$\lambda_1$ & $\left( m_h^2 - \frac{v_{2} v_{3}^2 \lambda_{5}}{2 v_{1}}
\right)/\left( 4 v_{1}^2 \right)$ \\ 
$\lambda_2$ & $\left[ (1 - \kappa), (1 + \kappa) \right] \times \lambda_{2p}$
\\ 
$\lambda_3$ & $\left[ (1 - \kappa), (1 + \kappa) \right] \times \lambda_{3p}$
\\ 
$\lambda_4$ & $\left[ (1 - \kappa), (1 + \kappa) \right] \times \lambda_{4p}$
\\ 
$\lambda_5$ & $4 v_{1} v_{2} \lambda_{3}/(v_{3})^2$ \\ 
$\lambda_6$ & $\left[ (1 - \kappa), (1 + \kappa) \right] \times \lambda_{6p}$
\\ 
$\lambda_7$ & $v_{2} \lambda_{5}/v_{1}$ \\ 
$\lambda_8$ & $\left[ (1 - \kappa), (1 + \kappa) \right] \times \lambda_{8p}$
\\ \hline
$M_{44}^{e}$ & $\left[ (1 - \kappa), (1 + \kappa) \right] \times M_{44p}^{e}$
\\ 
$M_{55}^{L}$ & $\left[ (1 - \kappa), (1 + \kappa) \right] \times M_{55p}^{L}$
\\ 
$\mu_{\func{sb}}$ & $\left[ (1 - \kappa), (1 + \kappa) \right] \times \mu_{%
\func{sb}p}$ \\ \hline
$y_{e}$ & $\sqrt{2} m_{e}/v_{2}$ \\ 
$y_{\mu}$ & $\sqrt{2} m_{\mu}/v_{2}$ \\ 
$y_{24}^{e} = y_2$ & $\left[ (1 - \kappa), (1 + \kappa) \right] \times y_{24p}^{e}$
\\ 
$y_{51}^{e} = y_1$ & $\left[ (1 - \kappa), (1 + \kappa) \right] \times y_{51p}^{e}$
\\ 
$x_{42}^{e} = x_2$ & $y_{\mu} M_{44}^{e} / \left( y_{24}^{e} v_3 \right)$ \\ 
$x_{15}^{L} = x_1$ & $y_{e} M_{55}^{L} / \left( y_{51}^{e} v_3 \right)$ \\ \hline
$\kappa$ & $0.1$ \\ \hline
\end{tabular}
\captionof{table}{Next parameter setup after the initial scan result} \label%
{tab:parameter_region_subsequent_scan}}
\end{center}

\subsection{A scanned result on the free parameters as well as observables
across over the first and second scan}

\label{subsec:scanned_result}

The best peaked value for each parameter is listed in Table \ref%
{tab:parameter_region_best_peaked} and energy scale is in unit of $\func{GeV}
$. Note that all cases are carried out independently and all points of plots
in each case are collected within $1\sigma$ constraint of each anomaly.
\begin{center}
{\renewcommand{\arraystretch}{1.5} 
\begin{tabular}{|c|c|c|c|c|c|}
\hline
\textbf{Parameter} & case A & case B & case C & case D & case E \\ \hline
$v_u = v_1$ & $245.925$ & $245.936$ & $245.951$ & $245.917$ & $245.948$ \\ 
$v_d = v_2$ & $6.086$ & $5.595$ & $4.921$ & $6.387$ & $5.077$ \\ 
$v_\phi = v_3$ & $-57.761$ & $-36.470$ & $-57.919$ & $-30.746$ & $-17.146$ \\ \hline
$\tan\beta = v_u/v_d$ & $40.410$ & $43.957$ & $49.977$ & $38.503$ & $48.441$ \\ \hline
$\lambda_1$ & $0.063$ & $0.064$ & $0.066$ & $0.064$ & $0.065$ \\ 
$\lambda_2$ & $-7.978$ & $8.414$ & $-2.000$ & $2.948$ & $10.382$ \\ 
$\lambda_3$ & $-6.344$ & $-2.675$ & $6.242$ & $-1.724$ & $-0.706$ \\ 
$\lambda_4$ & $1.859$ & $2.158$ & $-3.633$ & $10.837$ & $-2.796$ \\ 
$\lambda_5$ & $-11.384$ & $-11.070$ & $9.009$ & $-11.460$ & $-12.000$ \\ 
$\lambda_6$ & $2.888$ & $1.228$ & $0.866$ & $1.351$ & $1.324$ \\ 
$\lambda_7$ & $-0.282$ & $-0.252$ & $0.180$ & $-0.298$ & $-0.248$ \\ 
$\lambda_8$ & $-1.363$ & $-1.346$ & $-10.845$ & $-11.510$ & $7.033$ \\ \hline
$M_{44}^{e}$ & $1475.010$ & $1355.470$ & $1495.770$ & $1134.340$ & $1681.760$ \\ 
$M_{55}^{L}$ & $279.386$ & $211.263$ & $204.706$ & $323.292$ & $331.462$ \\ 
$\mu_{\func{sb}}$ & $424.618i$ & $443.435i$ & $480.993$ & $480.062i$ & $491.533$  \\ 
\hline
$y_{e} \left[ 10^{-4} \right]$ & $1.135$ & $1.234$ & $1.403$ & $1.081$ & $1.360$ \\ 
$y_{\mu} \left[ 10^{-2} \right]$ & $2.391$ & $2.600$ & $2.956$ & $2.278$ & $2.865$ \\ 
$y_{24}^{e} = y_2$ & $-3.161$ & $-3.101$ & $-2.942$ & $-1.548$ & $1.662$ \\ 
$y_{51}^{e} = y_1$ & $2.315$ & $2.164$ & $2.050$ & $1.352$ & $3.377$ \\ 
$x_{42}^{e} = x_2$ & $0.193$ & $0.312$ & $0.260$ & $0.543$ & $1.691$ \\ 
$x_{15}^{L} = x_1 \left[ 10^{-4} \right]$ & $2.371$ & $3.304$ & $2.419$ & $8.408$ & $7.787$ \\ \hline
$m_{H_1}$ & $213.390$ & $222.924$ & $212.147$ & $238.523$ & $205.477$ \\ 
$m_{H_2}$ & $911.585$ & $614.516$ & $891.413$ & $518.147$ & $354.709$ \\ 
$m_{A_1}$ & $741.343$ & $537.111$ & $807.268$ & $435.887$ & $282.964$ \\ 
$m_{A_2}$ & $1003.790$ & $939.553$ & $1035.800$ & $1006.240$ & $1015.760$ \\ 
$m_{H^\pm}$ & $938.259$ & $674.054$ & $987.625$ & $929.786$ & $504.684$ \\ \hline
$\Delta a_\mu \left[ 10^{-9} \right]$ & $2.734$ & $2.688$ & $2.935$ & $2.891$ & $2.393$ \\ 
$\Delta a_e \left[ 10^{-13} \right]$ & $-5.073$ & $-8.310$ & $-5.543$ & $%
-6.365$ & $-9.232$ \\ 
$a_{hWW}$ & $1.000$ & $1.000$ & $1.000$ & $1.000$ & $1.000$ \\ 
$R_{\gamma\gamma}$ & $0.999$ & $0.999$ & $0.999$ & $0.999$ & $0.999$ \\ \hline
$\chi^2$ & $1.794$ & $1.516$ & $1.870$ & $1.740$ & $1.579$ \\ \hline
\end{tabular}}
\captionof{table}{A best peaked value for each parameter at each case. All
energy scale is in $\func{GeV}$ units. \Steve{Notice that in all cases $v_3$ is smaller than the vector like mass parameters $M_{44}^{e}$ and $M_{55}^{L}$, which is consistent with the assumption made in section I, regarding the fact that the corresponding expansion parameter $v_3/M_{\psi}$ is less than unity.}} \label{tab:parameter_region_best_peaked}
\end{center}

Here, we put two constraints on the lightest vector-like mass and the
lightest non-SM scalar mass; the vector-like mass should be greater than $200%
\func{GeV}$ as well as the non-SM scalar mass\cite{Xu:2018pnq,Hernandez-Sanchez:2020vax}. After we
carry out second parameter scan based on the first scan result of Table \ref%
{tab:parameter_region_best_peaked}, range of the parameters are given in
Table \ref{tab:range_secondscan}.

\begin{center}
{\renewcommand{\arraystretch}{1.5} 
\resizebox{\textwidth}{!}{
\begin{tabular}{|c|c|c|c|c|c|}
\hline
\textbf{Parameter} & case A & case B & case C & case D & case E \\ \hline
$v_u = v_1$ & $\left[ 245.907 \rightarrow 245.938 \right]$ & $\left[ 245.921
\rightarrow 245.947 \right]$ & $\left[ 245.939 \rightarrow 245.959 \right]$
& $\left[ 245.898 \rightarrow 245.931 \right]$ & $\left[ 245.935 \rightarrow 245.957 \right]$ \\ 
$v_d = v_2$ & $\left[ 5.533 \rightarrow 6.761 \right]$ & $\left[ 5.087
\rightarrow 6.216 \right]$ & $\left[ 4.474 \rightarrow 5.468 \right]$ & $%
\left[ 5.807 \rightarrow 7.096 \right]$ & $\left[ 4.616 \rightarrow 5.641 \right]$ \\ 
$v_\phi = v_3$ & $\left[ -63.525 \rightarrow -51.985 \right]$ & $\left[
-40.117 \rightarrow -32.823 \right]$ & $\left[ -63.706 \rightarrow -52.128 %
\right]$ & $\left[ -33.820 \rightarrow -27.671 \right]$ & $\left[ -18.860 \rightarrow -15.438 \right]$ \\ \hline
$\tan\beta = v_u/v_d$ & $\left[ 36.371 \rightarrow 44.451 \right]$ & $\left[
39.561 \rightarrow 48.353 \right]$ & $\left[ 44.980 \rightarrow 54.975 %
\right]$ & $\left[ 34.653 \rightarrow 42.354 \right]$ & $\left[ 43.597 \rightarrow 53.284 \right]$ \\ \hline
$m_{H_1}$ & $\left[ 200.000 \rightarrow 242.653 \right]$ & $\left[ 201.520
\rightarrow 246.046 \right]$ & $\left[ 200.000 \rightarrow 230.754 \right]$
& $\left[ 215.523 \rightarrow 261.920 \right]$ & $\left[ 200.000 \rightarrow 220.017 \right]$ \\ 
$m_{H_2}$ & $\left[ 752.061 \rightarrow 1088.130 \right]$ & $\left[ 516.289
\rightarrow 724.997 \right]$ & $\left[ 735.831 \rightarrow 1059.900 \right]$
& $\left[ 441.371 \rightarrow 604.981 \right]$ & $\left[ 338.724 \rightarrow 374.424 \right]$ \\ 
$m_{A_1}$ & $\left[ 638.813 \rightarrow 853.637 \right]$ & $\left[ 442.527
\rightarrow 640.705 \right]$ & $\left[ 670.550 \rightarrow 945.705 \right]$
& $\left[ 357.697 \rightarrow 516.760 \right]$ & $\left[ 266.086 \rightarrow 297.589 \right]$ \\ 
$m_{A_2}$ & $\left[ 892.847 \rightarrow 1141.780 \right]$ & $\left[ 847.825
\rightarrow 1032.140 \right]$ & $\left[ 927.768 \rightarrow 1154.640 \right]$
& $\left[ 907.576 \rightarrow 1105.770 \right]$ & $\left[ 918.667 \rightarrow 1114.960 \right]$ \\ 
$m_{H^\pm}$ & $\left[ 783.823 \rightarrow 1111.600 \right]$ & $\left[
580.316 \rightarrow 779.945 \right]$ & $\left[ 842.585 \rightarrow 1143.880 %
\right]$ & $\left[ 856.237 \rightarrow 1007.360 \right]$ & $\left[ 478.900 \rightarrow 529.178 \right]$ \\ \hline
$M_{44}^e$ & $\left[ 1327.510 \rightarrow 1622.510 \right]$ & $\left[
1219.930 \rightarrow 1491.020 \right]$ & $\left[ 1346.190 \rightarrow
1645.330 \right]$ & $\left[ 1029.900 \rightarrow 1247.770 \right]$ & $\left[ 1513.590 \rightarrow 1849.930 \right]$ \\ 
$M_{55}^L$ & $\left[ 251.447 \rightarrow 307.323 \right]$ & $\left[ 200.000
\rightarrow 232.389 \right]$ & $\left[ 200.000 \rightarrow 225.176 \right]$
& $\left[ 290.963 \rightarrow 355.621 \right]$ & $\left[ 298.317 \rightarrow 364.604 \right]$ \\ 
$\lvert \mu_{\func{sb}} \rvert$ & $\left[ 382.158 \rightarrow 467.079 \right]
$ & $\left[ 399.091 \rightarrow 487.777 \right]$ & $\left[ 432.895
\rightarrow 529.091 \right]$ & $\left[ 432.059 \rightarrow 528.067 \right]$ & $\left[ 442.381 \rightarrow 540.679 \right]$
\\ \hline
$\Delta a_\mu \left[ 10^{-9} \right]$ & $\left[ 1.811 \rightarrow 3.410 %
\right]$ & $\left[ 1.810 \rightarrow 3.410 \right] $ & $\left[ 1.810
\rightarrow 3.410 \right]$ & $\left[ 1.810 \rightarrow 3.410 \right]$ & $\left[ 1.810 \rightarrow 3.410 \right]$ \\ 
$\Delta a_e \left[ 10^{-13} \right]$ & $\left[ -6.730 \rightarrow -5.200 %
\right]$ & $\left[ -11.142 \rightarrow -5.985 \right]$ & $\left[ -7.207
\rightarrow -5.200 \right]$ & $\left[ -8.721 \rightarrow -5.200 \right]$ & $\left[ -12.393 \rightarrow -5.442 \right]$
\\ 
$a_{hWW}$ & $\left[ 1.000 \rightarrow 1.000 \right]$ & $\left[ 1.000
\rightarrow 1.000 \right]$ & $\left[ 0.999 \rightarrow 1.000 \right]$ & $%
\left[ 1.000 \rightarrow 1.000 \right]$ & $\left[ 1.000 \rightarrow 1.000 \right]$ \\ 
$R_{\gamma\gamma}$ & $\left[ 0.999 \rightarrow 0.999 \right]$ & $\left[
0.999 \rightarrow 0.999 \right]$ & $\left[ 0.999 \rightarrow 1.000 \right]$
& $\left[ 1.000 \rightarrow 1.000 \right]$ & $\left[ 0.999 \rightarrow 1.000 \right]$ \\ \hline
$\chi^2$ & $\left[ 1.604 \rightarrow 2.750 \right]$ & $\left[ 1.501
\rightarrow 2.635 \right]$ & $\left[ 1.580 \rightarrow 2.761 \right]$ & $%
\left[ 1.509 \rightarrow 2.749 \right]$ & $\left[ 1.501 \rightarrow 2.720 \right]$ \\ \hline
\end{tabular}
}
\captionof{table}{A scanned range of each parameter at case A, B, C, D and E.
$H_{1,2}$ mean non SM CP-even scalars and $A_{1,2}$ are non SM CP-odd scalars and $H^\pm$ stand for non SM charged scalars in this model. All data of $\Delta
a_{\mu,e}$ are collected within the $1\sigma$ constraint of each anomaly.} %
\label{tab:range_secondscan}}
\end{center}

\subsection{The muon and electron anomalous magnetic moments}

\label{subsec:muon_electrong2_scalar}

\begin{figure}[]
\centering
\begin{subfigure}{0.48\textwidth}
\includegraphics[keepaspectratio,scale=0.45]{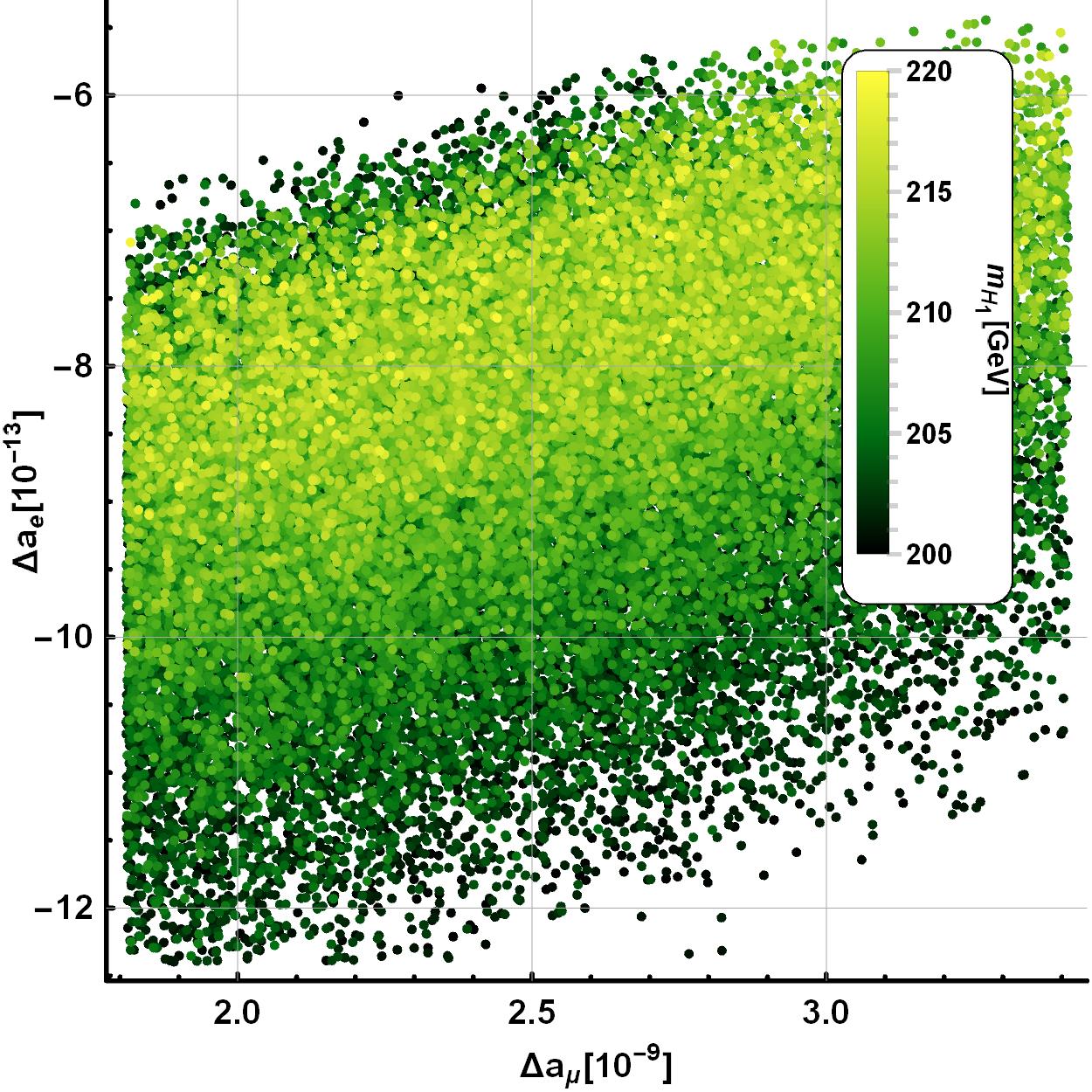} 
\end{subfigure}
\begin{subfigure}{0.48\textwidth}
\includegraphics[keepaspectratio,scale=0.45]{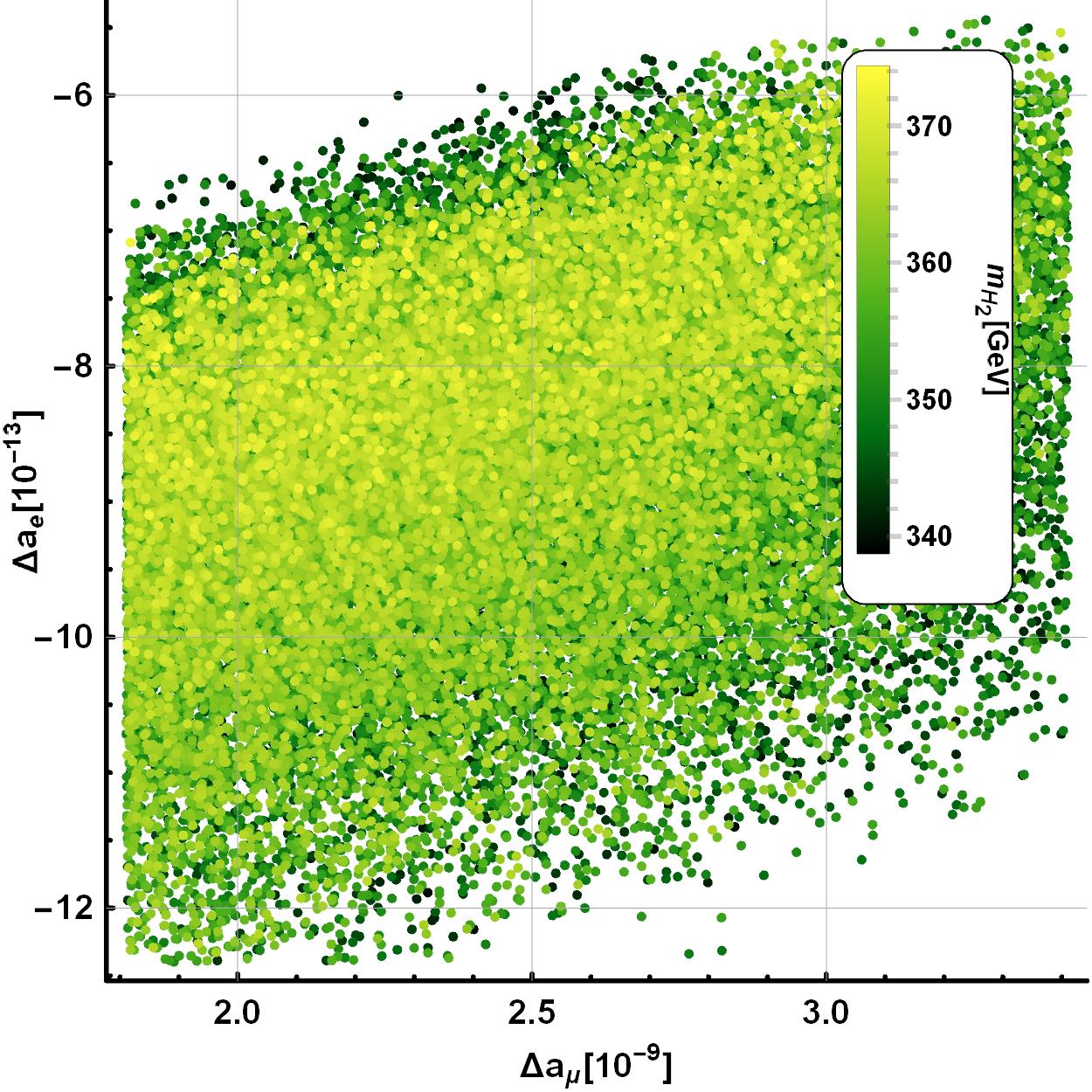} 
\end{subfigure} 
\par
\begin{subfigure}{0.48\textwidth}
\includegraphics[keepaspectratio,scale=0.45]{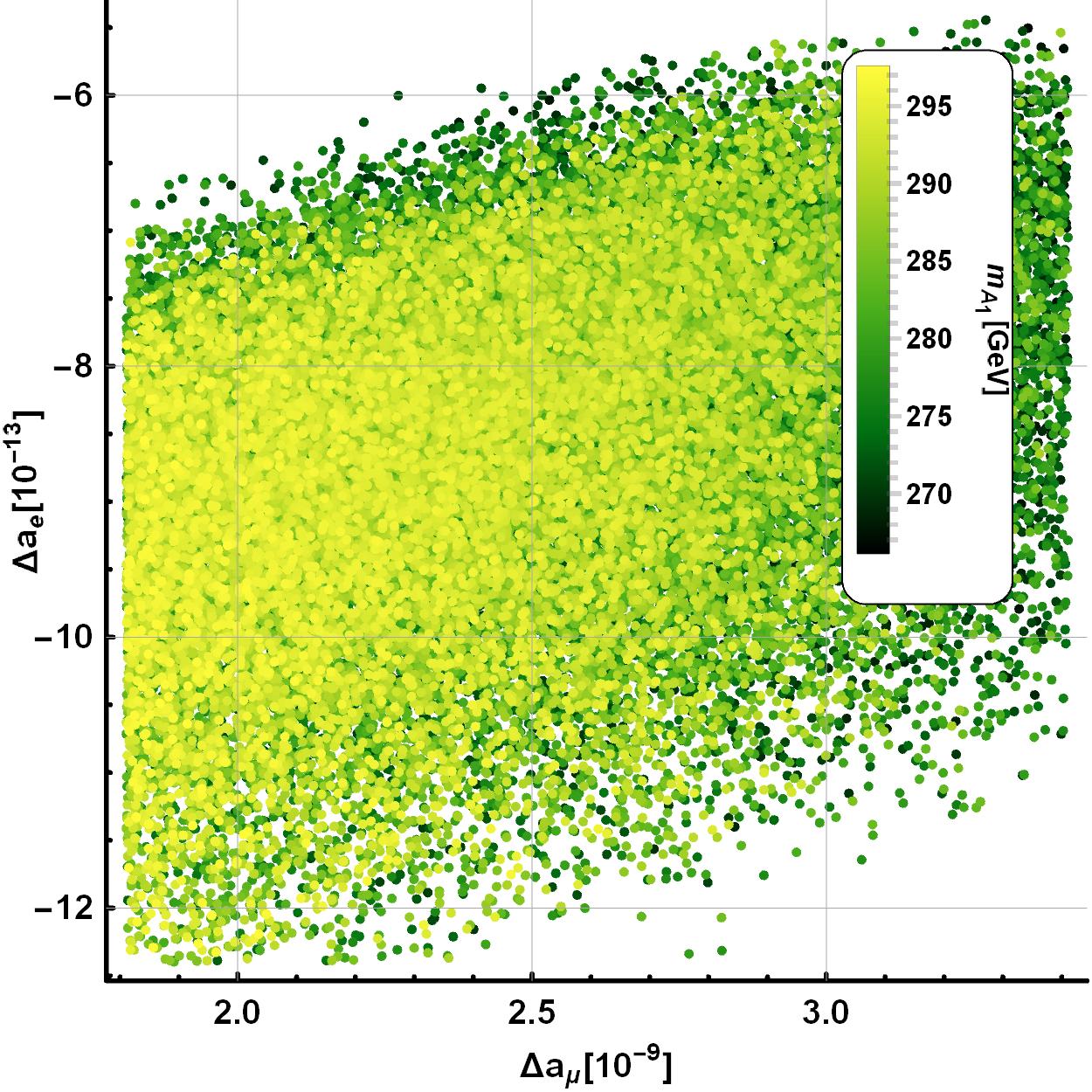} 
\end{subfigure}
\begin{subfigure}{0.48\textwidth}
\includegraphics[keepaspectratio,scale=0.45]{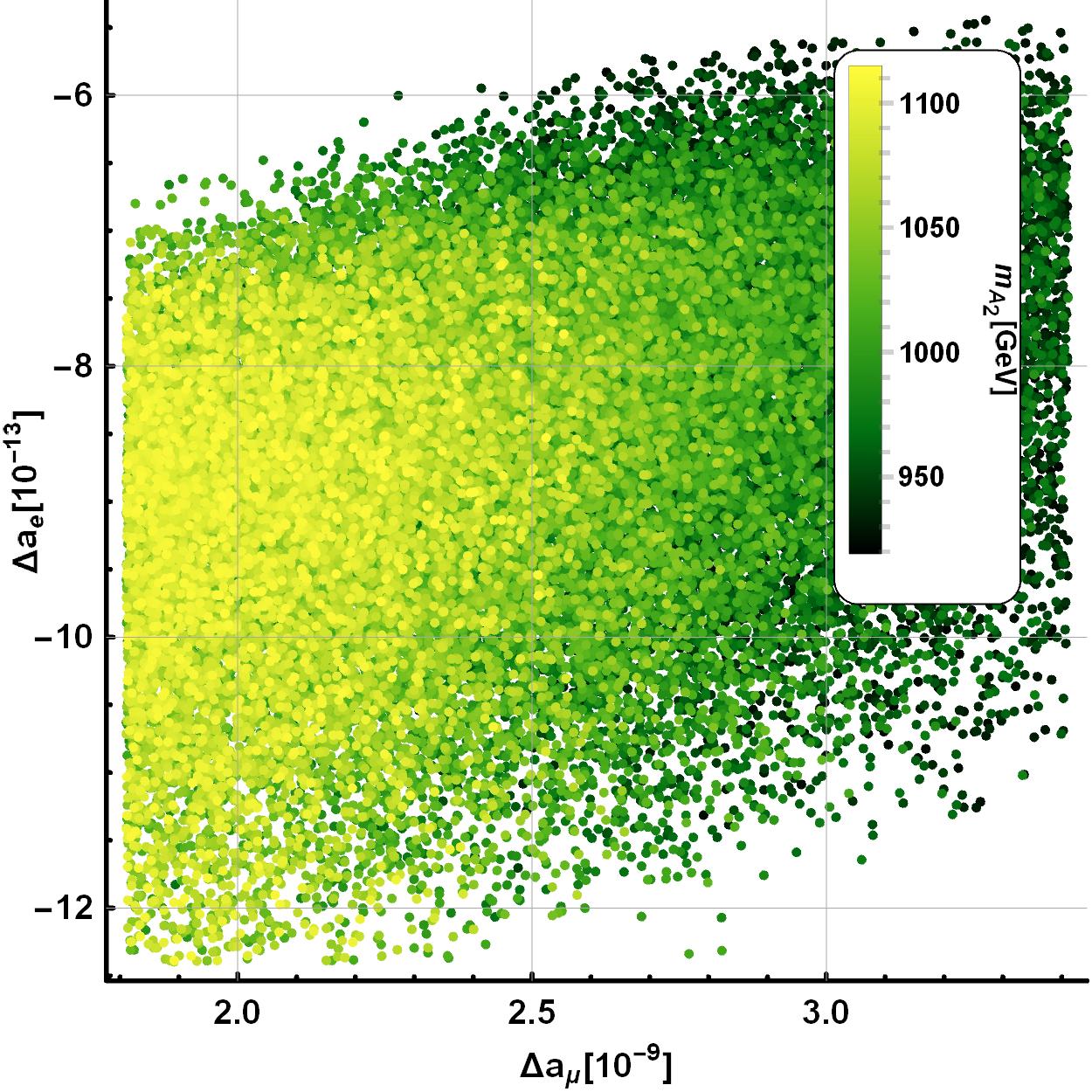} 
\end{subfigure}
\begin{subfigure}{0.48\textwidth}
\includegraphics[keepaspectratio,scale=0.45]{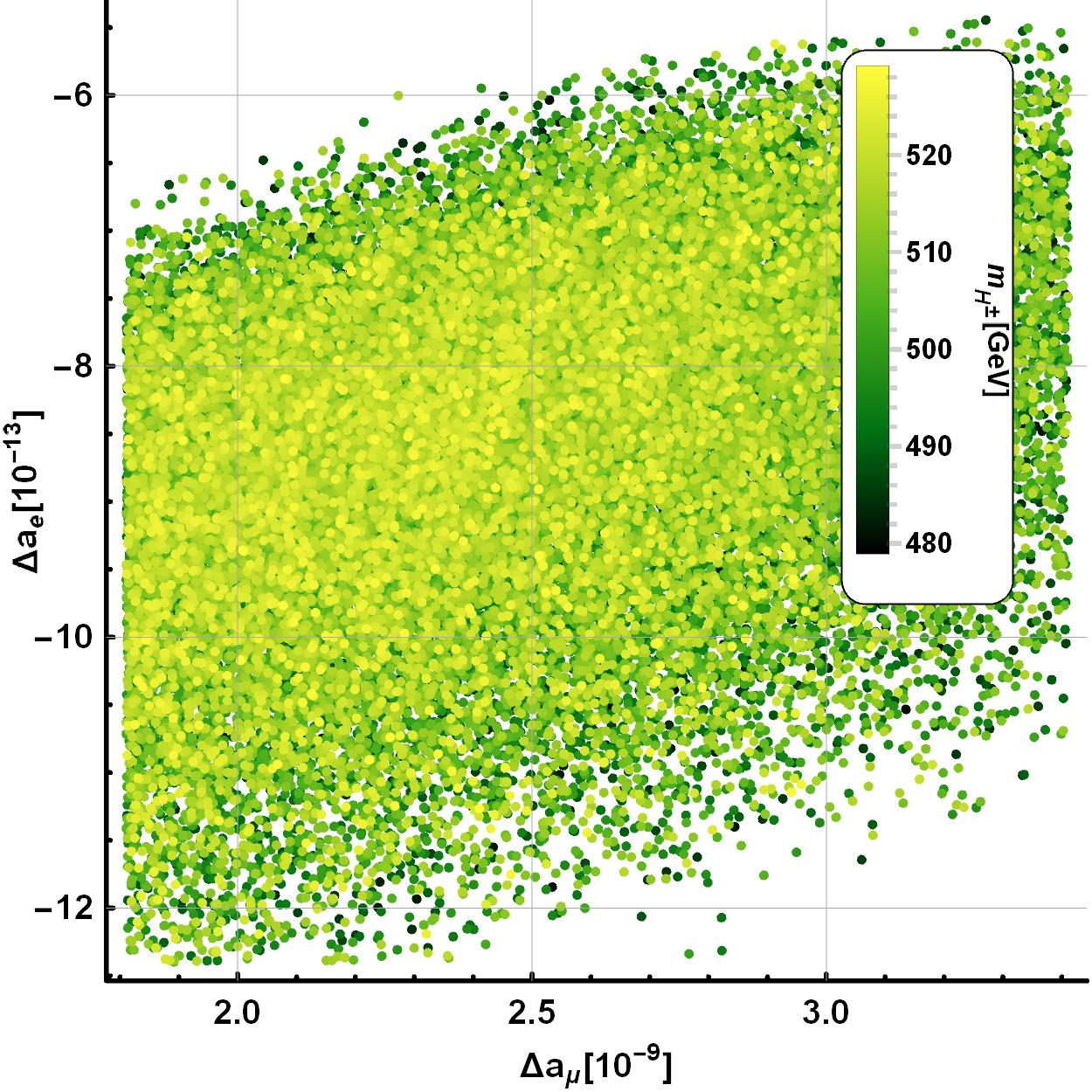} 
\end{subfigure}
\caption{Available parameter spaces for the muon anomaly versus electron
anomaly with a mass parameter which attends the both anomalies$%
(H_{1,2},A_{1,2})$ and does not$(H^\pm)$. $H_{1,2}$ are non-SM CP even scalars, $%
A_{1,2}$ are non-SM CP odd scalars and $H^{\pm}$ are non-SM charged scalars.
All points in each plot are collected within $1\protect\sigma$ constraint of
each anomaly.}
\label{fig:relevant_parameters_delamue}
\end{figure}

\begin{figure}[]
\centering
\begin{subfigure}{0.48\textwidth}
\includegraphics[keepaspectratio,scale=0.50]{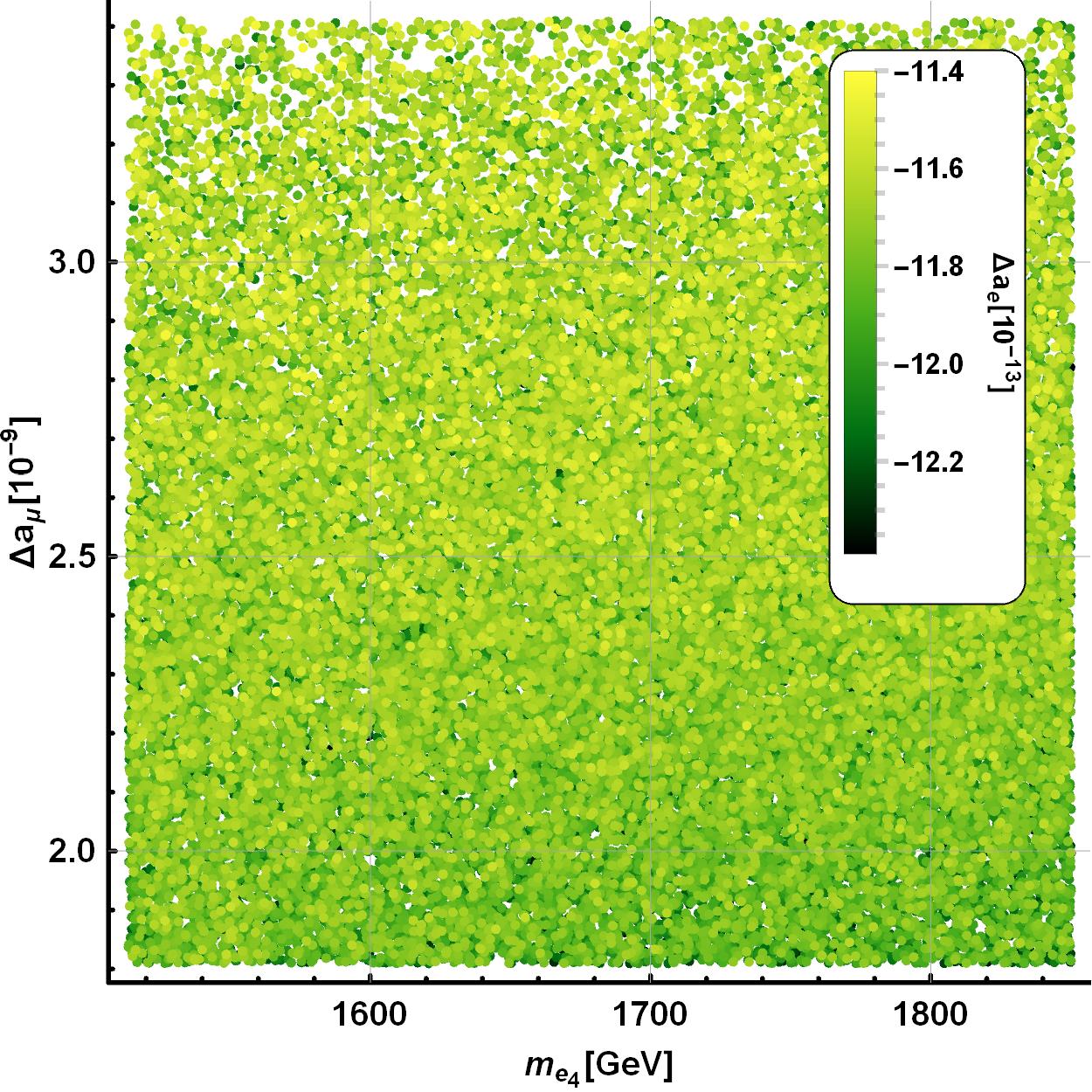} 
\end{subfigure}
\begin{subfigure}{0.48\textwidth}
\includegraphics[keepaspectratio,scale=0.50]{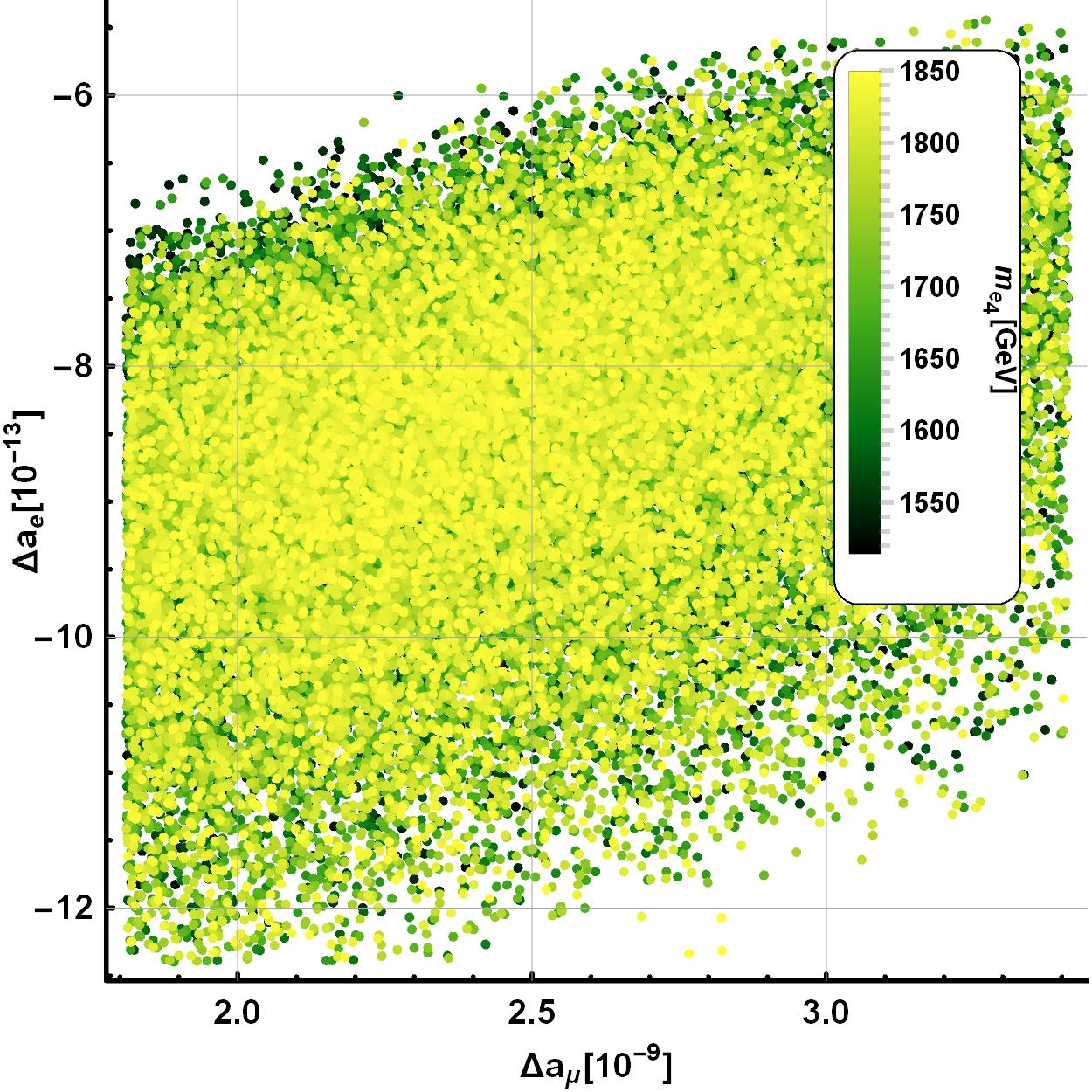} 
\end{subfigure} 
\par
\begin{subfigure}{0.48\textwidth}
\includegraphics[keepaspectratio,scale=0.50]{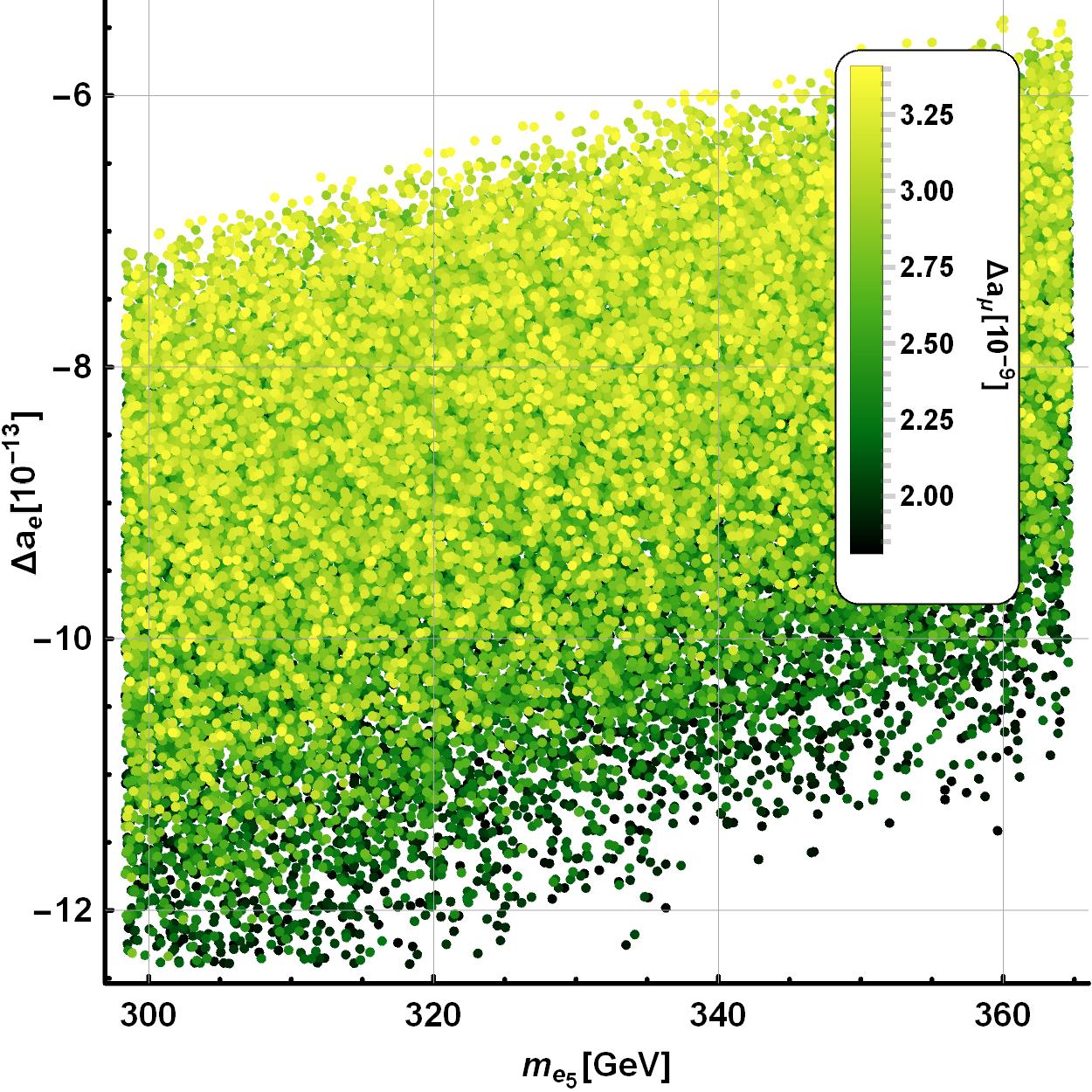} 
\end{subfigure}
\begin{subfigure}{0.48\textwidth}
\includegraphics[keepaspectratio,scale=0.50]{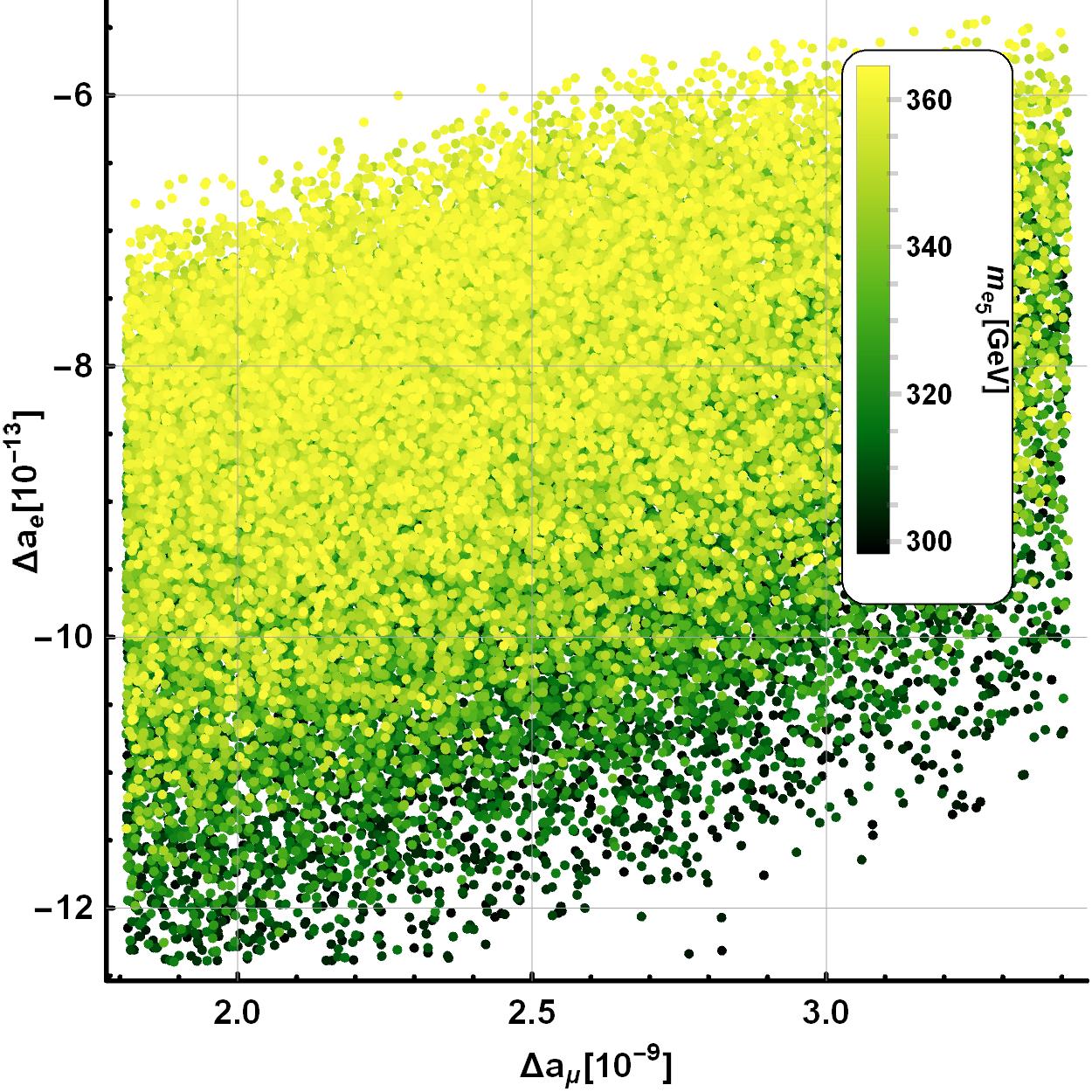} 
\end{subfigure}
\caption{Available parameter spaces for the muon anomaly(electron anomaly)
versus a relevant vector-like mass $m_{e_4}(m_{e_5})$ with another
anomaly(two left plots) in bar where $m_{e_4}(m_{e_5})$ is simplified
notation for $M_{44}^{e}(M_{55}^{L})$, while the two right plots for the
muon anomaly versus electron anomaly with a vector-like mass $%
m_{e_4}(m_{e_5})$}
\label{fig:relevant_parameters_vl_delamue}
\end{figure}

In order to confirm that our theoretical prediction for both anomalies can accommodate their constraints at $1\sigma$ and to analyze correlations between both anomalies and mass parameters, we consider cases B and E in Table \ref{tab:parameter_region_best_peaked} since their benchmark point have relative lower values of the $\chi^2$ function when compared to other cases. The reason that the cases B and E have the lower values of the $\chi^2$ function arises from the obtained value of the electron anomaly, which is very close to the central experimental value. All cases reveal nearly central value of muon anomaly constraint at $1\sigma$, whereas the other cases except B and E reveal nearly edge value of electron anomaly constraint at $1\sigma$. Therefore, the reason why the cases B and E are more converged is related to whether our theoretical prediction for both anomalies can gain access to their central value of each anomaly constraint at $1\sigma$. \Steve{More importantly, the case E is only one satisfying vacuum stability conditions and a detailed investigation for the vacuum stability of each case will be studied in a subsection}. For these reasons, we take the case E in Table \ref{tab:range_secondscan} to study the correlations. The relevant parameter spaces are listed in Figure \ref{fig:relevant_parameters_delamue} and \ref{fig:relevant_parameters_vl_delamue}.

To begin with, we consider the parameter spaces for the muon anomaly versus electron anomaly with a mass parameter which attends both anomalies $(H_{1,2}, A_{1,2})$ and does not $(H^\pm)$ in Figure \ref{fig:relevant_parameters_delamue}. Even thought the non-SM charged scalar does not attend both anomalies, the
similar pattern which the other scalars implement in Figure \ref%
{fig:relevant_parameters_delamue} is also appeared. We confirmed that mass of $H_2$ is nearly proportional to that of $H^\pm$, which causes the correlation identified in plots of the other non-SM scalars in Figure \ref{fig:relevant_parameters_delamue} is still maintained for the non-SM charged scalar. Interestingly, the cases A, B and C in Table \ref{tab:range_secondscan} reported $m_{H_2}$ is nearly proportional to $m_{H^\pm}$ one-to-one ratio, whereas the cases D and E revealed a fat proportion between them and still maintained the correlation.
\newline
As mentioned at the beginning of this section, we take the case E for the plots in Figure \ref{fig:relevant_parameters_delamue} and \ref{fig:relevant_parameters_vl_delamue} and a main distinction between the case E and others arises from the value of electron anomaly. If we take other cases instead of the case E to investigate the parameter spaces, the parameter region appeared in top-left plot of Figure \ref{fig:relevant_parameters_delamue} will be shifted upward by locating at the value of $-5$ or $-6 \times 10^{-13}$ for the electron anomaly. In other words, the whole colored region in Figure \ref{fig:relevant_parameters_delamue} is shifted upwards to meet the scanned value of electron anomaly constraint at $1\sigma$, holding the correlations. Therefore, the white region appeared in Figure \ref{fig:relevant_parameters_delamue} is not strictly excluded region and affected by how well a benchmark point is converged and by a factor of $\kappa$. However, these plots still tells a correlation between both anomalies and a tendency that the lighter mass of $H_1$ is located at edge region of the parameter space. Mass of the lightest non-SM scalar $H_1$ implied in top-left plot of Figure \ref{fig:relevant_parameters_delamue} is ranged from $200$ to $220\func{GeV}$~\cite{Hernandez-Sanchez:2020vax} and the cross section for this light non-SM scalar will be compared to that for SM Higgs in appendix. As for mass range of the other non-SM scalars confirmed in rest of other plots in Figure \ref{fig:relevant_parameters_delamue}, they all implied heavier mass than that of $H_1$ which can be flexible depending on how the parameters are converged as seen in each case of Table \ref{tab:range_secondscan}.
\newline~\newline
We investigate a correlation for an anomaly versus a relevant mass parameter with another anomaly in bar in Figure \ref{fig:relevant_parameters_vl_delamue}. Note that the fourth vector-like mass is relevant only for the muon anomaly, whereas the fifth is only for the electron anomaly. Even though the fourth (fifth) is irrelevant to the electron (muon) anomaly, it is good to express them together since we rearrange the mass parameters and the anomalies in bar for comparison. The top-left plot in Figure \ref{fig:relevant_parameters_vl_delamue} just fills in whole parameter region, thus no any correlation between the fourth vector-like mass and the muon anomaly is identified. After we rearranged the order of $m_{e_4}$ and $\Delta a_{\mu,e}$ from the top-left plot, we can confirm the similar correlation identified in Figure \ref{fig:relevant_parameters_delamue} from the top-right plot in Figure \ref{fig:relevant_parameters_vl_delamue}. The bottom-left plot identifies some correlation between the fifth vector-like mass and the electron anomaly contrary to the top-left plot. For the fifth vector-like mass, we put the constraint that the lightest vector-like mass should be greater than $200\func{GeV}$ \cite%
{Xu:2018pnq} and the mass region below $200\func{GeV}$ is all excluded. After rearranging the order of $m_{e_5}$ and $\Delta a_{\mu}$ as in the above plot, we confirmed the similar correlation appears in the bottom-right plot. Interestingly, the top-right and the bottom-right plots check the similar correlation.
\newline~\newline
We confirmed that the muon and electron anomalous magnetic moments with vector-like particles can be explained to within $1\sigma$ constraint of each anomaly in a unified way, which is based on two attributes; the first one is the extended scalar sector and the second one is related with the contributions of the vector-like leptons. The first one which is reflected in our prediction for both anomalies, consists of four non-SM scalars and these contributions play a crucial role for determining the magnitude of each anomaly. The second one is seen by two vertices of both anomaly diagrams. The other Yukawa interactions can take place at each vertex since the vector-like leptons come in the loop, which is differentiated by the case where the normal SM particles enter in the loop. To be more specific, the helicity flip mass caused by the vector-like fermions in the CP-even and CP-odd basis couples the initial particle inside the loop to another particle of different chirality, thus allowing different interactions at each vertex.  This means that the different sign problem can be solved by only considering multiplication of the Yukawa constants of each vertex and this property will be covered in detail in next subsection.

\Steve{\subsubsection{Vacuum stability}

An important feature of our extended 2HDM theory 
is that it predicts 
large values for the Yukawa coupling constants $y_{2,1}, x_{2,1}$ which can be ideally order of unity in our model. If the Yukawa coupling constants are much lower than unity, which means $y_{2,1},x_{2,1} \ll 1$, it will not cause any problem for stabilization of the scalar potential. 
However, large values of the leptonic Yukawa couplings are required in our model to successfully explain both $g-2$ anomalies within the $1\sigma$ experimentally allowed range 
and since they are somehow related with the electroweak sector parameters, it might be able to destabilize the Higgs potential. As 
previously mentioned, in our analysis of the scalar sector and $g-2$ anomalies we are restricting to the scenario of decoupling limit, 
which implies that the 
large values of the leptonic Yukawa couplings will have a very small impact in 
the stability of up-type Higgs $H_u$ potential, whereas the conditons for the stability of the down type Higgs $H_d$ potential need to be determined. 
To discuss the stability of the scalar potential, one has to analyze its
quartic terms because they will dominate the behaviour of the scalar
potential in the region of very large values of the field components. To
this end, the quartic terms of the scalar potential are written in terms of
the Hermitian bilinear combination of the scalar fields. To simplify our
analysis, we discuss the stability conditions of the resulting 2HDM scalar
potential arising after the gauge singlet scalar field $\phi $ acquires a vacuum expectation value. Such stability conditions have been analyzed in detail in the framework of 2HDM in \cite{Maniatis:2006fs,Bhattacharyya:2015nca}. In order to analyze the stability 
of the $H_d$ potential, what we need to check if 
the quartic scalar couplings in each case of Table~\ref{tab:parameter_region_best_peaked} fullfill the stability conditions to be determined below. Given that our Higgs potential corresponds to the one of an extended 2HDM with the flavon field $\phi$, in order to apply the stability conditions used in the reference~\cite{Bhattacharyya:2015nca} to our Higgs potential, we need to reduce the number of scalar degrees of freedom by considering the resulting 2HDM scalar potential arising after the gauge singlet scalar field $\phi$ is integrated out. 
From the scalar potential it follows that the relevant quartic coupling constant $\lambda_6$ must be positive, otherwise the vev $v_3$ would fall into negative infinity when the field $\phi$ value increases. For the same reason, the quartic coupling constants $\lambda_{1,2}$ must also be positive. 
From the aforementioned stability conditions 
we conclude that the cases A and C must be excluded since their corresponding quartic coupling constants $\lambda_2$ are negative. Assuming the flavon field $\phi$ develops its vev $v_3$, we can rewrite the Higgs potential in terms of $H_u$ and $H_d$ fields as follows:
\begin{equation}
\begin{split}
V =&\mu _{1}^{2}\left( H_{u}H_{u}^{\dagger }\right) +\mu _{2}^{2}\left(
H_{d}H_{d}^{\dagger }\right) +\mu _{3}^{2}\frac{v_3^{2}}{2}
+2\mu _{\func{sb}}^{2}\frac{v_{3}^{2}}{2}
+\lambda _{1}\left( H_{u}H_{u}^{\dagger }\right) ^{2}+\lambda _{2}\left(
H_{d}H_{d}^{\dagger }\right) ^{2} \\
&+\lambda _{3}\left( H_{u}H_{u}^{\dagger }\right) \left(
H_{d}H_{d}^{\dagger }\right) +\lambda _{4}\left( H_{u}H_{d}^{\dagger
}\right) \left( H_{d}H_{u}^{\dagger }\right) +\lambda _{5}\left( \varepsilon
_{ij}H_{u}^{i}H_{d}^{j}\frac{v_{3}^{2}}{2}+\func{h.c}\right)  \\
&+\lambda _{6}\frac{v_{3}^{4}}{4}+\lambda _{7}\frac{v_{3}^{2}}{2} \left( H_{u}H_{u}^{\dagger }\right) +\lambda _{8}\frac{v_{3}^{2}}{2} \left( H_{d}H_{d}^{\dagger }\right).
\end{split}
\end{equation}%
Dropping all numbers and combining same order terms, the Higgs potential becomes much simpler as follows:
\begin{equation}
\begin{split}
V =&\left( \mu _{1}^{2} + \lambda _{7}\frac{v_{3}^{2}}{2} \right)\left( H_{u}H_{u}^{\dagger }\right) +\left( \mu _{2}^{2} + \lambda _{8}\frac{v_{3}^{2}}{2} \right)\left(
H_{d}H_{d}^{\dagger }\right) + \lambda _{1} \left( H_{u}H_{u}^{\dagger }\right) ^{2}+\lambda _{2}\left(
H_{d}H_{d}^{\dagger }\right) ^{2} \\
&+\lambda _{3}\left( H_{u}H_{u}^{\dagger }\right) \left(
H_{d}H_{d}^{\dagger }\right) +\lambda _{5} \frac{v_{3}^{2}}{2} \left( \varepsilon
_{ij}H_{u}^{i}H_{d}^{j}+\func{h.c}\right)  \\
\end{split}
\end{equation}%
where it is worth mentioning that the 
$\lambda_4$ term can be safely removed in the Higgs potential since it does not play a role in the CP-even, odd but charged mass matrix (Now our focus is the neutral scalar sectors). Here, we can 
impose one 
extra condition for the stabilization check, which is that the redefined mass terms must be negative, otherwise we get zero vev as a global minimum.
\begin{equation}
\begin{split}
&\mu_1^2 + \lambda_7 \frac{v_3^2}{2} = -2\lambda_1 v_1^2 - \lambda_3 v_2^2 + 2\lambda_3 v_2^2 = -2\lambda_1 v_1^2 + \lambda_3 v_2^2 < 0 \\
&\mu_2^2 + \lambda_8 \frac{v_3^2}{2} = -2\lambda_2 v_2^2 + \lambda_3 v_1^2 - \frac{1}{2} \lambda_8 v_3^2 + \lambda_8 \frac{v_3^2}{2} = -2\lambda_2 v_2^2 + \lambda_3 v_1^2 < 0
\label{eqn:redefined_mass_parameters}
\end{split}
\end{equation}
We have used the decoupling limit of Equation~\ref{eqn:constraints_in_mass_matrix_CP_even} at the first equality of Equation~\ref{eqn:redefined_mass_parameters}. From this equation, it is possible to 
determine the appropriate sign for the quartic coupling constant $\lambda_3$. In our numerical analysis, the vev $v_1$ is much dominant than the vev $v_2$ so it leads to a negative sign for the quartic coupling constant $\lambda_3$, otherwise the below equation of Equation~\ref{eqn:redefined_mass_parameters} would become positive. The sign of the quartic coupling constant $\lambda_3$ also determines the one of $\lambda_{5,7}$ in the decoupling scenario, which means that $\lambda_{5,7}$ must also be negative. On top of that, the large Yukawa coupling constants $y,x$ can be understood in connection with the vev $v_3$. To this end, we consider the definition for the Yukawa coupling constants $x_1$ and $x_2$, which are given by: 
\begin{equation}
x_{2} = \left\lvert \frac{y_{\mu} M_{44}}{y_{2} v_3} \right\rvert, \quad x_{1} = \left\lvert \frac{y_{e} M_{55}}{y_{1} v_3} \right\rvert,
\end{equation}
where in order to successfully explain both $g-2$ anomalies within the $1\sigma$ experimentally allowed range, one has to rely on small values of $v_3$, which are $\mathcal{O}(10\func{GeV})$, and the small values of $v_3$ do not significantly spoil the down-type Higgs $H_d$ potential as seen in Equation~\ref{eqn:redefined_mass_parameters}. In other words, the mass parameters $\mu_{1,2}^2$ are much larger than the 
parameters $\lambda_{7,8}v_3^2/2$, 
thus allowing more freedom in the sign of $\lambda_8$. 
Then, we are now ready to match our simplified Higgs potential with the one given in the reference~\cite{Bhattacharyya:2015nca}. Taking into consideration that our Higgs alignment is different than the one of ~\cite{Bhattacharyya:2015nca}, our mass parameters can be redefined as follows:
\begin{gather}
m_{11}^2 = \mu_1^2+\lambda_7 \frac{v_3^2}{2}, \quad m_{22}^2 = \mu_2^2+\lambda_8 \frac{v_3^2}{2}, \quad m_{12}^2 = \lambda_5 \frac{v_3^2}{2} \\
\beta_1 = 2\lambda_1, \quad \beta_2 = 2\lambda_2, \quad \beta_3 = \lambda_3, \quad \beta_4 = 0, \quad \beta_5 = 0
\end{gather}
Then, following \cite{Maniatis:2006fs,Bhattacharyya:2015nca}, it is found that the scalar potential is stable, when the following relations are fullfilled:
\begin{eqnarray}
\beta _{1} &\geq &0,\hspace{1.5cm}\beta _{2}\geq 0,\hspace{1.5cm}\beta _{3}+\sqrt{\beta _{1}\beta _{2}}\geq 0
\label{eqn:stability_condition_A}
\end{eqnarray}
\begin{equation}
\beta_3 + \beta_4 + \sqrt{\beta_1 \beta_2} > \lvert \beta_5 \rvert \rightarrow \beta_3 + \sqrt{\beta_1 \beta_2} > 0,
\label{eqn:stability_condition_C}
\end{equation}
The last stability condition can be rewritten as shown on the right side since the $\beta_{4,5}$ are zero in our Higgs potential and the cases B and D must be excluded by this 
last condition shown in Equation~\ref{eqn:stability_condition_C}. The conditions given in Eqs. (\ref{eqn:stability_condition_A}) and (\ref{eqn:stability_condition_C}) are crucial to guarantee the stability of the electroweak vacuum. Furthermore, one has to require that the squared masses for the
physical scalars are positive. Besides that, according to \cite{Bhattacharyya:2015nca}, the minimum of the scalar potential is a global
minumum when the following condition is fulfilled:
\begin{equation}
m_{12}^2 \left( m_{11}^2 - m_{22}^2 \sqrt{\frac{\beta_1}{\beta_2}} \right) \left( \tan\beta - \sqrt[4]{\frac{\beta_1}{\beta_2}} \right) > 0 \rightarrow m_{12}^2 \left( m_{11}^2 - m_{22}^2 \sqrt{\frac{\beta_1}{\beta_2}} \right) > 0
\end{equation}
where the latter condition on the left hand side is always successfully fulfilled for all cases, so we can simply drop off the condition as shown on the right side. Then, it is enough to confirm whether each case satisfies the reduced global minimum condition and the case E successfully fulfills that requirement as shown below: 
\begin{gather}
m_{12}^2 = -1763.9\func{GeV}^2, \quad m_{11}^2 = -7896.5\func{GeV}^2, \quad m_{22}^2 = -43258.8\func{GeV}^2, \quad \sqrt{\frac{\beta_1}{\beta_2}} = 0.0791994\\
m_{12}^2 \left( m_{11}^2 - m_{22}^2 \sqrt{\frac{\beta_1}{\beta_2}} \right) \approx 7.886 \times 10^6\func{GeV}^4 > 0
\end{gather}
Thus, we have numerically checked that the best fit point corresponding to the case E obtained in the numerical analysis of the scalar potential and $g-2$ muon and electron anomalies is consistent
with the above given stability conditions of the scalar potential and at the same time ensure positive values for the squared masses of the physical scalars, consistent with the current experimental data. Finally, to close this section, it is worth mentioning that  
the large Yukawa coupling constants $y,x$ involve 
the small vev $v_3$ in our model and 
this ensures that not only the $H_u$ potential is stable in the decoupling scenario but also the $H_d$ potential successfully fullfill the requirements of vacuum stability for 
both the small vev $v_3$ and appropriate values of the quartic scalar couplings.}

\subsubsection{How is the scalar exchange possible to accommodate both
anomalies at $1\protect\sigma$ constraint analytically?}

In order to 
analyze how the scalar exchange is able 
to explain both anomalies within the $1\sigma$ range, 
we revisit the analytic expressions for both muon and electron anomalous
magnetic moments:

\begin{equation}
\begin{split}
\Delta a_{\mu }=y_{24}^{e}x_{42}^{e}\frac{m_{\mu }^{2}}{8\pi ^{2}}\Big[&
\left( R_{e}^{T}\right) _{22}\left( R_{e}^{T}\right) _{32}I_{S}^{(\mu
)}\left( m_{e_{4}},m_{H_{1}}\right) +\left( R_{e}^{T}\right) _{23}\left(
R_{e}^{T}\right) _{33}I_{S}^{(\mu )}\left( m_{e_{4}},m_{H_{2}}\right) \\
& -\left( R_{o}^{T}\right) _{22}\left( R_{o}^{T}\right) _{32}I_{P}^{(\mu
)}\left( m_{e_{4}},m_{A_{1}}\right) -\left( R_{o}^{T}\right) _{23}\left(
R_{o}^{T}\right) _{33}I_{P}^{\left( \mu \right) }\left(
m_{e_{4},}m_{A_{2}}\right) \Big], \\
\Delta a_{e}=y_{51}^{e}x_{15}^{L}\frac{m_{e}^{2}}{8\pi ^{2}}\Big[& \left(
R_{e}^{T}\right) _{22}\left( R_{e}^{T}\right) _{32}I_{S}^{(e)}\left(
m_{e_{5}},m_{H_{1}}\right) +\left( R_{e}^{T}\right) _{23}\left(
R_{e}^{T}\right) _{33}I_{S}^{(e)}\left( m_{e_{5}},m_{H_{2}}\right) \\
& -\left( R_{o}^{T}\right) _{22}\left( R_{o}^{T}\right)
_{32}I_{P}^{(e)}\left( m_{e_{5}},m_{A_{1}}\right) -\left( R_{o}^{T}\right)
_{23}\left( R_{e}^{T}\right) _{33}I_{P}^{\left( E\right) }\left(
m_{e_{5}},m_{A_{2}}\right) \Big],
\label{eqn:muon_electron_anomaly_prediction}
\end{split}%
\end{equation}

where

\begin{equation}
I_{S\left( P\right) }^{\left( e,\mu \right) }\left( m_{E_{4,5}},m_{S}\right)
=\int_{0}^{1}\frac{x^{2}\left( 1-x\pm \frac{m_{E_{4,5}}}{m_{e,\mu }}\right) }{%
m_{e,\mu }^{2}x^{2}+\left( m_{E_{4,5}}^{2}-m_{e,\mu }^{2}\right) x+m_{S,P}^{2}\left(
1-x\right) }dx  \label{eqn:loop_integrals}
\end{equation}
with $S(P)$ corresponding to scalar (pseudoscalar) and $E_{4,5}$ standing for the vector-like family. Furthermore, $E_{4}$ and $E_{5}$ only contribute to the muon and electron anomalous magnetic moments, respectively.

First of all, we focus on the sign 
of each anomaly. The different signs of each anomaly indicated by the $%
1\sigma$ experimentally allowed range 
can be understood at the level of Yukawa constants apart from the loop
structures. As seen in Table \ref{tab:parameter_region_initial_scan}, the
Yukawa coefficient $y$ can be either positive or negative, while $x$ only
remains positive since we take the absolute value to the $x$. 
We also considered the case where the coefficients $x,y$ are purely
positive, assuming $v_3$ is positive, without taking absolute value and the
multiplication of the Yukawa coefficients $x \times y$ cannot change the
sign of each anomaly since the denominator of $x$ includes $y$ and they are
cancel out. Then, the sign problem depends on summing over loop functions and we found that the order of the muon anomaly prediction is
suitable, 
whereas the corresponding to the electron anomaly is about $10^{-16}$ which
is too small to be accommodated within the 
$1\sigma$ experimentally allowed range. Therefore, we found that taking an
absolute value to one of the Yukawa coefficients is an appropriate strategy
for the sign and allows to reproduce the correct order of magnitude of each
anomaly 
allowed by the $1\sigma$ experimentally allowed range, for an appropiate
choice of the model parameters. This feature is a crucial difference compared with the $W$ or $Z^\prime$ gauge boson exchange\cite{CarcamoHernandez:2019ydc}. The $W$ gauge boson exchange covered in the main body of this work keeps the same coupling constant at each vertex, therefore it is completelly different from 
the scalar exchange with vector-like leptons. For the $Z^\prime$ exchange covered in \cite{CarcamoHernandez:2019ydc}, it has the common property that the coupling constant of each vertex is different to each other, whereas the coupling constants of the $Z^\prime$ are more constrained by the mixing angle between $i$th chiral family and fourth vector-like family, so it is impossible to explain both anomalies at the same time. As a result, allowing different Yukawa constants with appropiate signs enables both anomalies to be explained in a unified way. 
\newline
~\newline
Next we turn our attention to the order of magnitude of our predictions for both anomalies.
Considering that the sign problem is solved by having each Yukawa constant $%
y $ either positive or negative, it can be easily understood that inside
the structure in parentheses of Equation \ref%
{eqn:muon_electron_anomaly_prediction} should imply the same direction, which is 
is determined by the contribution of all loop functions in
parentheses. Since the mass difference among non-SM scalars and vector-like
particles is not so big, we have to consider their masses in the computation of muon and electron anomalous magnetic moments, as follows from Equation \ref%
{eqn:muon_electron_anomaly_prediction}. 
For an easy analysis, we take the case E reported in Table \ref%
{tab:parameter_region_best_peaked} and suppose that

\begin{equation}
\begin{split}
\left( R_e^T \right)_{22} \left( R_e^T \right)_{32} = c_1, \quad \left(
R_e^T \right)_{23} \left( R_e^T \right)_{33} = -c_1, \quad &\left( R_o^T
\right)_{22} \left( R_o^T \right)_{32} = c_2, \quad \left( R_o^T
\right)_{23} \left( R_o^T \right)_{33} = -c_2, \\
I_S^{\mu} \left( m_{e_4}, m_{H_1} \right) = d_1, \quad I_S^{\mu} \left(
m_{e_4}, m_{H_2} \right) = d_2, \quad &I_P^{\mu} \left( m_{e_4}, m_{A_1}
\right) = -d_3, \quad I_P^{\mu} \left( m_{e_4}, m_{A_2} \right) = -d_4, \\
I_S^{e} \left( m_{e_5}, m_{H_1} \right) = e_1, \quad I_S^{e} \left( m_{e_5},
m_{H_2} \right) = e_2, \quad &I_P^{e} \left( m_{e_5}, m_{A_1} \right) =
-e_3, \quad I_P^{e} \left( m_{e_5}, m_{A_2} \right) = -e_4, \\
d_1 > d_3 > d_2 > d_4, &\quad e_1 > e_3 > e_2 > e_4
\end{split}%
\end{equation}

where $c_{1,2}$ are arbitrary constant between $0$ and $1$ either positive
or negative and mass ordering among $d(e)_i, (i=1,2,3,4)$ can be easily
understood by considering mass difference between non-SM scalars and
vector-like particles. The muon and electron anomaly prediction can be
rewritten in terms of these redefined constants:

\begin{equation}
\begin{split}
\Delta a_\mu = y_2 x_2 \frac{m_\mu^2}{8\pi^2} \left[ c_1 d_1 - c_1 d_2 + c_2
d_3 - c_2 d_4 \right] = y_2 x_2 \frac{m_\mu^2}{8\pi^2} \left[ c_1 \left( d_1
- d_2 \right) + c_2 \left( d_3 - d_4 \right) \right] = y_2 x_2 \frac{m_\mu^2%
}{8\pi^2} \left[ c_1 d_{12} + c_2 d_{34} \right] \\
\Delta a_e = y_1 x_1 \frac{m_e^2}{8\pi^2} \left[ c_1 e_1 - c_1 e_2 + c_2 e_3
- c_2 e_4 \right] = y_1 x_1 \frac{m_e^2}{8\pi^2} \left[ c_1 \left( e_1 - e_2
\right) + c_2 \left( e_3 - e_4 \right) \right] = y_1 x_1 \frac{m_e^2}{8\pi^2}
\left[ c_1 e_{12} + c_2 e_{34} \right]
\end{split}%
\end{equation}

where $y_2,x_2,y_1,x_1$ are simplified notation for $%
y_{24}^e,x_{42}^e,y_{51}^e,x_{15}^L$, respectively, and $d(e)_{ij} \equiv
d(e)_i - d(e)_j$ and $d(e)_{ij}$ are positive. Since the inside structure in
parentheses depends on relative magnitude of both $c_{1,2}$ and $d(e)_{ij}$
at this stage where no more analytic simplication is possible, it is good to
implement a specific value for them. Referring the values used to derive the
result of case E, they are

\begin{equation}
\begin{split}
y_2 x_2 \frac{m_\mu^2}{8 \pi^2} c_1 d_1 = -4.629 \times 10^{-7}, &\quad y_1
x_1 \frac{m_e^2}{8 \pi^2} c_1 e_1 = -8.532 \times 10^{-12} \\
-y_2 x_2 \frac{m_\mu^2}{8 \pi^2} c_1 d_2 = 4.520 \times 10^{-7}, &\quad
-y_1 x_1 \frac{m_e^2}{8 \pi^2} c_1 e_2 = 6.808 \times 10^{-12} \\
y_2 x_2 \frac{m_\mu^2}{8 \pi^2} c_2 d_3 = 7.984 \times 10^{-8}, &\quad y_1
x_1 \frac{m_e^2}{8 \pi^2} c_2 e_3 = 1.323 \times 10^{-12} \\
-y_2 x_2 \frac{m_\mu^2}{8 \pi^2} c_2 d_4 = -6.659 \times 10^{-8}, &\quad -y_1
x_1 \frac{m_e^2}{8 \pi^2} c_2 e_4 = -5.217 \times 10^{-13}
\label{eqn:middle_values_for_prediction}
\end{split}%
\end{equation}

and summing over all values in left or right column of Equation \ref%
{eqn:middle_values_for_prediction} yields the prediction for muon and
electron anomaly at $1\sigma$

\begin{equation}
\begin{split}
\Delta a_\mu &= y_2 x_2 \frac{m_\mu^2}{8\pi^2} \left[ c_1 d_1 - c_1 d_2 +
c_2 d_3 - c_2 d_4 \right] = 2.393 \times 10^{-9} \\
\Delta a_e &= y_1 x_1 \frac{m_e^2}{8\pi^2} \left[ c_1 e_1 - c_1 e_2 + c_2
e_3 - c_2 e_4 \right] = -9.232 \times 10^{-13}.
\end{split}%
\end{equation}

\section{CONCLUSION}

\label{sec:Conclusion}
We have proposed a model to account for the hierarchical structure of the SM Yukawa couplings.
In our approach 
the SM is an effective theory arising from a theory with extended particle
spectrum and symmetries.
The considered model includes 
an extension of the 2HDM where the particle spectrum is enlarged by the
inclusion of two vector-like fermion families, right handed Majorana neutrinos and a
gauge singlet scalar field, together with 
the inclusion of a global $U(1)^\prime$ symmetry spontaneously broken at the 
$\func{TeV}$ scale. 
Since the $U(1)^\prime$ symmetry is global, this model does not feature 
a $Z^\prime$ boson and it is softly broken in the 2HDM potential to avoid a Goldstone boson. 
Its main effect is to forbid SM Yukawa interactions due
to the $U(1)^\prime$ charge conservation. 
Besides that, this model has the
property of the 2HDM type II where one Higgs doublet couples with the up-type fermions whereas the remaining one has Yukawa interactions with down-type fermions, where such couplings are allowed between chiral fermions and vector-like fermions due to the choice of 
$U(1)^\prime$ charges (chiral fermions having zero charges while vector-like fermions, Higgs and flavons have charges $\pm 1$).
Below the mass scale of the vector-like fermions, such couplings result in effective Yukawa couplings suppressed by a factor 
$\left\langle \phi \right\rangle/M$ where the numerator is the vev of the flavon
and the denominator is the vector-like mass. This factor naturally determines
the magnitude of SM interactions and the mass scale for the vector-like
fermions under a suitable choice of the flavon vev. 
We have developed a mixing formalism based on $7 \times 7$ mass matrices to describe the mixing 
of the three chiral families with the two vector-like
families. 
\newline~\newline
Within the above proposed model, 
we have focused on accommodating the long-established muon and less established electron anomalous
magnetic moments at one-loop level. A main difficulty arises from the sign of each anomalous deviation of the experimental value from its SM prediction. Generally, the Feynman diagrams for the muon
and electron anomalous magnetic moments have the same structure except from the fact that the external particles are different, which makes it difficult to flip the sign of each contribution. Specifically we have required that both deviations in 
Equation \ref{eqn:deltaamu_deltaae_at_1sigma}) 
at one-loop should be accommodated within the $1\sigma$ experimentally allowed range,
which is a challenging requirement. %
\newline
~\newline
We first considered in detail the $W$ boson exchange contributions to the muon and electron 
anomalous magnetic moments at one-loop.
The relevant sector for the $W$ boson exchange is that of the neutrino and we analyzed a novel
operator that generates the masses of the light active neutrinos in this model. 
The well-known five dimensional Weinberg
operator which we refer as type Ia seesaw mechanism does not work in this model 
since it is forbiden by the $U(1)$ symmetry due to the fact that both $SU(2)$ scalar doublets are negatively charged under this symmetry. 
For this reason, we made use of the Weinberg-like operator known as
type Ib seesaw mechanism allowed in this model. With the type Ib seesaw
mechanism, we built the neutrino mass matrix with two vector-like neutrinos and ignored fifth vector-like neutrinos since they are too heavy to
contribute to the phenomenology. The deviation of unitarity $\eta$ derived
from the heavy vector-like neutrinos plays a crucial role for enhancing the
sensitivity of the CLFV $\mu \rightarrow e \gamma$ decay to the observable
level. Furthermore, the Yukawa constants of Dirac neutrino mass matrix can be connected to the observables measured in neutrino oscillation experiments. 
One of the neutrino Yukawa constants is defined with a
suppression factor $\epsilon$. Therefore, the effective $3 \times 3$
neutrino mass matrix tells that the tiny masses of the light active neutrinos depend on the mass scale of vector-like neutrinos as well as on the suppression factor $\epsilon$. %
This implies that mass scale of vector-like neutrinos is not required to be of the order 
of $10^{14}$ GeV, as in 
the conventional type Ia seesaw mechanism. 
In our proposed model, the vector-like neutrinos can have masses at the $\func{TeV}$ scale, thus allowing to test our model at colliders. Those vector-like neutrinos can be pair produced at the LHC via Drell-Yan annihilation mediated by a virtual $Z$ gauge boson. They can also be produced in association with a SM charged lepton via Drell-Yan annihilation mediated by a $W$ gauge boson. These heavy vector like sterile neutrinos can decay into a SM charged lepton and light active neutrinos. Thus, the heavy neutrino pair production at a proton-proton collider will give rise to an opposite sign dilepton final state, which implies that the observation of an excess of events in this final state over the SM background can be a smoking gun signature
of this model, whose observation will be crucial to assess its viability. 
It is confirmed that the branching ratio of $\mu \rightarrow e \gamma$ decay
can be expressed in terms of the deviation of unitarity $\eta$ as shown in 
\cite{Hernandez-Garcia:2019uof,Calibbi:2017uvl} and our prediction
for the muon and electron anomalous magnetic moments can also be written in
terms of non-unitarity. We derived the analytic expression for the anomalies
and found that the order of magnitude of these predictions is too small to accommodate the
experimental bound within the $1\sigma$ range and the sign of each prediction also points
out in the same direction. Therefore, we concluded that the $W$ boson exchange at
one-loop is not enough to explain both anomalies at $1\sigma$ and this
conclusion has been a good motivation to search for another possibility such
as scalar exchange, which is one of the main purposes of this work.  
\newline
~\newline
We then turned our attention to the 2HDM contributions (inclusion also of the singlet scalar $\phi$) to the muon and electron 
anomalous magnetic moments, assuming by a choice of parameters a diagonal charged lepton mass matrix to 
suppress the branching ratio of $\mu \rightarrow e \gamma$. In our analysis
we considered in detail the scalar sector of our model, which is composed of 
two $SU(2)$ scalar doublets $H_u$ and $H_d$ and one electrically neutral complex scalar $\phi$ by studying the corresponding scalar potential, deriving the squared mass matrices for the CP-even, CP-odd neutral and electrically charged scalars and determining the resulting scalar mass spectrum.
We have restricted to the scenario corresponding to the decoupling limit where 
no mixing between the physical SM Higgs $h$ and the physical non-SM scalars $H_{1,2}$
arise and within this scenario we have imposed the restrictions arising from 
the Higgs diphoton decay rate, the $hWW$ coupling, the $125$ GeV mass of the SM-like Higgs and the experimental lower bounds on non SM scalar masses, to determine the allowed parameter space consistent with the muon and electron anomalous magnetic moments. To this end, we have constructed a $\chi^2$ fitting function, which measures the deviation of the values of the physical observables obtained in the model, i.e., $(g-2)_{e,\mu}$, the $125$ GeV SM-like Higgs mass, the Higgs diphoton signal strength, the $hWW$ coupling, with respect to their experimental values. Its minimization allows to determine the values of the model parameters consistent with the measured experimental values of these observables. 
After saturating the $\chi^2$ value less than or nearly
2, we obtained five independent benchmark points and carried out second scan with
the benchmark points to find a correlation between observables and mass
parameters. For the plots, we took an appropriate case which is more converged when compared to other
ones and satisfying the vacuum stability conditions. We found that our prediction for both anomalies can be explained
within the $1\sigma$ constraint of each anomaly and a correlation 
proportional for muon versus electron anomaly is appeared in Figure \ref%
{fig:relevant_parameters_delamue} and \ref%
{fig:relevant_parameters_vl_delamue}. Here, we put two constraints on mass
of the lightest non-SM scalar and of the lightest vector-like family; $%
m_{H_1}, m_{e_5} > 200\func{GeV}$ based on references. The second scan result tells that the available parameter space is not significantly constrained by current experimental
results on non-SM scalar mass and vector-like mass, while keeping
perturbativity for quartic couplings and Yukawa constants. \Steve{An important feature of our BSM model is it predicts the large Yukawa coupling constants $y,x$, which might be able to destabilize the Higgs potential. The up-type Higgs $H_u$ potential is not significantly affected by the large Yukawa coupling constants in the decoupling scenario, whereas there is no safe condition for the down-type Higgs $H_d$ potential which can be worsen by mixing with the flavon field $\phi$. The large Yukawa coupling constants $x$ introduces small values for the vev $v_3$ in the definition of $x$ and the energy scale is confirmed by order of $10\func{GeV}$ in our numerical analysis. On top of that, we also identified the appropriate sign of quartic coupling constants can make the Higgs potential stable. Therefore, the down type $H_d$ Higgs potential is stable by both the small vev $v_3$ and the appropriate quartic coupling constants in our BSM model.} Lastly, we
discussed how we were able to explain both $(g-2)_{e,\mu}$ anomalies at $1\sigma$ constraint
and impact of the light non-SM scalar $H_1$. For the former, we first
simplified the prediction for both anomalies and used some numerical values at
the stage where no more analytic simplication is possible. For the latter,
we compared the cross section for the SM process $pp \rightarrow h$ and BSM process $pp \rightarrow H_1$ and included this comparison in Appendix~\ref{B}. 

We conclude that the proposed model of fermion mass hierarchies 
is able to successfully accommodate both the muon and electron anomalous magnetic moments within the $1\sigma$ experimentally allowed ranges, with the dominant contributions arising from one loop diagrams involving the 2HDM scalars and vector-like leptons.
The resulting model parameter space consistent with the $(g-2)_{e,\mu}$ anomalies requires masses of non-SM scalars and vector-like particles in the sub TeV and TeV ranges, thus making these particles accessible at the LHC and future colliders.

\section*{Acknowledgements}
We would like to thank Simon King for discussions.
This research has received funding from Fondecyt (Chile), Grant No.~1170803. SFK acknowledges the STFC
Consolidated Grant ST/L000296/1 and the European Union's Horizon 2020
Research and Innovation programme under Marie Sk\l {}odowska-Curie grant
agreements Elusives ITN No.\ 674896, HIDDeN European ITN project (H2020-MSCA-ITN-2019//860881-HIDDeN) and InvisiblesPlus RISE No.\ 690575.


\appendix

\section{Quark mass matrices in two bases}
\label{A}

As the lepton mass matrix is constructed in main body of this work, the
quark sector can be built in a similar way. Like the lepton sector, we make
use of two approaches to an effective lepton mass matrix, one of which is a
convenient basis and the other is a decoupling basis.

\subsection{A convenient basis for quarks}

Consider the $7 \times 7$ quark mass matrix rotated as in the lepton sector.

\begin{equation}
\begin{split}
M^{u}& =\left( 
\begin{array}{c|ccccccc}
& u_{1R} & u_{2R} & u_{3R} & u_{4R} & u_{5R} & \widetilde{Q}_{4R} & 
\widetilde{Q}_{5R} \\ \hline
\overline{Q}_{1L} & 0 & 0 & 0 & 0 & y_{15}^{u}v_{u} & 0 & x_{15}^{Q}v_{\phi }
\\ 
\overline{Q}_{2L} & 0 & 0 & 0 & y_{24}^{u}v_{u} & y_{25}^{u}v_{u} & 0 & 
x_{25}^{Q}v_{\phi } \\ 
\overline{Q}_{3L} & 0 & 0 & 0 & y_{34}^{u}v_{u} & y_{35}^{u}v_{u} & 
x_{34}^{Q}v_{\phi } & x_{35}^{Q}v_{\phi } \\ 
\overline{Q}_{4L} & 0 & 0 & y_{43}^{u}v_{u} & 0 & 0 & M_{44}^{Q} & M_{45}^{Q}
\\ 
\overline{Q}_{5L} & y_{51}^{u}v_{u} & y_{52}^{u}v_{u} & y_{53}^{u}v_{u} & 0
& 0 & 0 & M_{55}^{Q} \\ 
\overline{\widetilde{u}}_{4L} & 0 & x_{42}^{u}v_{\phi } & x_{43}^{u}v_{\phi }
& M_{44}^{u} & 0 & 0 & 0 \\ 
\overline{\widetilde{u}}_{5L} & x_{51}^{u}v_{\phi } & x_{52}^{u}v_{\phi } & 
x_{53}^{u}v_{\phi } & M_{54}^{u} & M_{55}^{u} & 0 & 0 \\ 
&  &  &  &  &  &  & 
\end{array}%
\right)  \\
M^{d}& =\left( 
\begin{array}{c|ccccccc}
& d_{1R} & d_{2R} & d_{3R} & d_{4R} & d_{5R} & \widetilde{Q}_{4R} & 
\widetilde{Q}_{5R} \\ \hline
\overline{Q}_{1L} & 0 & 0 & 0 & y_{14}^{d}v_{d} & y_{15}^{d}v_{d} & 0 & 
x_{15}^{Q}v_{\phi } \\ 
\overline{Q}_{2L} & 0 & 0 & 0 & y_{24}^{d}v_{d} & y_{25}^{d}v_{d} & 0 & 
x_{25}^{Q}v_{\phi } \\ 
\overline{Q}_{3L} & 0 & 0 & 0 & y_{34}^{d}v_{d} & y_{35}^{d}v_{d} & 
x_{34}^{Q}v_{\phi } & x_{35}^{Q}v_{\phi } \\ 
\overline{Q}_{4L} & 0 & 0 & y_{43}^{d}v_{d} & 0 & 0 & M_{44}^{Q} & M_{45}^{Q}
\\ 
\overline{Q}_{5L} & y_{51}^{d}v_{d} & y_{52}^{d}v_{d} & y_{53}^{d}v_{d} & 0
& 0 & 0 & M_{55}^{Q} \\ 
\overline{\widetilde{d}}_{4L} & 0 & x_{42}^{d}v_{\phi } & x_{43}^{d}v_{\phi }
& M_{44}^{d} & 0 & 0 & 0 \\ 
\overline{\widetilde{d}}_{5L} & x_{51}^{d}v_{\phi } & x_{52}^{d}v_{\phi } & 
x_{53}^{d}v_{\phi } & M_{54}^{d} & M_{55}^{d} & 0 & 0 \\ 
\end{array}%
\right) 
\end{split}%
\label{quark}
\end{equation}

Notice that the same rotations operated in the lepton sector is applied to
both up- and down-type quark sector except for $y_{14}^{d}$ since quark
doublet rotation is already used in the up-type quark sector. These two mass
matrices clearly tells that this model is an extended 2HDM in that the
up-type SM Higgs $H_u$ corresponds to up-type quark sector, while the
down-type SM Higgs $H_d$ corresponds for down-type quark sector. 
\subsection{A basis for decoupling heavy fourth and fifth vector-like family}

In this section, we treat the decoupling basis with quarks holding
an assumption $\left\langle \phi \right\rangle \approx M_{44}^Q$.
As in the charged lepton mass matrix, we can obtain the Yukawa matrix from
the $5\times 5$ upper blocks of Equation \ref{quark},

\begin{equation}
\widetilde{y}_{\alpha\beta}^{u} = 
\begin{pmatrix}
0 & 0 & 0 & 0 & y_{15}^{u} \\ 
0 & 0 & 0 & y_{24}^{u} & y_{25}^{u} \\ 
0 & 0 & 0 & y_{34}^{u} & y_{35}^{u} \\ 
0 & 0 & y_{43}^{u} & 0 & 0 \\ 
y_{51}^{u} & y_{52}^{u} & y_{53}^{u} & 0 & 0%
\end{pmatrix}%
, \hspace{0.1cm} \widetilde{y}_{\alpha\beta}^{d} = 
\begin{pmatrix}
0 & 0 & 0 & y_{14}^{d} & y_{15}^{d} \\ 
0 & 0 & 0 & y_{24}^{d} & y_{25}^{d} \\ 
0 & 0 & 0 & y_{34}^{d} & y_{35}^{d} \\ 
0 & 0 & y_{43}^{d} & 0 & 0 \\ 
y_{51}^{d} & y_{52}^{d} & y_{53}^{d} & 0 & 0%
\end{pmatrix}%
\end{equation}

where $\alpha$ and $\beta$ run from $1$ to $5$. The Yukawa matrices $%
\widetilde{y}_{\alpha \beta}^{u,d}$ can be diagonalized by the unitary
rotations $V$

\begin{equation}
V_Q = V_{45}^Q V_{35}^Q V_{25}^Q V_{15}^Q V_{34}^Q V_{24}^Q V_{14}^Q, 
\hspace{0.5cm} V_{u} = V_{45}^{u} V_{35}^{u} V_{25}^{u} V_{15}^{u}
V_{34}^{u} V_{24}^{u} V_{14}^{u}, \hspace{0.5cm} V_{d} = V_{45}^{d}
V_{35}^{d} V_{25}^{d} V_{15}^{d} V_{34}^{d} V_{24}^{d} V_{14}^{d}
\label{eqn:unitary_mixing_matrix}
\end{equation}

where each of the unitary matrices $V_{i4,5}$ are parameterized by a single
angle $\theta_{i4,5}$ featuring the mixing between the $i$th SM chiral quark
and the $4,5$th vector-like quark. In the rotated mass matrix, we need $%
(3,4), (1,5), (2,5), (3,5)$ mixing in the $Q$ sector and $(2,4), (3,4),
(1,5), (2,5), (3,5)$ mixing in the $u, d$ sectors to go to the decoupling
basis therefore the unitary mixing matrices $V$ are defined to be

\begin{equation}
\begin{split}
V_Q &= V_{35}^{Q} V_{25}^{Q} V_{15}^{Q} V_{34}^{Q} \\
&= 
\begin{pmatrix}
1 & 0 & 0 & 0 & 0 \\ 
0 & 1 & 0 & 0 & 0 \\ 
0 & 0 & c_{35}^Q & 0 & s_{35}^Q \\ 
0 & 0 & 0 & 1 & 0 \\ 
0 & 0 & -s_{35}^Q & 0 & c_{35}^Q%
\end{pmatrix}
\begin{pmatrix}
1 & 0 & 0 & 0 & 0 \\ 
0 & c_{25}^Q & 0 & 0 & s_{25}^Q \\ 
0 & 0 & 1 & 0 & 0 \\ 
0 & 0 & 0 & 1 & 0 \\ 
0 & -s_{25}^Q & 0 & 0 & c_{25}^Q%
\end{pmatrix}
\begin{pmatrix}
c_{15}^Q & 0 & 0 & 0 & s_{15}^Q \\ 
0 & 1 & 0 & 0 & 0 \\ 
0 & 0 & 1 & 0 & 0 \\ 
0 & 0 & 0 & 1 & 0 \\ 
-s_{15}^Q & 0 & 0 & 0 & c_{15}^Q%
\end{pmatrix}
\begin{pmatrix}
1 & 0 & 0 & 0 & 0 \\ 
0 & 1 & 0 & 0 & 0 \\ 
0 & 0 & c_{34}^Q & s_{34}^Q & 0 \\ 
0 & 0 & -s_{34}^Q & c_{34}^Q & 0 \\ 
0 & 0 & 0 & 0 & 1%
\end{pmatrix}%
, \\
&\approx 
\begin{pmatrix}
1 & 0 & 0 & 0 & s_{15}^Q \\ 
0 & 1 & 0 & 0 & s_{25}^Q \\ 
0 & 0 & 1 & s_{34}^Q & s_{35}^Q \\ 
0 & 0 & -s_{34}^Q & 1 & 0 \\ 
-s_{15}^Q & -s_{25}^Q & -s_{15}^Q & 0 & 1%
\end{pmatrix}%
, \\
&s_{34}^Q = \frac{x_{34}^Q \left\langle \phi \right\rangle}{\sqrt{\left(
x_{34}^Q \left\langle \phi \right\rangle \right)^2 + \left( M_{44}^Q
\right)^2}}, \hspace{0.5cm} s_{15}^Q = \frac{x_{15}^Q \left\langle \phi
\right\rangle}{\sqrt{\left( x_{15}^Q \left\langle \phi \right\rangle
\right)^2 + \left( M_{55}^Q \right)^2}}, \\
&s_{25}^Q = \frac{x_{25}^Q \left\langle \phi \right\rangle}{\sqrt{\left(
x_{25}^Q \left\langle \phi \right\rangle \right)^2 + \left( M_{55}^{\prime
Q} \right)^2}}, \hspace{0.5cm} s_{35}^Q = \frac{x_{35}^{\prime Q}
\left\langle \phi \right\rangle}{\sqrt{\left( x_{35}^{\prime Q} \left\langle
\phi \right\rangle \right)^2 + \left( M_{55}^{\prime\prime Q} \right)^2}}, \\
& x_{35}^{\prime Q} \left\langle \phi \right\rangle = c_{34}^{Q} x_{35}^{Q}
\left\langle \phi \right\rangle + s_{34}^{Q} M_{45}^{Q}, \hspace{0.5cm}
M_{45}^{\prime Q} = -s_{34}^{Q} x_{35}^{Q} \left\langle \phi \right\rangle +
c_{34}^{Q} M_{45}^{Q} \\
&\widetilde{M}_{44}^{Q} = \sqrt{\left( x_{34}^Q \left\langle \phi
\right\rangle \right)^2 + \left( M_{44}^Q \right)^2}, \\
&M_{55}^{\prime Q} = \sqrt{\left( x_{15}^Q \left\langle \phi \right\rangle
\right)^2 + \left( M_{55}^Q \right)^2}, \hspace{0.1cm} M_{55}^{\prime\prime
Q} = \sqrt{\left( x_{25}^Q \left\langle \phi \right\rangle \right)^2 +
\left( M_{55}^{\prime Q} \right)^2}, \hspace{0.1cm} \widetilde{M}_{55}^{Q} = 
\sqrt{\left( x_{35}^{\prime Q} \left\langle \phi \right\rangle \right)^2 +
\left( M_{55}^{\prime\prime Q} \right)^2}
\end{split}%
\end{equation}

\begin{equation}
\begin{split}
V_{u} &= V_{35}^{u} V_{25}^{u} V_{15}^{u} V_{34}^{u} V_{24}^{u} \\
&= 
\begin{pmatrix}
1 & 0 & 0 & 0 & 0 \\ 
0 & 1 & 0 & 0 & 0 \\ 
0 & 0 & c_{35}^{u} & 0 & s_{35}^{u} \\ 
0 & 0 & 0 & 1 & 0 \\ 
0 & 0 & -s_{35}^{u} & 0 & c_{35}^{u}%
\end{pmatrix}
\begin{pmatrix}
1 & 0 & 0 & 0 & 0 \\ 
0 & c_{25}^{u} & 0 & 0 & s_{25}^{u} \\ 
0 & 0 & 1 & 0 & 0 \\ 
0 & 0 & 0 & 1 & 0 \\ 
0 & -s_{25}^{u} & 0 & 0 & c_{25}^{u}%
\end{pmatrix}
\begin{pmatrix}
c_{15}^{u} & 0 & 0 & 0 & s_{15}^{u} \\ 
0 & 1 & 0 & 0 & 0 \\ 
0 & 0 & 1 & 0 & 0 \\ 
0 & 0 & 0 & 1 & 0 \\ 
-s_{15}^{u} & 0 & 0 & 0 & c_{15}^{u}%
\end{pmatrix}
\\
& \times 
\begin{pmatrix}
1 & 0 & 0 & 0 & 0 \\ 
0 & 1 & 0 & 0 & 0 \\ 
0 & 0 & c_{34}^{u} & s_{34}^{u} & 0 \\ 
0 & 0 & -s_{34}^{u} & c_{34}^{u} & 0 \\ 
0 & 0 & 0 & 0 & 1%
\end{pmatrix}
\begin{pmatrix}
1 & 0 & 0 & 0 & 0 \\ 
0 & c_{24}^{u} & 0 & s_{24}^{u} & 0 \\ 
0 & 0 & 1 & 0 & 0 \\ 
0 & -s_{24}^{u} & 0 & c_{24}^{u} & 0 \\ 
0 & 0 & 0 & 0 & 1%
\end{pmatrix}
\approx 
\begin{pmatrix}
1 & 0 & 0 & 0 & \theta_{15}^{u} \\ 
0 & 1 & 0 & \theta_{24}^{u} & \theta_{25}^{u} \\ 
0 & 0 & 1 & \theta_{34}^{u} & \theta_{35}^{u} \\ 
0 & -\theta_{24}^{u} & -\theta_{34}^{u} & 1 & 0 \\ 
-\theta_{15}^{u} & -\theta_{25}^{u} & -\theta_{35}^{u} & 0 & 1%
\end{pmatrix}%
, \\
&s_{24}^{u} \approx \frac{x_{42}^{u} \left\langle \phi \right\rangle}{%
M_{44}^{u}}, \hspace{0.5cm} s_{34}^{u} \approx \frac{x_{43}^{u} \left\langle
\phi \right\rangle}{M_{44}^{\prime u}}, s_{15}^{u} \approx \frac{x_{51}^{u}
\left\langle \phi \right\rangle}{M_{55}^{u}}, \hspace{0.5cm} s_{25}^{u}
\approx \frac{x_{52}^{\prime u} \left\langle \phi \right\rangle}{%
M_{55}^{\prime u}}, \hspace{0.5cm} s_{35}^{u} \approx \frac{x_{53}^{u}
\left\langle \phi \right\rangle}{M_{55}^{\prime\prime u}}, \\
& x_{52}^{\prime u} \left\langle \phi \right\rangle = c_{24}^{u} x_{52}^{u}
\left\langle \phi \right\rangle + s_{24}^{u} M_{54}^{u}, \hspace{0.5cm}
M_{54}^{\prime u} = -s_{24}^{u} x_{52}^{u} \left\langle \phi \right\rangle +
c_{24}^{u} M_{54}^{u}, \\
& x_{53}^{\prime u} \left\langle \phi \right\rangle = c_{34}^{u} x_{53}^{u}
\left\langle \phi \right\rangle + s_{34}^{u} M_{54}^{\prime u}, \hspace{0.5cm%
} M_{54}^{\prime\prime u} = -s_{34}^{u} x_{53}^{u} \left\langle \phi
\right\rangle + c_{34}^{u} M_{54}^{\prime u}, \\
&M_{44}^{\prime u} = \sqrt{\left( x_{42}^{u} \left\langle \phi \right\rangle
\right)^2 + \left( M_{44}^{u} \right)^2} \hspace{0.5cm}, \widetilde{M}%
_{44}^{u} = \sqrt{\left( x_{43}^{u} \left\langle \phi \right\rangle
\right)^2 + \left( M_{44}^{u} \right)^2}, \\
&M_{55}^{\prime u} = \sqrt{\left( x_{51}^{u} \left\langle \phi \right\rangle
\right)^2 + \left( M_{55}^{u} \right)^2}, \hspace{0.1cm} M_{55}^{\prime%
\prime u} = \sqrt{\left( x_{52}^{\prime u} \left\langle \phi \right\rangle
\right)^2 + \left( M_{55}^{\prime u} \right)^2}, \hspace{0.1cm} \widetilde{M}%
_{55}^{u} = \sqrt{\left( x_{53}^{\prime u} \left\langle \phi \right\rangle
\right)^2 + \left( M_{55}^{\prime\prime u} \right)^2}.
\end{split}%
\end{equation}

With the defined unitary mixing matrices in place, the $5 \times 5$ Yukawa
matrices in a mass basis (primed) are transformed by

\begin{equation}
\widetilde{y}_{\alpha\beta}^{\prime u} = V_Q \widetilde{y}_{\alpha\beta}^{u}
V_{u}^{\dagger}, \hspace{0.1cm} \widetilde{y}_{\alpha\beta}^{\prime d} = V_Q 
\widetilde{y}_{\alpha\beta}^{d} V_{d}^{\dagger},
\label{eqn:effective_Yukawa_constant}
\end{equation}

where tilde with prime means interaction basis whereas tilde alone corresponds to the mass
basis. The effective SM Yukawa couplings for the quarks then correspond to
the $3 \times 3$ upper block of $\widetilde{y}_{\alpha\beta}^{\prime u},%
\widetilde{y}_{\alpha\beta}^{\prime d}$, namely

\begin{equation}
y_{ij}^{u} \widetilde{H}_u \overline{Q}_{iL} u_{jR}, \hspace{0.1cm} y_{ij}^{d} 
\widetilde{H}_d \overline{Q}_{iL} d_{jR}, \hspace{0.1cm} \text{ with } y_{ij}^{u}
\equiv \widetilde{y}_{ij}^{\prime u}, \hspace{0.1cm} y_{ij}^{d} \equiv \widetilde{y}%
_{ij}^{\prime d}, \hspace{0.1cm} (i,j=1,2,3).
\end{equation}

The $3 \times 3$ SM Yukawa matrices for up- and down-type quark sector read:

\begin{equation}
\begin{split}
y_{ij}^u &= 
\begin{pmatrix}
s_{15}^Q y_{51}^u + y_{15}^u \theta_{15}^{u} & s_{15}^Q y_{52}^u + y_{15}^u
\theta_{25}^{u} & s_{15}^Q y_{53}^u + y_{15}^u \theta_{35}^{u} \\ 
s_{25}^Q y_{51}^u + y_{25}^u \theta_{15}^{u} & s_{25}^Q y_{52}^u + y_{24}^u
\theta_{24}^{u} + y_{25}^u \theta_{25}^{u} & s_{25}^Q y_{53}^u + y_{24}^u
\theta_{34}^{u} + y_{25}^u \theta_{35}^{u} \\ 
s_{35}^Q y_{51}^u + y_{35}^u \theta_{15}^{u} & s_{35}^Q y_{52}^u + y_{34}^u
\theta_{24}^{u} + y_{35}^u \theta_{25}^{u} & s_{34}^Q y_{43}^u + s_{35}^Q
y_{53}^u + y_{34}^u \theta_{34}^{u} + y_{35}^u \theta_{35}^{u}%
\end{pmatrix}
\\
y_{ij}^d &= 
\begin{pmatrix}
s_{15}^Q y_{51}^d + y_{15}^d \theta_{15}^{d} & s_{15}^Q y_{52}^d + y_{14}^d
\theta_{24}^{d} + y_{15}^d \theta_{25}^{d} & s_{15}^Q y_{53}^d + y_{14}^d
\theta_{34}^{d} + y_{15}^d \theta_{35}^{d} \\ 
s_{25}^Q y_{51}^d + y_{25}^d \theta_{15}^{d} & s_{25}^Q y_{52}^d + y_{24}^d
\theta_{24}^{d} + y_{25}^d \theta_{25}^{d} & s_{25}^Q y_{53}^d + y_{24}^d
\theta_{34}^{d} + y_{25}^d \theta_{35}^{d} \\ 
s_{35}^Q y_{51}^d + y_{35}^d \theta_{15}^{d} & s_{35}^Q y_{52}^d + y_{34}^d
\theta_{24}^{d} + y_{35}^d \theta_{25}^{d} & s_{34}^Q y_{43}^d + s_{35}^Q
y_{53}^d + y_{34}^d \theta_{34}^{d} + y_{35}^d \theta_{35}^{d}%
\end{pmatrix}%
\end{split}%
\end{equation}

\section{Heavy scalar production at a proton-proton collider}
\label{B}

We have confirmed that the mass of the non-SM CP even scalar $H_{1}$ is ranged from $%
200$ to $240\func{GeV}$ in Table \ref{tab:range_secondscan} and this light
mass of $H_{1}$ has not been observed at CERN or other experiments so far.
In order to see how big an impact of $H_{1}$ is when compared to that of SM
Higgs $h$, we studied a total cross section for the SM process $pp\rightarrow h$ and for BSM process $pp \rightarrow H_1$. The SM cross section for $pp\rightarrow h$ process is

\begin{equation}
\sigma_{\func{SM}} = \frac{\alpha_S^2 m_h^2}{64\pi v^2} \left( L\left( \frac{%
m_h^2}{m_t^2} \right) \right)^2 \frac{1}{S} \int_{\ln \sqrt{m_h^2/S}}^{-\ln 
\sqrt{m_h^2/S}} \func{PDF}(0,x_1(y),m_h) \func{PDF}(0,x_2(y),m_h) dy
\label{eqn:total_cross_section_pptoh}
\end{equation}

where $L$ is a loop integral

\begin{equation}
\begin{split}
L(a) &= \lvert \big[ 2a + (-4+a) \func{PolyLog}\left( 2,1/2\left( -\sqrt{-4+a%
} \sqrt{a} + a \right) \right) \\
&+ (-4+a) \func{PolyLog}\left( 2,1/2\left( \sqrt{-4+a} \sqrt{a} + a \right)
\right) \big]/a^2 \rvert,
\end{split}%
\end{equation}

$\alpha _{S}$ is the strong coupling constant, $v$ is the conventional SM Higgs vev $246.22\func{%
GeV}$, $m_{h}$ is the Higgs mass $125\func{GeV}$, $m_{t}$
is the top quark mass $173\func{GeV}$, $S$ is the squared LHC center of mass
energy $\left( 14\func{TeV}\right) ^{2}$, $\func{PDF}$ corresponds to the
parton distribution function where $0$ means 0th parton -
gluon, $x$ is the momentum fraction of the proton carried out by the gluon. 
Here the factorization scale has been taken to be equal to the SM like Higgs
boson mass $m_{h}$ 
and $x_{1,2}(y)$ are defined as follows:

\begin{equation}
x_1(y) = \frac{\sqrt{m_h^2/S}}{S} \exp(y), \quad x_2(y) = \frac{\sqrt{m_h^2/S%
}}{S} \exp(-y).
\end{equation}

With these defined functions and values, the total cross section for $pp
\rightarrow h$ is

\begin{equation}
\sigma_{\func{SM}} \simeq 18 \func{pb}.
\end{equation}

Next, the total cross section for $pp \rightarrow H_1$ process is

\begin{equation}
\sigma \left( pp\rightarrow H_{1}\right) =\frac{\alpha
_{S}^{2}m_{H_{1}}^{2}a_{hbb}^{2}}{64\pi v_2^2}\left( L\left( \frac{%
m_{H_{1}}^{2}}{m_{b}^{2}}\right) \right) ^{2}\frac{1}{S}\int_{\ln \sqrt{%
m_{H_{1}}^{2}/S}}^{-\ln \sqrt{m_{H_{1}}^{2}/S}}\func{PDF}(0,x_{1}^{\prime
}(y),m_{H_{1}})\func{PDF}(0,x_{2}^{\prime }(y),m_{H_{1}})dy
\label{eqn:total_cross_section_pptoH1}
\end{equation}

where $m_{H_1}$ is mass of non-SM CP even scalar $H_1$, and $x_{1,2}^\prime$
are defined in a similar way:

\begin{equation}
x_1^\prime (y) = \frac{\sqrt{m_{H_1}^2/S}}{S} \exp(y), \quad x_2^\prime (y)
= \frac{\sqrt{m_{H_1}^2/S}}{S} \exp(-y)
\end{equation}

\begin{figure}[H]
\centering
\includegraphics[keepaspectratio,scale=0.5]{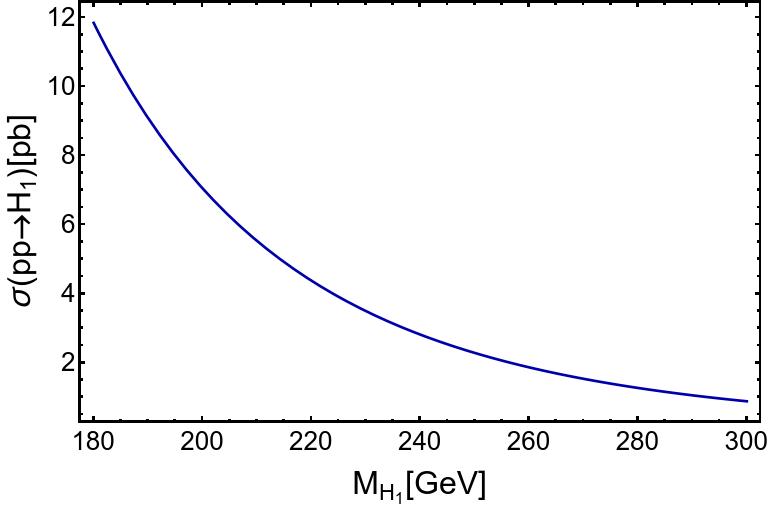}
\caption{The total cross section for $pp \rightarrow H_1$ at $14\func{TeV}$}
\label{fig:pptoH1at14TeV}
\end{figure}

One main distinction between Equation \ref{eqn:total_cross_section_pptoh}
and Equation \ref{eqn:total_cross_section_pptoH1} is the non-SM scalar $H_1$
only interacts with down-type quark pair $b\bar{b}$ since it is a mixed
state between $h_d^0$ and $\phi$ while the SM Higgs $h$ can interact with
top-quark pair $t\bar{t}$. According to the mass range of $H_1$ reported in
Table \ref{tab:range_secondscan}, the total cross section for $pp
\rightarrow H_1$ is given in Figure \ref{fig:pptoH1at14TeV}.

The total cross section for $pp \rightarrow H_1$ runs
from nearly $8\func{pb}$ at $200 \func{GeV}$ to smaller values as mass of $%
H_1$ increases. The order of magnitude of this cross section for $pp \rightarrow H_1$ is
compatible to that of the SM process $pp \rightarrow h$, however the BSM
process is strongly suppressed since its single LHC production via gluon fusion mechanism is dominated by the triangular bottom quark loop as mentioned in Section \ref{sec:Numerical_analysis_of_scalars}. Therefore, our prediction with the
light non-SM scalar $H_1$ is possible to accommodate each anomaly constraint
at $1\sigma$.


\end{document}